\documentclass[11pt,prd,showpacs,nofootinbib,preprint]{revtex4}
\usepackage[a4paper, total={7in, 10in}]{geometry}
\usepackage{color}
\usepackage{float}
\usepackage{booktabs}
\usepackage{multirow}
\usepackage{amsmath}
\usepackage{graphicx}
\usepackage{esint}
\usepackage{caption}
\usepackage{subcaption}
\usepackage{hyperref}

\makeatletter

\DeclareFontEncoding{LGR}{}{}

\ProvideTextCommand{\~}{LGR}[1]{\char126#1}

\newcommand{\lyxmathsym}[1]{\ifmmode\begingroup\def\b@ld{bold}
  \text{\ifx\math@version\b@ld\bfseries\fi#1}\endgroup\else#1\fi}

\providecommand{\tabularnewline}{\\}

\@ifundefined{textcolor}{}
{%
 \definecolor{BLACK}{gray}{0}
 \definecolor{WHITE}{gray}{1}
 \definecolor{RED}{rgb}{1,0,0}
 \definecolor{GREEN}{rgb}{0,1,0}
 \definecolor{BLUE}{rgb}{0,0,1}
 \definecolor{CYAN}{cmyk}{1,0,0,0}
 \definecolor{MAGENTA}{cmyk}{0,1,0,0}
 \definecolor{YELLOW}{cmyk}{0,0,1,0}
}

\usepackage{xcolor}\usepackage{epsfig}\usepackage{slashed}\usepackage{appendix}
\usepackage{tabularx}
\usepackage{pdfpages}
\usepackage{epstopdf}
\usepackage{multirow}
\usepackage{siunitx}
\usepackage{soul}

\makeatother

\begin{document}
\begin{center}
{\bf\Large\boldmath Probing New Physics in light of recent developments in $b \rightarrow c \ell \nu$ transitions}
\\[5mm]
\par
\end{center}

\begin{center}
\setlength{\baselineskip}{0.2in}{Tahira Yasmeen$^{a,}$\footnote{tahira709@gmail.com (corresponding author)}, Ishtiaq Ahmed$^{b,}$\footnote{ishtiaqmusab@gmail.com}, Saba Shafaq$^{a,}$\footnote{saba.shafaq@iiu.edu.pk}, Muhammad Arslan$^{c,}$\footnote{arslan.hep@gmail.com} and Muhammad Jamil Aslam$^{c,}$\footnote{jamil@qau.edu.pk}}\\[5mm] 
$^{a}$~\textit{Department of Physics, International Islamic University, Islamabad 44000, Pakistan.}\\
 $^{b}$~\textit{National Center for Physics, Islamabad 44000, Pakistan.}\\
 $^{c}$~\textit{Department of Physics, Quaid-i-Azam University, Islamabad 45320, Pakistan.}\\[5mm] 
\end{center}

\begin{abstract}
The experimental studies of the observables associated with the $b \rightarrow c$ transitions in the semileptonic $B-$ meson decays at BaBar, Belle and LHCb have shown some deviations from the Standard Model (SM) predictions, consequently, providing a handy tool to probe the possible new physics (NP). In this context, we have first revisited the impact of recent measurements of $R({D^{(*)}})$ and $R(\Lambda_c)$ 
on the parametric space of the NP scenarios. 
In addition, we have included the $R(J/\psi)$ data in the analysis  
and found that their influence on the best-fit points and the parametric space is mild. Using the recent HFLAV data, after validating the well established sum rule of $R(\Lambda_c)$, we derived the similar sum rule for $R(J/\psi)$. Furthermore, according to the updated data, we have modified the correlation among the different observables, giving us their interesting interdependence. Finally, to discriminate the various NP scenarios, we have plotted the different angular observables and their ratios for $B \to D^* \tau\nu_\tau$ against the transfer momentum square $\left(q^2\right)$, using the $1\sigma$ and $2\sigma$ parametric space of considered NP scenarios.  By implementing the collider bounds on NP Wilson coefficients, we find that, in the parametric space of some NP WCs is significantly restrained. To see the clear influence of NP on the amplitude of the angular observables, we have also calculated their numerical values in different $q^2$ bins and shown them through the bar plots.  We hope their precise measurements will help to discriminate various NP scenarios.
\end{abstract}

\maketitle

\section{Introduction}
The SM of particle physics successfully explained most of the experimental measurements; however, in semi-leptonic $B$-meson decays, $2 \sigma$- $4\sigma$ level deviations from SM predictions have been observed in recent measurements of $R({D^{(*)}})$, $R({J/\psi})$ and the $\tau$ polarization asymmetry \cite{Fedele:2022xyz, Blanke:2018yud,Blanke:2019qrx, Dutta:2013qaa,Dutta:2017xmj,Dutta:2017wpq,Dutta:2018jxz,Azatov:2018kzb,Heeck:2018ntp,Li:2016vvp,Aaij:2014ora,Caprini:1997mu,Bailey:2014tva,Sakaki:2013bfa}. These observables belong to the $b \rightarrow c \ell \nu_{\ell}$ transitions occurring through flavor-changing-charged-current (FCCC). Therefore, the observables belonging to the FCCC transitions are an excellent tool to check the SM predictions and to hunt for physics beyond it, i.e., the New Physics (NP). 

As we know that the theoretical predictions for the decay rates of semileptonic decays bear hadronic uncertainties mainly arising due to the form factors (non-perturbative quantities) and from Cabibbo–Kobayashi–Maskawa (CKM) matrix elements. However, in the ratios such as  $R({D^{(*)}})$ \cite{Ligeti:2016npd, Abdesselam:2019wbt, Abdesselam:2019dgh},  $R({J/\psi})$ \cite{Aaij:2017tyk,Murphy:2018sqg}, $R({X_c})$ \cite{Kamali:2018fhr,Grossman:1994ax, Colangelo:2016ymy}, and $\mathcal{R}({\Lambda_c})$ \cite{Aaij:2022abc,Detmold:2015aaa},
the dependence on the CKM elements and on the form factor cancels out.
From the experimental point of view, the ratios $R({D^{(*)}})$, are measured at the BaBar ~\cite{Lees:2012xj,Lees:2013uzd}, Belle~\cite{Huschle:2015rga,Sato:2016svk,Abdesselam:2016xqt,G.Caria:2019ff,Belle:2019rba} and LHCb~\cite{Aaij:2015yra,Aaij:2017uff} collaborations and the latest values of HFLAV world average \cite{HFLAV:2023link} shows an approximately $3.3\sigma$ deviations from their SM predictions \cite{Lees:2012xj,Lees:2013uzd,Fajfer:2012vx,Kamenik:2008tj, Amhis:2016xyh,Bailey:2015xy,Bailey:2012jg,Aoki:2016frl}. The other such observables are the ratios of decay rates of $B$ meson decaying to a polarized and unpolarized final state meson, i.e., $P_{\tau}( D^{*})$ and  $F_{L}(D^*)$. Their measurements at the Belle reported $1.5 - 2 \,\sigma$ deviation from their SM results \cite{Hirose:2016wfn,Hirose:2017dxl,Tanaka:2012nw,Asadi:2018sym,Alok:2016qyh,Tanaka:2010se}. Particularly,  $F_{L}(D^*)$ is important in probing different NP scenarios because the $D^{*}$ polarizations help us to distinguish between different Lorentz structures (scalar, vector and tensor operators) which influence its value \cite{Blanke:2018yud}. Similarly, the observable $R({J/\psi})$ has around $2\sigma$ deviation from its SM value $\approx0.23-0.29$ \cite{LHCbRJpsi,LHCbStatus,Watanabe:2017mip,Chauhan:2017uil,Cohen:2018dgz,Tran:2018kuv,Issadykov:2023afe,Azizi:2019aaf}. Although, its form factors are not precisely known, but to see its current impact on the parametric space, we have included this observable in our analysis. 
Additionally, like $R({D^{(*)}})$, the LHCb   \cite{Aaij:2022abc} collaboration has recently measured $\mathcal{R}(\Lambda_c)$ in $\Lambda_b \rightarrow \Lambda_{c}\tau^{-}\bar{\nu}_{\tau}$ decays.  Its opposite behavior compared to the $R({D^{(*)}})$ triggered a lot of theoretical interest, see e.g., \cite{Fedele:2022xyz} for an updated discussion. Finally, because of the lack of accurate measurement of the branching ratio of $B_c^- \rightarrow\tau^- \nu$ decay, the  lifetime of $B_c$ meson put some stringent constraints of the possible NP parameters \cite{Iguro:2018vqb,Adamczyk,Celis:2016azn,Asadi:2018wea,Asadi:2018sym}.
In this work, for the $\chi^2$ analysis we have used the recent measurements, given in TABLE \ref{Table1}, of the six observables discussed above, i.e.,  $R({D^{(*)}})$, $P_{\tau}( D^{*})$, $F_{L}(D^*)$, $R({J/\psi})$ and $\mathcal{R}(\Lambda_c)$. For the unobserved decay $B_c^- \rightarrow\tau^- \nu$, we use the $10\%$, $30\%$ and $60\%$ upper limits on its branching ratio  ~\cite{Alonso:2016oyd,Akeroyd:2017mhr,Gershtein:1994jw,Bigi:1995fs,Beneke:1996xe,Chang:2000ac,Kiselev:2000pp} in our analysis. 

Perhaps, it is useful to mention that in the earlier attempts the constraint on the parametric space of NP WC's are obtained by considering only the vector or scalar contributions separately \cite{Shi:2019gxi, Kamali:2018bdp, Gomez:2019xfw, Bhattacharya:2011qm, Antonelli:2008jg, Becirevic:2019tpx,Cardozo:2020abc}. However, Blanke \textit{et al.} have done a comprehensive analysis by considering scalar, vector and tensor couplings, but by using only the four experimentally measured observables, namely, $R({D^{(*)}})$, $P_\tau(D^*)$ and $F_{L}(D^*)$  \cite{Blanke:2018yud}. Including the recent measurement of $R(\Lambda_c)$ this analysis was revised by Fedele \textit {et al.} \cite{Fedele:2022xyz}. 

\begin{table}[H]
\centering{}%
\begin{tabular}{|c|c|c|}
\toprule  
\hline
\hline
Observables  & SM Predictions  & Experimental Measurements  \tabularnewline
\midrule 
\hline
\multirow{1}{*}{$R({D})$} & \multirow{1}{*}{$0.298\pm0.004$} \cite{HFLAV:2023link} & $0.344\pm0.026$ HFLAV \cite{HFLAV:2023link} 
 \tabularnewline
 \hline
\midrule 
\multirow{1}{*}{$R({D}^{*})$} & \multirow{1}{*}{$0.254\pm0.005$} \cite{HFLAV:2023link} & $0.285\pm0.012$ HFLAV \cite{HFLAV:2023link}  \tabularnewline
\midrule 
\hline
$P_{\tau}\left(D^{*}\right)$  & $-0.497\pm0.007$ \cite{Tanaka:2012nw}   & $-0.38\pm0.51_{-0.16}^{+0.21}$~\cite{Hirose:2017dxl,Hirose:2016wfn}  \tabularnewline
\midrule 
\hline
 $F_{L}\left(D^{*}\right)$  & $0.464\pm0.003$ \cite{Iguro:2022yzr}  & $0.60\pm0.08\pm0.04$\cite{Abdesselam:2019wbt} \tabularnewline
\midrule 
\hline
$R\left(J/\psi\right)$  & $0.258\pm0.038$ \cite{RJSHI:12,Harrison:2020gvo}, & $0.71\pm0.17\pm0.18$ \cite{Aaij:2017tyk} \tabularnewline
\midrule 
\hline
$R\left(\Lambda_{c}\right)$  & $0.324\pm0.004$  \cite{Fedele:2022xyz}& $0.242\pm0.026\pm0.040\pm0.059$ \cite{LHCb:2022piu} \tabularnewline
\bottomrule
\hline
\hline

\end{tabular}\caption{Different Physical observables, with their experimental measurements and the SM predictions.
} 
\label{Table1} 
\end{table}

The situation is robustly changing; theoretically we have a better control over the uncertainties of the form factors of $B\to D^{(*)}$ \cite{Cui:2023bzr,Iguro:2022yzr} decay, and experimentally after the recent measurements of Belle \cite{belle:2} and LHCb \cite{LHCB:1,LHCB:2} the HFLAV \cite{HFLAV:2023link} updated their earlier results accordingly. In addition, it will be interesting to redo the analysis by considering these updated values along with the measurements of $R(J/\psi)$ and $R(\Lambda_c)$. For this purpose, we include all these observables, mentioned above, together with their updated measurements in our fit analysis which are absent in previous studies \cite{Blanke:2018yud,Shi:2019gxi, Kamali:2018bdp, Gomez:2019xfw, Bhattacharya:2011qm, Antonelli:2008jg, Becirevic:2019tpx,Cardozo:2020abc,Fedele:2022xyz}. 

With this motivation, the main purpose of this work is not only to explore the allowed parametric space according to the current situation regarding $b\to c$ transitions but also to see the sensitivity of some angular observables to the NP models which may provide a tool to discriminate among different NP scenarios. To achieve this, we analyse the CP-even angular observables in $B \to D^*\ell \nu_\ell$ decays and to see their sensitivity on the NP couplings, we plotted them against the invariant dilepton mass $q^2$. 
We would like to mention here that LHC analysis of high-$pT$ mono-$\tau$ searches with missing transverse energy searches has also given the upper limits
on the NP WC's, so we have imposed these upper limits on the considered NP scenarios. We have also calculated their numerical values both in different $q^2$ bins and in the the full $q^2$ region.\\  
\textbf{Scheme of our Analysis:}
Some benchmarks of current analysis are described as:
\begin{itemize}
\item To accomplish the goal discussed above, we extend the SM weak effective Hamiltonian (WEH) for the charged current $b\rightarrow c \tau \nu$ by adding the new scalar, vector and tensor type contributions.
 \item  In the current study, the analysis has been done at 2 TeV by using the latest data of all available observables $R({D^{(*)}})$, $R(J/\psi),\; F_L(D^*),\; P_\tau(D^*)$ and $\mathcal{R}(\Lambda_c)$. In addition, for the comparison with Blanke \textit{et al.}, plots at 1TeV with updated measurements are also shown. The recipe of the analysis is similar to the Blanke \textit{et al.} \cite{Blanke:2018yud}.
\item Based on the choice of observables used for the fitting analysis for NP couplings, we consider the following cases:
\begin{itemize}
\item Fit A: $R({D^{(*)}})$, $F_L(D^*)$ , $P_\tau(D^*)$
\item Fit B: $R({D^{(*)}})$, $F_L(D^*)$, $P_\tau(D^*)$, $R(J/\psi)$,
\item Fit C: $R({D^{(*)}})$, $F_L(D^*)$, $P_\tau(D^*)$, $R(J/\psi)$, $\mathcal{R}(\Lambda_c)$
\end{itemize}
\item We validate the sum rule of $R(\Lambda_c)$ \cite{Blanke:2018yud,Fedele:2022xyz}, and update it by including the recent theoretical and experimental developments. Similarly, there is a large uncertainty in the value of $R(J/\psi)=0.71\pm0.17\pm0.18$ measured by LHCb collaboration \cite{LHCb:2017vlu} and to support the future experimental value with its theoretical predicted value $0.23-0.29$, we have also discussed the sum rule of $R(J/\psi)$ in terms of  $R({D^{(*)}})$. 
\item Furthermore, to see the discriminatory power of the observables under consideration, we have also found the correlation among different observables as a function of $R({D^{(*)}})$ in different two-dimensional (2D) NP scenarios.
\item Finally, using the $1\sigma$ and $2\sigma$ intervals of the NP couplings, we will calculate the numerical values of various CP-even observables in $B\to D^{*}\tau\nu_\tau$ decay \cite{Alok:2016qyh,Mandal:2020htr,Becirevic:2019tpx,Zhang:2020dla,Faustov:2022ybm} and discuss their potential to segregate different NP scenarios. We have also shown bar plots of these angular observables in different bins.
\end{itemize}
This paper is organized as follows: In Section \ref{TF}, after giving the effective Hamiltonian, we have listed the analytical expressions for the considered observables as a function of NP WCs. The fit procedure which is used to get the allowed values of different NP WCs has also been discussed in the same section. Section \ref{Fit} discusses 1D and 2D NP scenarios and their phenomenological analysis of the parametric space with and without considering the collider LHC bounds. The correlation among the observables and the sum rules are discussed in Section \ref{correlation}. In Section \ref{Senseang}, we check the sensitivity of CP even angular observables for different NP scenarios  and compare their values with the corresponding SM predictions by plotting them against $q^2$ by taking into account the LHC bounds. Finally, the bar plots are drawn to show their numerical values in different $q^2$ bins.
\section{Theoretical Formulation}\label{TF}
At quark level, we consider the following the WEH for $b \to c \ell \bar{\nu}_\ell$ transitions
\begin{equation}
{\cal H}_{\text{eff}}^{b\to c \ell \bar\nu_\ell} =  \frac{4G_F}{\sqrt{2}}V_{cb}\big [ \left(  1 + C_{V}^{L} \right){\cal O}_{V}^{L} +C_{V}^{R} {\cal O}_{V}^{R}
+C_{S}^{R} {\cal O}_{S}^{R} +C_{S}^{L} {\cal O}_{S}^{L}+C_{T} {\cal O}_T \big ]  + \text{h.c.} .
\label{eq:effH}
\end{equation}
where $\ell=\mu$, $\tau$, $G_F$ is the Fermi constant, $V_{cb}$ is the CKM matrix element and $C^{X}_{i}$ are the new WCs with $i=V,S,T$ and $X=L,R$. The corresponding quark level operators $\mathcal{O}^X_i$ are
\begin{eqnarray}
\mathcal{O}_{S}^{L}&=&\left(\bar{c}P_{L}b\right)\left(\bar{\ell} P_{L}\nu\right),\hspace{2cm}
\mathcal{O}_{S}^{R}=\left(\bar{c}P_{R}b\right)\left(\bar{\ell}P_{L} \nu\right),\nonumber\\ 
\mathcal{O}_{V}^{L}&=&\left(\bar{c}\gamma^{\mu}P_{L}b\right)\left(\bar{\ell}\gamma_{\mu} P_{L}\nu\right),\hspace{1.2cm}
\mathcal{O}_{V}^{R}=\left(\bar{c}\gamma^{\mu}P_{R}b\right)\left(\bar{\ell}\gamma_{\mu}P_{L} \nu\right),\nonumber\\ 
\mathcal{O}_{T}&=&\left(\bar{c}\sigma^{\mu\nu}P_{L}b\right)\left(\bar{\ell}\sigma_{\mu\nu}P_{L}\nu\right),\label{opt}
\end{eqnarray}
with $P_L = \frac{1-\gamma_5}{2}$ and $P_R = \frac{1+\gamma_5}{2}$. We know that experimentally no new states beyond the SM have been found so far up to an energy scale of approximately $1 \text{TeV}$. Also,the measurements of the Higgs couplings are all consistent with the SM expectations, therefore, the right-handed operators do not contribute in the SM \cite{Murgui:2019czp} making the coupling $C_V^R$ to be universal, which is strongly constrained from $b\to c(e,\mu)\bar\nu_{(e,\mu)}$ data. However, if the assumption of linearity of EWSB is relaxed then one can consider a non-universal $C_V^R$ coupling in the analysis, therefore, we included it and discuss this case separately in our analysis (see detail in ref. \cite{Jung:2018a,Murgui:2019czp}.
Moreover, in the absence of the experimental evidence of deviations from the SM in the tree-level transitions involving light leptons, it is assumed that the NP effects generally supposed to appear in the third generation of leptons \cite{Murgui:2019czp}.

The new WCs present in Eq. (\ref{eq:effH}) are calculated at $2\;\text{TeV}$, and these are related to $\mu$=$m_{b}$ scale as follows
\cite{Gonzalez-Alonso:2017iyc}:
\begin{align}
C_{V}^{L}(m_{b}) & = 1.12C_{V}^{L}(2\text{TeV}),\quad C_{V}^{R}(m_{b}) = 1.07C_{V}^{R}(2\text{TeV}),\quad C_{S}^{R}(m_{b})  = 2 C_{S}^{R}(2\text{TeV}),\nonumber \\
&\left(\begin{array}{c}
C_{S}^{L}(m_{b})\\
C_{T}(m_{b})
\end{array}\right)  =\left(\begin{array}{cc}
1.91 & -0.38\\
0 & 0.89
\end{array}\right)\left(\begin{array}{c}
C_{S}^{L}(2\text{TeV})\\
C_{T}(2\text{TeV})
\end{array}\right).\label{eq3}
\end{align}

\subsection{Analytical Expressions of the Observables}

By sandwiching the WEH given in Eq. (\ref{eq:effH}), the analytical expressions of the ratios $R({D^{(*)}})$, $R({J/\psi})$ and the observables depend on the polarization of final state particles, $F_L(D^*)$, $P_\tau(D)$ and $P_\tau(D^*)$ can be parameterized in terms of NP WCs as follows \cite{Asadi:2018sym,Iguro:2018vqb, Asadi:2018wea, Watanabe:2017mip, Gomez:2019xfw, Datta:2019abc, Murgui:2019czp, Detmold:2015aaa, Cardozo:2020abc, Iguro:2022yzr}: 
\begin{eqnarray}
R({D})&=&R_{D}^{SM}\Bigl\{\big|1+C_{V}^L+C_{V}^R\big|^{2}
+1.01\big|C_{S}^R+C_{S}^L\big|^2+1.49Re\big[\left(1+C_{V}^L+C_{V}^R\right)\left(C_{S}^R+C_{S}^L\right)^{*}\big]\notag \\ 
&+&0.84\big|C_{T}\big|^2+1.08Re\big[\left(1+C_{V}^L+C_{V}^R\right)\left(C_{T}\right)^{*}\big]\Bigl\}, \label{eq4}\\ 
R({D^*})&=&R_{D^*}^{SM}\Bigl\{\big| 1+C_{V}^L\big|^2+ \big| C_{V}^R\big|^2+0.04 \big|C_{S}^L-C_{S}^R\big|^2+16.0\big|C_{T}\big|^2-1.83Re\big[\left(1+C_{V}^L\right)C_{V}^{R^{*}}\big]\notag\\  &+&6.60Re[\left(C_{V}^R\right)C_{T}^{*}\big]-5.17Re\big[\left(1+C_{V}^L\right)C_{T}^{*}\big]+0.11Re \big[ \left( 1+C_{V}^L-C_{V}^R \right) \left( C_{S}^R-C_{S}^L \right)^{*} \big] \Bigl\},\label{eq5}\\ 
F_{L}(D^*)&=&F_{L}(D^*)^{SM}\left(\dfrac{R_{D^*}}{R_{D^*}^{SM}}\right)^{-1}\Bigl\{ \big|1+ C_{V}^L-C_{V}^R\big|^{2}+0.08\big| C_{S}^L-C_{S}^R\big|^{2}+6.90\big| C_{T}\big|^{2}\notag\\
&-&0.25Re\big[\left(1+C_{V}^L-C_{V}^R\right)\left(C_{S}^{L}-C_{S}^{R}\right)^{*}\big]-4.30 Re\big[\left(1+C_{V}^L-C_{V}^R\right)C_{T}^{*}\big]\Bigl\}, \label{eq6}\\
P_{\tau}(D)^{*}&=&{P_{\tau}( D^{*})^{SM}}\left(\dfrac{R_{D^*}}{R_{D^*}^{SM}}\right)^{-1}
\Bigl\{\big| 1+C_{V}^L\big|^2+ \big| C_{V}^R\big|^2-0.07\left(\big| C_{S}^R-C_{S}^L\big|^2\right)-1.85\big| C_{T}\big|^2 \notag\\ 
&+&0.23 Re\big[\left(1+C_{V}^L-C_{V}^R\right)\left(C_{S}^L-C_{S}^R\right)^{*}\big]-1.79Re \big[\left(1+C_{V}^L\right)C_{V}^{R^*}]-3.47Re \big[\left(1+C_{V}^L\right)C_{T}^{*}\big]\notag\\ 
&+& 4.41Re \big[\left(C_{V}^R\right)C_{T}^{*}
\big]\Bigl\},\label{eq7}\\
P_{\tau}(D)&=&{P_{\tau}( D)^{SM}}\left(\dfrac{R_{D^*}}{R_{D^*}^{SM}}\right)^{-1}
\Bigl\{\big| 1+C_{V}^L+ C_{V}^R\big|^2+3.04\left(\big| C_{S}^R+C_{S}^L\big|^2\right)+0.17\big| C_{T}\big|^2 \notag\\ 
&+&4.50 Re\big[\left(1+C_{V}^L+C_{V}^R\right)\left(C_{S}^L+C_{S}^R\right)^{*}\big]-1.09Re \big[\left(1+C_{V}^L+C_{V}^{R^*}\right)C_{T}^{*}],\label{eq8}
\end{eqnarray}
\begin{eqnarray}
R(J/\psi)&=&R(J/\psi)^{SM}\Bigl\{\big| 1+C_{V}^L\big|^{2}+\big| C_{V}^R\big|^{2}  -1.82Re\big[\left(1+C_{V}^L\right)C_{V}^{R*}\big]+0.04 \left(\big| C_{S}^L-C_{S}^R\big|^{2} \right)\notag\\
&+&0.10Re\big[\left(1+C_{V}^L-C_{V}^R\right)\left(C_{S}^R-C_{S}^{L}\right)^*\big]+14.7\big| C_{T}\big|^{2}-5.39Re\big[\left(1+C_{V}^L\right)C_{T}^{*}\big] \notag\\ 
&+&6.57Re\big[\left(C_{V}^R\right)C_{T}^{*}\big]\Bigl\},\label{eq9}\\
\mathcal{R}(\Lambda_c)&=&\mathcal{R}(\Lambda_c)^{SM}\Bigl\{\big|1+C_V^L\big|^2+\big| C_V^R\big|^2+0.491Re\big[\left(1+C_{V}^L\right)C_S^{R*}\big]
+0.316\big[\left(1+C_V^L\right)C_S^{L*}\big]\notag\\  
&+&0.484\big[\left(C_S^L\right)C_S^{R*}\big]
+0.31 \left( \big| C_S^L \big|^2 +\big|C_S^R\big|^2\right)-2.96Re\big[\left(1+C_V^L\right)C_T^{*}\big]+10.52\big|C_T\big|^2 \notag\\ 
&-&0.678Re\big[\left(1+C_V^L\right)C_V^{R*}\big]+0.316Re\big[\left(C_S^R\right)C_V^{R*}\big]+0.491Re\big[\left(C_S^L\right)C_V^{R*}\big]\notag\\
&+&4.85Re\big[\left(C_V^R\right)C_T^{*}\big]\Bigl\}.
\label{eq10}
\end{eqnarray}
Similarly, the branching ratio of $B_c \rightarrow \tau \nu$ decay can take the form \cite{Li:2016vvp, Alonso:2016oyd, Celis:2016azn, Akeroyd:2017mhr}:
\begin{equation}
\frac{Br (B_c \rightarrow \tau \nu )}{Br (B_c \rightarrow \tau \nu )|_{\rm SM}} = \left|1+ \left( C_V^{L}  - C_V^{R}\right)  +   \frac{m_{B_c}^2}{m_\tau (m_b+m_c)} \left( C_S^{R}  - C_S^{L}\right)  \right|^2.
\label{eq12}
\end{equation}
\subsection{Fit Procedure}

The standard $\chi^{2}$ analysis of the aforementioned observables for the decays governed by $b \rightarrow c \tau{\nu}$ transitions can be done by using 
\begin{equation*}
\chi^{2}(C_{i}^{X})=\sum_{l,m}^{N_{obs}}\left[O_{l}^{exp}-O_{l}^{th}(C_{i}^{X})\right]C_{lm}^{-1}\left[O_{m}^{exp}-O_{m}^{th}(C_{i}^{X})\right],
\end{equation*}
where $N_{obs}$ represents the number of observables, $O_{l}^{exp(th)}$ are the experimental (theoretical) values of the observables, and $C^{X}_{i}$ are the NP WCs. $C_{lm}$ is the covariance matrix incorporating the theoretical and experimental uncertainties.
However, instead of using covariance matrix, the $\chi^{2}$ function can be written in the form of pulls, i.e.,
$\chi^{2}$=$\sum_{i}^{N_{obs}}\left(\text{pull}_i\right)^2$,
where $\text{pull}_i=(O_\text{exp}^i-O_\text{th}^i)/\sqrt{\sigma_{\text{exp}}^{i 2}+\sigma_{\text{th}}^{i 2}}$. Here $\sigma_{\text{exp(th)}}^{i}$ shows the experimental (theoretical) error which are added in quadrature. The correlation of $R({D})$ and $R({D^{*}})$ has been taken into account by using the following relation
\[
\chi^2_{R(D)-R(D^{\ast})}= \dfrac{\text{pull}(R(D))^2+\text{pull}(R(D^{\ast}))^2-2\rho \text{pull}(R(D))\text{pull}(R(D^{\ast}))}{(1-\rho^2)}.\notag
\] 
The latest value of $R(D)-R(D^{\ast})$ correlation reported in ref. \cite{HFLAV:2023link} is $\rho= -0.39$, and for the uncorrelated observables, this value is zero.

For analysis, using latest data reported by HFLAV (c.f. TABLE \ref{Table1}), we first calculate the best fit points by minimizing the $\chi^{2}$ function ($\chi^{2}_{min}$) in the region of parameteric space that is compatible with the upper bound of $BR(B_c \rightarrow \tau \nu)$ $< 60\%$,  $< 30\%$ and $< 10\%$ \cite{Cardozo:2020abc}. $\chi^{2}_{min}$ is thus used to evaluate the p-values, which are the measure of goodness of fit and allows us to quantify the level of agreement between the data and the NP scenarios \cite{  Shi:2019gxi,Blanke:2018yud,Blanke:2019qrx}. The number of degrees of freedom (dof), $N_{dof}$=$N_{obs}-N_{par}$, where $N_{par}=1(2)$ for 1D (2D) scenarios while number of observables, $N_{obs}$ is the number of observables used in the fitting, i.e., $N_{obs}$ is $4,5$ and $6$ for the Fits A, B and C, respectively.  The SM Pull is defined as $pull_{SM}=\chi^{2}_{SM}-\chi^{2}_{min}$, where $\chi^{2}_{SM}$=$\chi^{2}(0)$,  which can be converted into an equivalent significance in units of standard deviations $(\sigma)$. 
\subsection{ Specific NP scenarios influenced by Leptoquark (LQ) Models} \label{NPSce}
Among the different NP models, the LQ models have recently gained attention to solve the $B-$Physics anomalies, therefore, in this study, we consider different 1D and 2D scenarios of LQ models as discussed in refs.
\cite{Alonso:2015sja,Calibbi:2015kma,Fajfer:2015ycq,Barbieri:2015yvd,Barbieri:2016las,Hiller:2016kry,Bhattacharya:2016mcc,Buttazzo:2017ixm,Kumar:2018kmr,Assad:2017iib,DiLuzio:2017vat,Calibbi:2017qbu,Bordone:2017bld,Barbieri:2017tuq,Blanke:2018sro,Greljo:2018tuh,Bordone:2018nbg,Matsuzaki:2018jui,Crivellin:2018yvo,DiLuzio:2018zxy,Biswas:2018snp,Deshpande:2012rr,Tanaka:2012nw,Sakaki:2013bfa,
Bauer:2015knc,Cai:2017wry,Crivellin:2017zlb,Altmannshofer:2017poe,Marzocca:2018wcf, He:2012zp,Greljo:2015mma,Boucenna:2016wpr,He:2017bft,Kalinowski:1990ba,Hou:1992sy,Kosnik:2012dj,Biswas:2018iak,Crivellin:2012ye,Crivellin:2013wna,Celis:2012dk,Ko:2012sv,Crivellin:2015hha,Dhargyal:2016eri,Chen:2017eby,Iguro:2017ysu,Martinez:2018ynq,Biswas:2018jun}:
\begin{itemize}
    \item For 1D scenarios: $C_{V}^{L}$, $C_{S}^L$, $C_{S}^R$, $C_{S}^{L}=4C_T$,
    \item For 2D scenarios: $(C_{V}^{L},C_{S}^{L}=-4C_T)$, $(C_{S}^L,C_{S}^R)$, $(C_{V}^L,C_{S}^R)$, $(\text{Re}[C_{S}^L=4C_T],\text{Im}[C_{S}^L=4C_T])$.
\end{itemize}
The combinations arise from the $C_V^R$ term will be discussed in the upcoming section separately.  
\section{Allowed parametric space in 1D and 2D LQ scenarios}\label{Fit}
In this section, we perform the $\chi^2$ analysis of above mentioned 1D and 2D LQ scenarios with the latest HFLAV data, and by using the fitting procedure discussed in previous section.
\subsection{1D Scenarios}
\begin{figure}[H]
  \begin{subfigure}[b]{1\textwidth}
  \centering
\includegraphics[width=\textwidth]{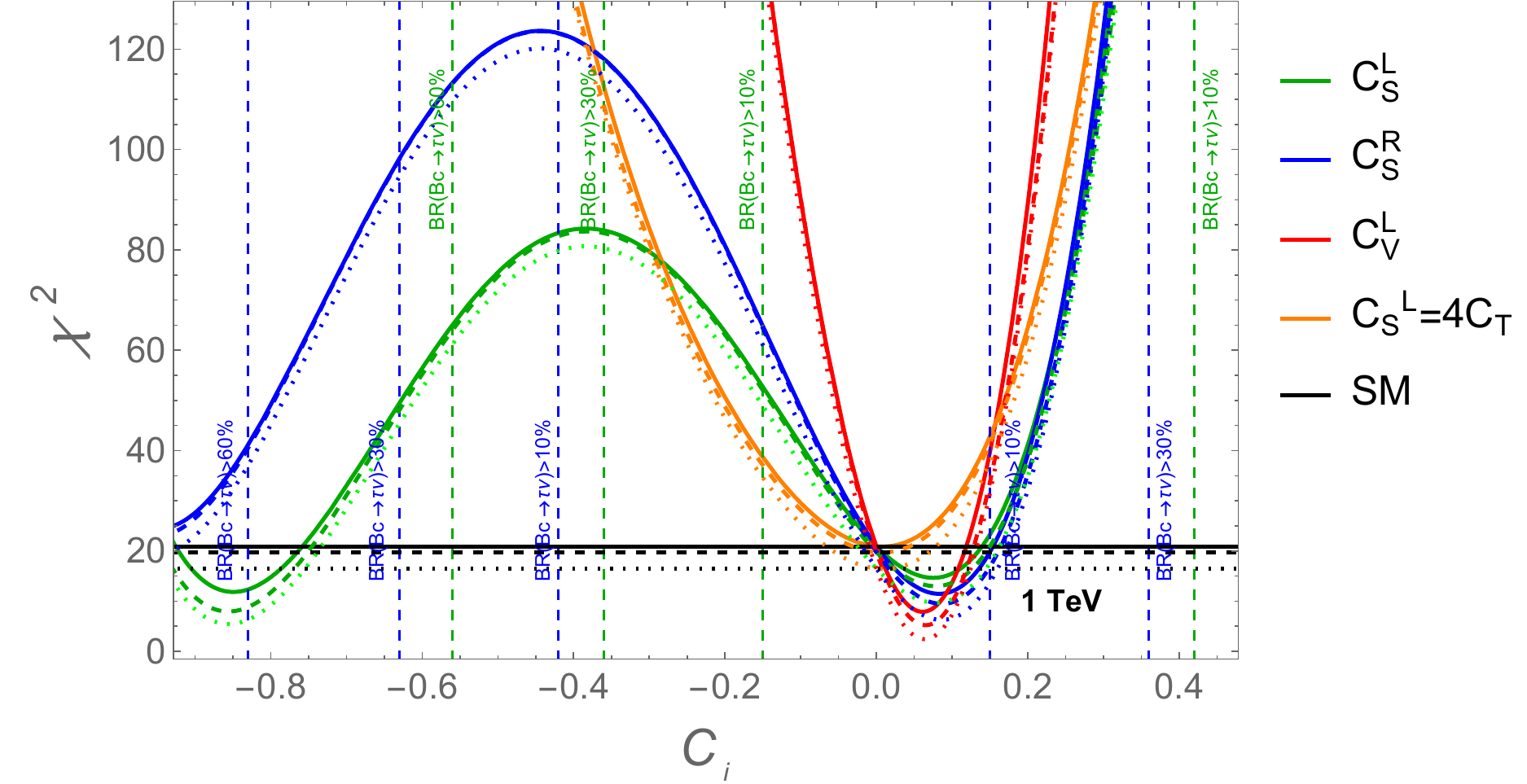}
\includegraphics[width=\textwidth]{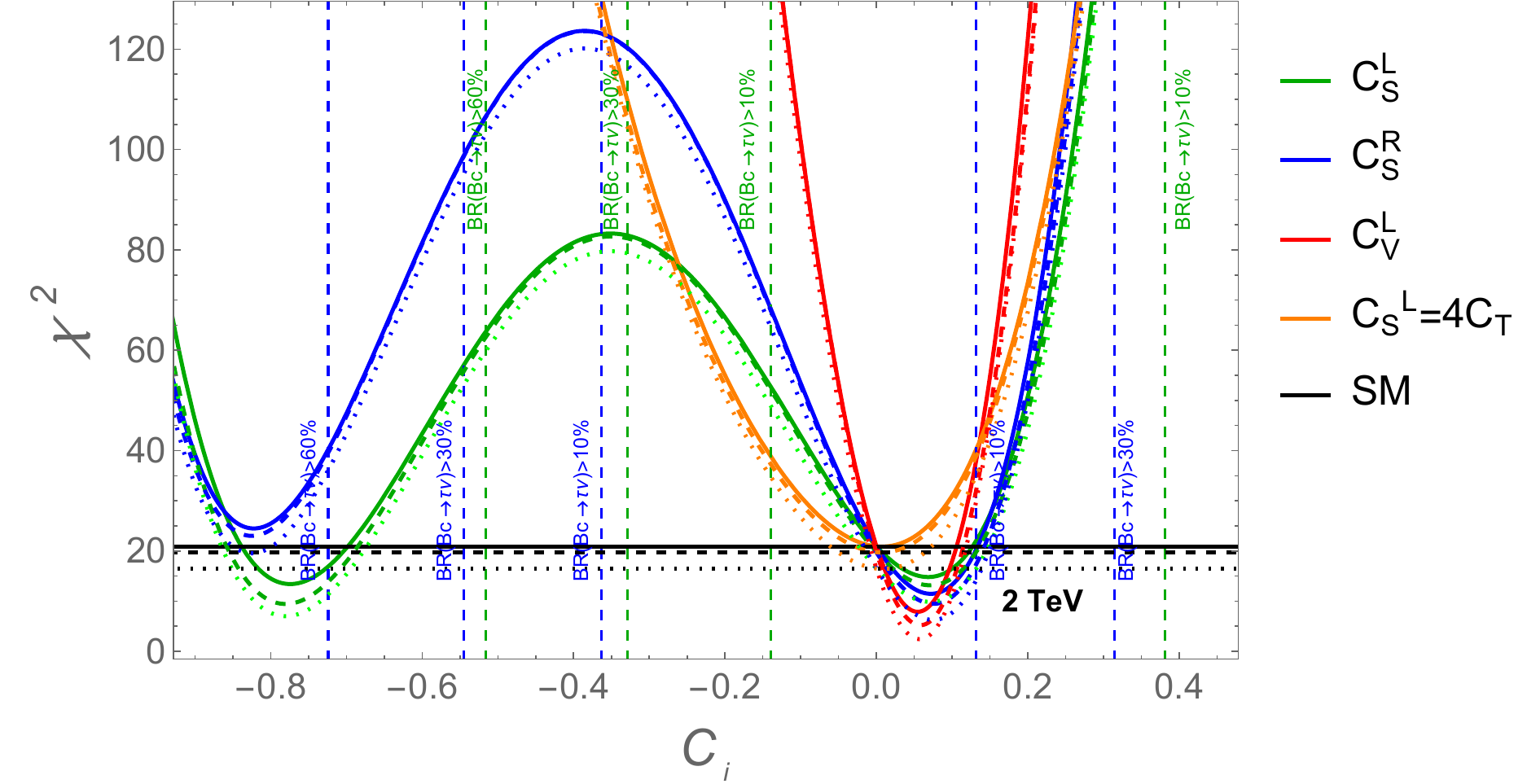}
\label{Chi-square-HF}
\end{subfigure}
\hfill
\caption{Dependence of $\chi^{2}$ with one Wilson coefficient active at a time for different Fits at 1 and 2 TeV scales. The colored dotted, dashed and solid lines represent Fits A, B and C, respectively, while the black lines represent their SM values. The dashed vertical lines correspond to the constraint for $C_S^{L,R}$ from $B_c \rightarrow \tau \nu$ assuming the values of $10\%$ and $30\%$ and $60\%$.}
\label{couplingchi2}
\end{figure}

In FIG. (\ref{couplingchi2}), we have shown the $\chi^{2}$ dependence on the SM and NP WCs for 1D scenarios at 1 and 2 TeV scales. The dashed vertical lines correspond to the constraints on $C_i^X$ from the different upper bounds of $BR(B_c \rightarrow \tau \nu)$. The dotted, dashed and solid curves represent the cases of Fits A,  B, and C, respectively. From this FIG. \ref{couplingchi2}, one can notice that the positive best fit points of the NP models are not significantly changed with respect to the scales of new WC's, while the negative best fit points and the vertical lines of $BR(B_c \rightarrow \tau \nu)$ are slightly shifted to the right side.  It can also be noticed that the updated data is indicating that the negative solutions of best fit points for $C_{S}^L$ and $C_{S}^R$ are still excluded by the maximum upper limit of $BR(B_c \rightarrow \tau \nu)\leq60\%$ as reported by Blanke \textit{et al.} \cite{Blanke:2018yud, Blanke:2019qrx}, and this is also disfavored with respect to their SM values. One can also see from the plot that near the positive best fit point, the $C_{V}^L$ and $C_{S}^R$ still favourable to explain the data, while the favourable situation of $C_{S}^L$  is further improved with the new data.

In columns 2 - 5 of TABLE \ref{1D}, we have listed the numerical values of the Best fit points, $\chi^2_{min}$,  p-value and $Pull_{SM}$ in different Fits for 1D scenarios. Here the first, second and third rows represent Fit A, B and C, respectively. The last seven columns show the predictions of different observables at the best fit point with the $\sigma$ deviation. $\chi^2_{SM}$ and p-value are also given at the top of the TABLE \ref{1D} for all Fits. It is worth mentioning here that we have calculated these numerical values on both 1 and 2 TeV scales and found that these values are not significantly changed.
The slight dependence of best fit point of $C_S^R$ (see TABLE I of \cite{Blanke:2018yud}) on $BR(B_c \rightarrow \tau \nu)$  has also disappeared for the new data. The p-values, $Pull_{SM}$ and predictions of different observables for 1D scenarios are also tabulated in TABLE \ref{1D}. The 1$\sigma$ and 2$\sigma$ intervals for WCs have also been calculated and are listed in TABLE \ref{2sigma}. One can see that the p-values for all 1D scenarios are improved by considering the new experimental data, particularly, for $C_V^L$ ($C_S^R$). These two scenarios are significantly enhanced and moved up to $48.8\%$ ($9.96\%$) for the Fit A ($N_{obs}=4$) which previously found to be $35\%$ $(\simeq2\%)$ \cite{Blanke:2018yud}. This shows that $C_V^L$ ($C_S^R$) are still favorable. Despite, the p-value of $C_{S}^L=4C_T=(0.06-0.09)\%$ slightly improved, this scenario still describes the data poorly. 

We can see from TABLE \ref{1D} that with increasing the number of observables, $\chi^2_{min}$ (p-value) is increased (decreased), consequently, the pull is reduced with all Fit cases under consideration except $C_V^{R}$ . This translates that p-value decreases, i.e., the goodness of the fit is reduced when we include the data of $R(J/\psi)$ and $R(\Lambda_c)$ in the analysis, except for the scenarios ($C_{S}^{L}$,$C_{S}^{L}=4C_{T}$). This is attributed to the large experimental uncertainty in the measurement of $R(J/\psi)$ and inconsistency of the measurement of $R(\Lambda_c)$ with respect to $R({D^{(*)}})$.
\begin{table}[H]
\centering{}%
\begin{tabular}{|c|c|c|c|c|c|c|c|c|c|c|c|}
\toprule
\multicolumn{12}{c}{$\chi^{2}_{SM}=(16.50,19.76,20.92)_{A,B,C}$, $p-value=(2.41,1.38,1.89)_{A,B,C}\times10^{-3}$, 
}\tabularnewline
\hline
\hline
Scenarios & Best fit & $\chi^2_{min}$ & $p-value\%$ & pull$_{SM}$  & $R(D)$  & $R(D^{*})$  & $R(J/\psi)$   &  $F_L (D^{*})$  & $P_{\tau}(D^{*})$& $P_{\tau}(D)$&$R(\Lambda_c)$
\tabularnewline
\midrule
\midrule 
\hline
\multirow{3}{*}{$C_{S}^{L}$} & \multirow{3}{*}{0.07}&9.85 &1.98 &2.58   &\multirow{2}{*}{$0.362$} &\multirow{2}{*}{$0.250$} & \multirow{2}{*}{$0.254$} &\multirow{2}{*}{ $0.455$} &\multirow{2}{*}{ $-0.518$} & \multirow{2}{*}{0.489}&\multirow{2}{*}{$0.339$}
\tabularnewline
 &  &13.16 &1.03 &2.56   &$0.69\sigma$ & $-2.91\sigma$ & $-1.84\sigma$ & $-1.62\sigma$ & $-0.25\sigma$  & & $1.27\sigma$
 \tabularnewline
  &  & 14.80&1.12 &2.47 &   & &  &   &  &  & 
\tabularnewline
\hline
\multirow{3}{*}{$C_{S}^{R}$} & \multirow{1}{*}{0.07} &6.26 &9.96 &3.20  &\multirow{2}{*}{$0.366$} & \multirow{2}{*}{$0.258$} & \multirow{2}{*}{$0.261$} &\multirow{2}{*}{ $0.473$} & \multirow{2}{*}{$-0.472$} & \multirow{2}{*}{0.498}&\multirow{2}{*}{$0.348$}
\tabularnewline
& \multirow{1}{*}{0.08} & 9.46& 5.05&3.21 &  $0.85\sigma$ & $-2.25\sigma$ & $-1.83\sigma$ & $-1.42\sigma$ & $-0.17\sigma$  & $ $ & $1.39\sigma$
\tabularnewline
& \multirow{1}{*}{0.07}  & 11.46&4.29 &3.07 &  &  &  & &  &   & 
\tabularnewline
\hline
\multirow{3}{*}{$C_{V}^{L}$} & \multirow{1}{*}{0.06} &2.43 &48.8 &3.75 &  \multirow{2}{*}{$0.332$} & \multirow{2}{*}{$0.283$} & \multirow{2}{*}{$0.287$} & \multirow{2}{*}{$0.464$} & \multirow{2}{*}{$-0.497$} & \multirow{2}{*}{$0.331$} & \multirow{2}{*}{$0.361$}
\tabularnewline
& \multirow{1}{*}{0.06}   & 5.20&26.7 &3.82 & $-0.46\sigma$ & $-0.17\sigma$ & $-1.71\sigma$ & $-1.52\sigma$ & $-0.21\sigma$ &  & $1.57\sigma$
 \tabularnewline
 &  \multirow{1}{*}{0.05}  & 7.87&16.4 &3.61 &  &   &  &  &  &  &
\tabularnewline
\hline
\multirow{3}{*}{$C_{V}^{R}$} & \multirow{1}{*}{-0.04} &13.62 &0.35 &1.69&  \multirow{2}{*}{$0.273$} & \multirow{2}{*}{$0.274$} & \multirow{2}{*}{$0.278$} & \multirow{2}{*} {$0.467$} & \multirow{2}{*}{$-0.496$} & \multirow{2}{*}{$0.331$}&\multirow{2}{*}{$0.333$}
\tabularnewline
& \multirow{1}{*}{-0.05} & 16.54&0.24 &1.79 &  $-2.73\sigma$ & $-0.92\sigma$ & $-1.74\sigma$ & $-1.49\sigma$ & $-0.21\sigma$ &  & $1.19\sigma$
 \tabularnewline
  & \multirow{1}{*}{-0.04} & 18.05&0.28 &1.69 &  &   &  &  &  &  &
  \tabularnewline
\hline
\multirow{3}{*}{$C_{S}^{L}=4C_T$} & \multirow{1}{*}{0.009} &16.41 &0.09 &0.30 &  \multirow{2}{*}{$0.304$} & \multirow{2}{*}{$0.251$} & \multirow{2}{*}{$0.255$} & \multirow{2}{*}{$0.463$} &\multirow{2}{*}{ $-0.500$} & \multirow{2}{*}{$0.341$}&\multirow{2}{*}{$0.323$}
\tabularnewline
&  \multirow{1}{*}{0.007}  &19.71 & 0.06&0.22 &  $-1.53\sigma$ & $-2.83\sigma$ & $-1.83\sigma$ & $-1.53\sigma$ & $-0.22\sigma$ & & $1.07\sigma$
 \tabularnewline
 &  \multirow{1}{*}{0.007} & 20.87&0.08&0.22 &  &  & &  &  &    &
  \tabularnewline
  \hline
\hline
\end{tabular}
\caption{Fit results for 1D scenarios by using all available data. The best fit points, $\chi^2_{min}$,  p-value, $Pull_{SM}$ are given. The columns (2-5) represent the results for different parameters: First, second and third rows in these columns represent Fits A, B and C, respectively. The last seven columns show the predictions of different observables at the best fit point with the $\sigma$ deviation. $\chi^2_{SM}$ and p-value are also given at the top of the table for all Fits.}
\label{1D}
\end{table}  
For the best fit points of the NP scenario the theoretically predicted values of the observables are given in TABLE \ref{1D} (last seven columns). Using the relation \cite{Blanke:2018yud}
\begin{equation}
	\text{d}_{O_i} =  \frac{O^\text{NP}_i -O^\text{exp}_i}{\sigma^{O^\text{exp}_i}},\
	\label{discrepancy}
\end{equation}
we have also tabulated their discrepancies from the corresponding experimental values.
The results can be concluded from the above table as:  
 For $R(D)$ and $P_\tau(D^*)$  the deviations are found to be less than $1\sigma$ for all NP scenarios except $R(D)$ in  $C_S^L=4C_T$ ($C_V^R$) scenario, where it is found to be $-1.53\sigma$ ($-2.73\sigma$). On the other hand, for the observables, $R({D^*})$, $R(J/\psi)$, $F_L(D^*)$ and $R(\Lambda_c)$, the deviations are between $1 - 3\sigma$ , except for $R{(D^*)}$ in $C_V^L$ which is $-0.17\sigma$. It is also important to mention here that the values of these observables are mildly effected by changing $N_{obs}$ in the analysis.
\subsection{2D scenarios}

In this section, we will perform the phenomenological analysis of 2D NP scenarios that are  defined in section \ref{NPSce}, which are generated by exchanging a single new LQ or a Higgs particle. In this case, for the $\chi^2$ analysis, we have considered $N_{par}=2$.

\begin{figure}[H]
 \centering{}
     \begin{subfigure}[b]{0.35\textwidth}
         \centering
         \includegraphics[width=7cm,height=6cm]{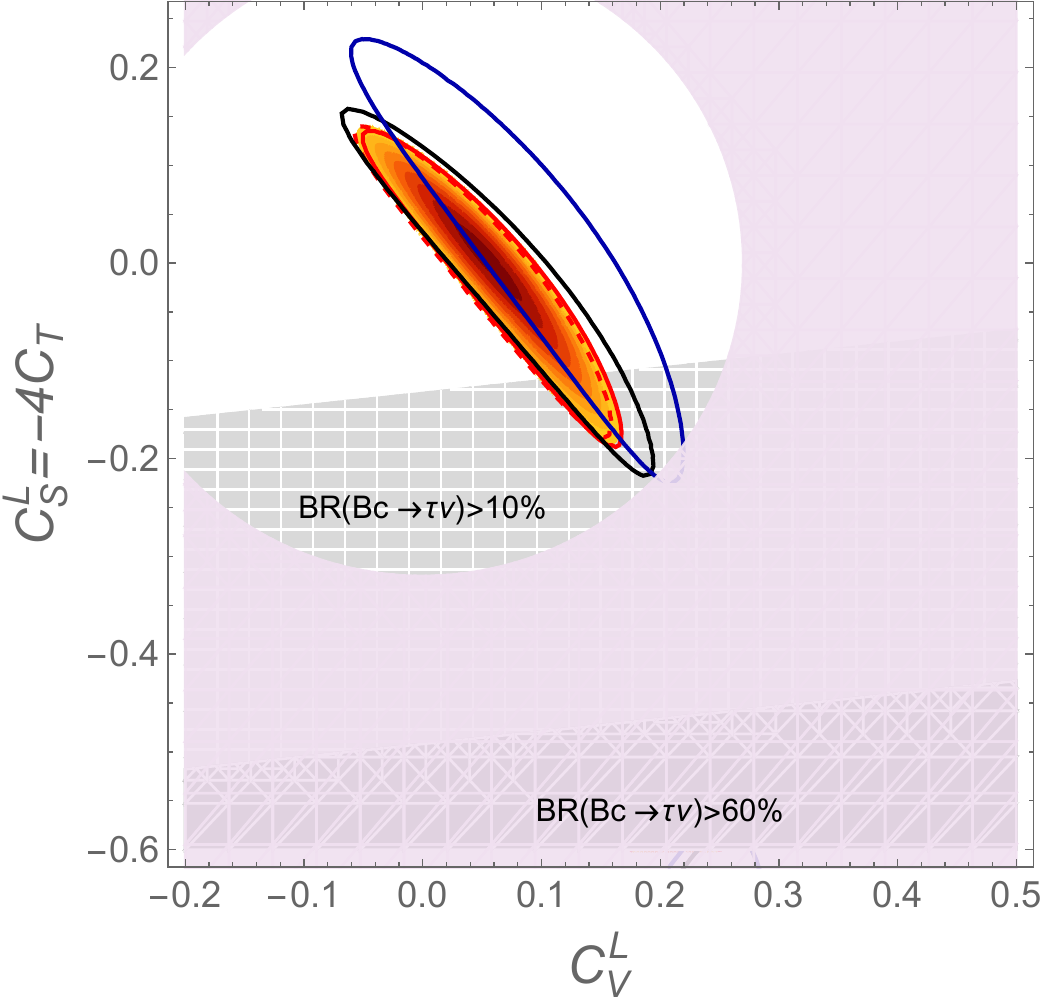}
         \caption{}
     \end{subfigure}\hspace{1.5cm}
\centering{}
     \begin{subfigure}[b]{0.35\textwidth}
         \centering{}
         \includegraphics[width=7cm,height=6cm]{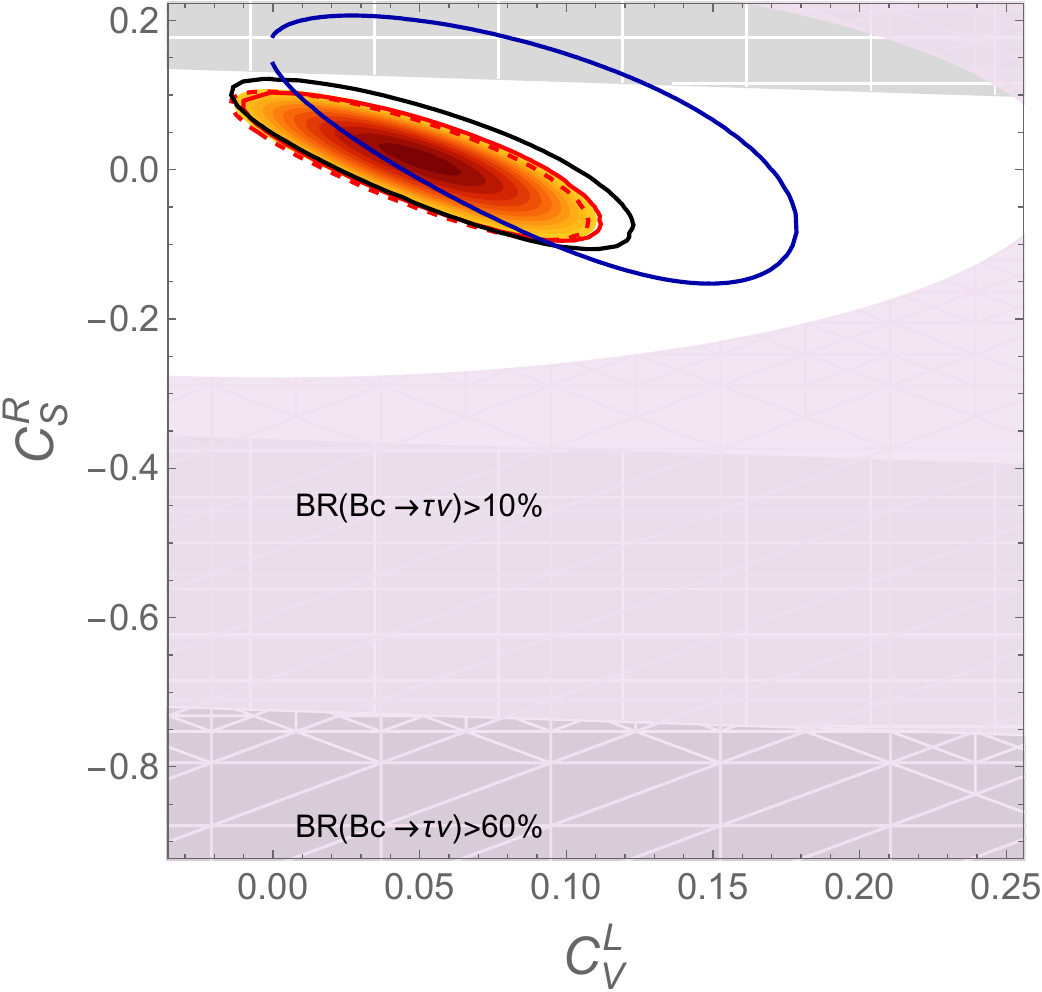}\includegraphics[width=0.7cm,height=6cm]{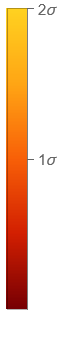}
        \caption{}
     \end{subfigure}\\

 \centering{}
     \begin{subfigure}[b]{0.35\textwidth}
         \centering{}
\includegraphics[width=7cm,height=6cm]{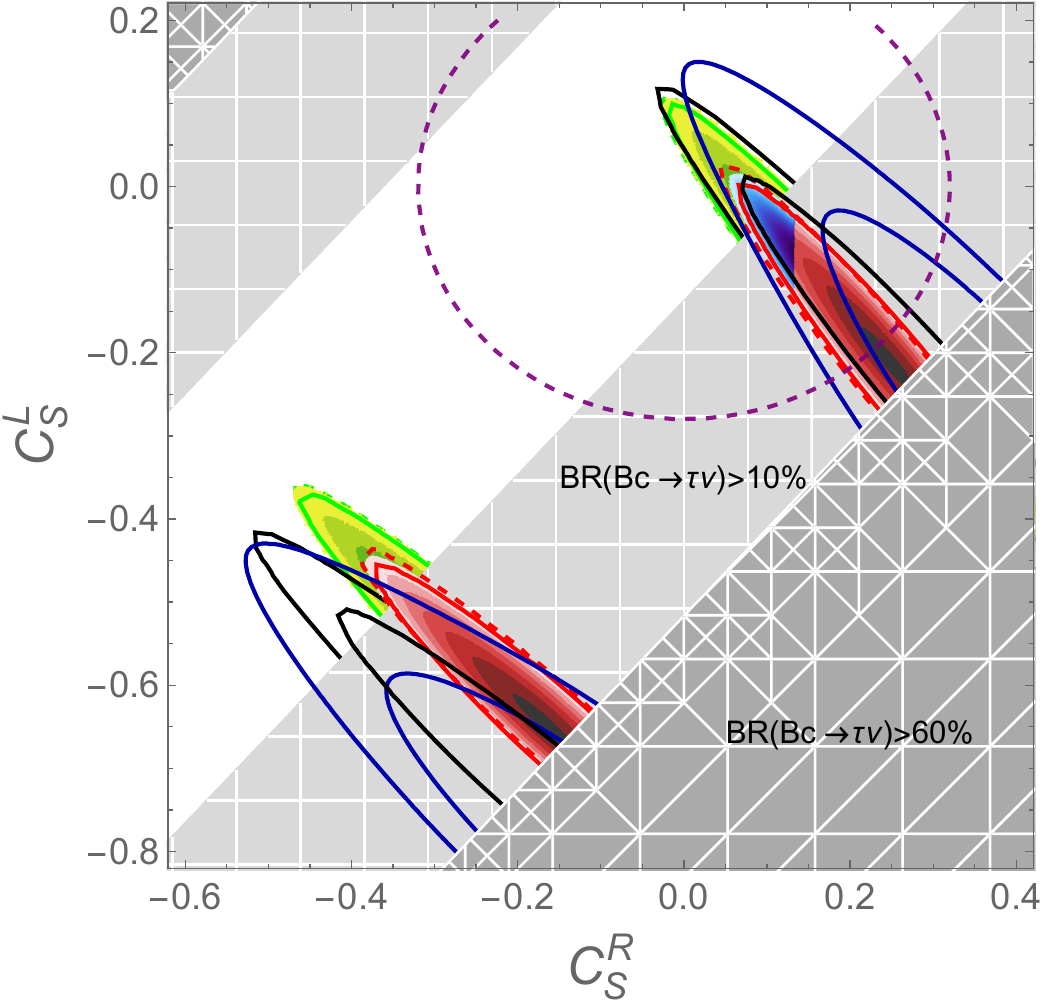}
         \caption{}
     \end{subfigure}\hspace{1.5cm}
\centering{}
     \begin{subfigure}[b]{0.35\textwidth}
         \centering{}
         \includegraphics[width=7cm,height=6cm]{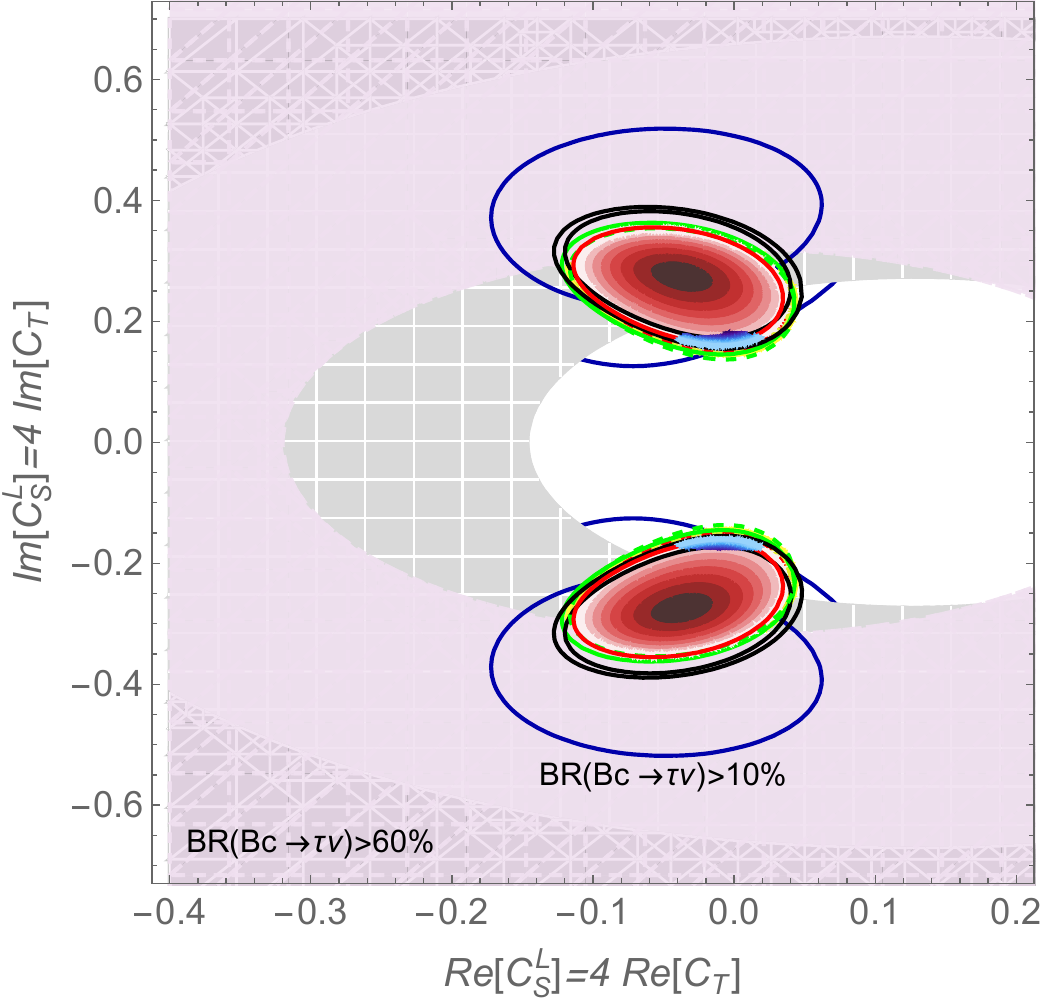}\includegraphics[width=1.5cm,height=6cm]{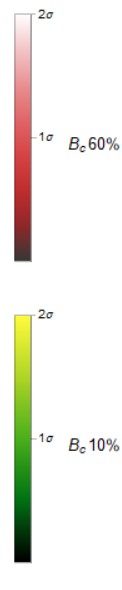}
         \caption{}
     \end{subfigure}
     \caption{\label{Monika fig2}Results of the fits for 2D NP scenarios. The light and dark gray regions show the $10\%$ and $60\%$ branching ratio constraints. The light (dark) shaded contours represent the $1\sigma$ $(2\sigma)$ intervals. FIGS. (a), (b), (c) and (d) show the $(C_{V}^{L},C_{S}^{L}=-4C_{T})$, $(C_{V}^{L},C_{S}^{R})$, $(C_{S}^{R},C_{S}^{L})$ and 
$(Re[C_S^L=4C_T],\text{Im}[C_S^L=4C_T])$  scenarios, respectively. The shaded contours represent Fit A, solid contours represent Fit B and the dashed contours represent Fit C. The red and green colors in FIGS. 1c and 1d show the allowed parametric region when the branching ratio constraints are taken to be $60\%$ and $10\%$, respectively. The blue (black) contours show the results by using the previous (new) data of $R(D^{(*)})$ at 1TeV. The purple shaded regions shows the excluded current collider bounds at the 2$\sigma$ level. The dashed purple circle in plot(c) represents the collider constraint on the charged Higgs scenario and the shaded blue color shows the LHC future bounds.}
\end{figure}

In FIG. \ref{Monika fig2}, we have plotted the $1\sigma$ and $2\sigma$ allowed parametric space in the 2D NP scenarios planes. The shaded colored regions (the black contours) represent at 2 TeV (1 TeV) allowed parametric space for the Fit A, while the solid red and dashed contours are for Fits B and C, respectively. Moreover in FIG. \ref{Monika fig2}, for the comparison with ref. \cite{Blanke:2018yud}, we have also shown the allowed parametric regions by the blue contours for NP scenarios for Fit A where the authors of ref. \cite{Blanke:2018yud} considered the old data of the observables. The gray hatched regions are excluded by the $60\%$ and $10\%$ upper limits on the branching ratio of $B_c \rightarrow \tau \nu$. From the plots of FIG. \ref{Monika fig2} , one can easily see the change in the allowed parametric space by changing the scale of NP WC's from 1 TeV to 2 TeV and by the updated data. We can observe that the updated data and shifting the scale of NP WC's from low to high, squeeze the allowed parametric space of NP scenarios. 

It can be noted from FIG. \ref{Monika fig2}a that the allowed parametric space for scenario $(C_{V}^L,C_{S}^L=-4C_{T})$ for Fit A (orange shaded region) is not much effected whether we include the $R(J/\psi)$  (Fit B (red solid contour)) or by inclusion of $R(J/\psi)$ and $R(\Lambda_c)$ together (Fit C (red dashed contour)) in the analysis. On the other hand, in the scenario $(C_{V}^L,C_{S}^R)$, the allowed parametric space is neither effected by the $B_c \rightarrow \tau \nu$ nor by the number of observables and remained approximately same for Fit A, B and C. FIGS. \ref{Monika fig2}c and \ref{Monika fig2}d depict the allowed parametric space for scenarios $(C_{S}^L,C_{S}^R)$ and $(\text{Re}[C_{S}^L=4C_{T},\text{Im}[C_{S}^L=4C_{T}])$, respectively, where red (green) shaded region represents with $B_c \rightarrow \tau \nu$ $\leq60\%$ ($\leq10\%$). From these figures, one can see that for both these scenarios the allowed parametric space at $B_c \rightarrow \tau \nu$ $\leq10\%$ are almost remain the same for Fits A, B and C. On the other hand at $B_c \rightarrow \tau \nu$ $\leq60\%$), the parametric space get elongated for Fit C (red dashed contour).

By applying the upper limits of $BR(B_c \rightarrow \tau \nu)$ $< 60\%$,  $< 30\%$ and $< 10\%$,  the results for the different parameters of 2D scenarios for the Fits A, B and C are given in TABLE \ref{2D}. 
One can observe that for all the fitting cases, the best fit point of scenarios $(C_V^L, C_S^L=-4C_T)$ and $(C_V^L,C_S^R)$ are not effected by the $BR(B_c \rightarrow \tau \nu)$, where as it is not the case for the other two scenarios, $(C_S^L,C_S^R)$ and $(\text{Re}[C_{S}^L=4C_{T},\text{Im}[C_{S}^L=4C_{T}])$ as mentioned in \cite{Blanke:2018yud} also. However, by using the updated data for Fit A, the goodness of the fit (p-value) increases $\sim4\%$ $(22\%\to29.8\%)$ for $(C_V^L, C_S^L=-4C_T)$ and $\sim2\%$ $(30.8\%\to31.8\%)$ for $(C_V^L,C_S^R)$. Here, one can also notice from FIG. \ref{Monika fig2}c  that the NP scenario $(C_S^L,C_S^R)$ is significantly effected by the WC's scale as compared to other three scenarios. Therefore, for $(C_S^L,C_S^R)$ by setting $BR(B_c \rightarrow \tau \nu)< \left(10, 30, 60\right)\%$ the p-values are increased $\sim \left(9, 34, 3\right)\%$ at 2 TeV and $\sim \left(9, 28, 7\right)\%$ at 1 TeV, respectively. 

Therefore, the updated data indicate that the scenario $(C_S^L,C_S^R)$ with hard cut $BR(B_c \rightarrow \tau \nu)< 30\%$ is also favorable. The variation in p-value for the scenario $(\text{Re}[C_{S}^L=4C_{T},\text{Im}[C_{S}^L=4C_{T}])$ got improved, as for $BR(B_c \rightarrow \tau \nu)$ $< 60\%$ and  $< 30\%$, the p-value is increased from $22\%\to26.7\%$ while for $BR(B_c \rightarrow \tau \nu)$ $< 10\%$, the p-value is increased from $0.3\%\to14.1\%$. Hence, the latest data show that the impact of $BR(B_c \rightarrow \tau \nu)$ on the p-values became more crucial, therefore, an accurate measurement of $BR(B_c \rightarrow \tau \nu)$ will be helpful to clear the smog for the most favourable NP scenario. 
  
Apart from the dependence of $BR(B_c \rightarrow \tau \nu)$, we also analyze the impact of $N_{obs}$ on the parameters of NP scenarios.  For this purpose, we compare the values of different parameters of Fits A, B and C, which are listed in TABLE \ref{2D}.  
The data of $R(J/\psi)$ and $R(\Lambda_c)$ directly effect the parametric space of NP scenarios as can be seen from the given values of different parameters in TABLE \ref{2D}. 
These values show that the goodness of fit (the p-value) is decreased when we increased the $N_{obs}$. For instance, the p-value of most favorable scenario $(C_S^L,C_S^R)$ at $BR(B_c \rightarrow \tau \nu)$ $< 60\%$  is $71.5\%$ for Fit A is reduced up to $32.2\%$ for Fit B and even further reduced for Fit C, where it is $16.9\%$ only.
\begin{itemize}
    \item Interestingly, for the Fit C, at $BR(B_c \rightarrow \tau \nu)$ $<10\%$, the most favorable scenario $(C_S^L,C_S^R)$ becomes less favorable (p-value=$3.8$) in comparison with some other NP scenarios, as we can see from the p-values given in TABLE \ref{2D}. Although, there is a large uncertainty in the experimental value of $R{(J/\psi)}$ and similarly, the recent experimental measurement of  $R(\Lambda_c)$ is under debate \cite{Fedele:2022xyz}, the behavior of analysis that discussed above is indicating that the future data on different observables is also valuable to decide which NP scenario is more suitable to explain the various anomalies.
\end{itemize}
\begin{table}[H]
\centering{}%
\scalebox{0.75}{
\begin{tabular}{|c|c|c|c|c|c|c|c|c|c|c|c|}
\toprule
\multicolumn{12}{c}{$\chi^{2}_{SM}=(16.50,19.76,20.92)_{A,B,C}$, $p-value=(2.41,1.38,1.89)_{A,B,C}\times10^{-3}$, 
}\tabularnewline
\hline
\hline
Scenarios & Best fit & $\chi^2_{min}$ & P-Value\% & pull$_{SM}$ & $R(D)$  & $R(D^{*})$  & $R(J/\psi)$   &  $F_L (D^{*})$  & $P_{\tau}(D^{*})$ & $P_{\tau}(D)$ &$R(\Lambda_c)$
\tabularnewline
\midrule
\midrule 
\hline
\multirow{3}{*}{$(C^L_V,C^L_S=-4C_T)$} &\multirow{1}{*}{(0.05,0.003)} &2.42 &29.8 &3.75 & \multirow{2}{*}{$0.336$} & \multirow{2}{*}{$0.284$} &\multirow{2}{*}{ $0.289$} & \multirow{2}{*}{$0.462$} & \multirow{2}{*}{$-0.497$} & \multirow{2}{*}{$0.342$} &\multirow{2}{*}{$0.363$}
\tabularnewline
 & \multirow{1}{*}{(0.06,0.00)} &5.20 &15.7 &3.82 & $-0.31\sigma$ & $-0.08\sigma$ & $-1.70\sigma$ & $-1.54\sigma$ & $-0.21\sigma$ &   & $1.59\sigma$
 \tabularnewline
  &\multirow{1}{*}{(0.05,0.01)}  & 7.86&9.68 &3.61 &  &  & &  &  &  &   
\tabularnewline
\hline
\multirow{3}{*}{$(C^L_S,C^R_S)|_{60\%}$} & (-0.65,-0.17),(-0.21,0.25)&0.67 &71.5 &3.98 &\multirow{2}{*}{$0.339$} &\multirow{2}{*}{ $0.285$} &\multirow{2}{*}{ $0.287$} & \multirow{2}{*}{$0.524$} &\multirow{2}{*} {$-0.335$} & \multirow{2}{*}{$0.441$}\multirow{2}{*}{} &\multirow{2}{*}{$0.369$}
\tabularnewline
 &(-0.65,-0.16),(-0.22,0.25)  & 3.49& 32.2&4.03 &  $-0.19\sigma$ & $0\sigma$ & $-1.71\sigma$ & $-0.84\sigma$ & $0.08\sigma$ &  & $1.67\sigma$
 \tabularnewline
 &(-0.63,-0.18),(-0.20,0.24)   & 6.43&16.9 &3.81 &  &  & &  &    &  &
\tabularnewline
\hline
\multirow{3}{*}{$(C^L_S,C^R_S)|_{30\%}$} & \multirow{3}{*}{(-0.58,-0.24),(-0.14,0.18)} &1.57 &45.6 &3.86 &\multirow{2}{*}{$0.341$} & \multirow{2}{*}{$0.275$} &\multirow{2}{*}{ $0.278$} & \multirow{2}{*}{$0.508$} & \multirow{2}{*}{$-0.379$} & \multirow{2}{*}{$0.435$} &\multirow{2}{*}{$0.358$}
\tabularnewline
&  & 4.53&20.9 &3.90 &$-0.11\sigma$ & $-0.83\sigma$ & $-1.74\sigma$ & $-1.02\sigma$ & $0.001\sigma$ &  & $1.52\sigma$
 \tabularnewline
 &  & 7.08&13.1 &3.72 &  &  & &  &   &  &
\tabularnewline
\hline
\multirow{3}{*}{$(C^L_S,C^R_S)|_{10\%}$} & \multirow{3}{*}{(-0.49,-0.34),(-0.03,0.09)} &4.84 &8.8 &3.41 & \multirow{2}{*}{$0.366$} &\multirow{2}{*}{ $0.261$} & \multirow{2}{*}{$0.265$} & \multirow{2}{*}{$0.481$} &\multirow{2}{*}{ $-0.451$} &\multirow{2}{*}{$0.502$} & \multirow{2}{*}{$0.370$}
\tabularnewline
&  & 7.99&4.6 &3.43 &$0.84\sigma$ & $-2\sigma$ & $-1.79\sigma$ & $-1.33\sigma$ & $-0.12\sigma$  & & $1.68\sigma$
 \tabularnewline
  &  & 10.11&3.8 &3.29 &  &  & &   &  &  &
  \tabularnewline
\hline
\multirow{3}{*}{$(C^L_V,C^R_S)$} & \multirow{3}{*}{(0.05,0.01)} &2.29 &31.8 &3.77 & \multirow{2}{*}{$0.341$} & \multirow{2}{*}{$0.283$} & \multirow{2}{*}{$0.288$} & \multirow{2}{*}{$0.465$} & \multirow{2}{*}{$-0.493$} & \multirow{2}{*}{$0.350$}& \multirow{2}{*}{$0.364$}
\tabularnewline
&  &5.12 & 16.3&3.83 &$-0.11\sigma$ & $-0.16\sigma$ & $-1.70\sigma$ & $-1.51\sigma$ & $-0.20\sigma$  &  & $1.61\sigma$
 \tabularnewline
 &  & 7.77&10 &3.63 &  &  & &  &    &  &
  \tabularnewline
\hline
\multirow{3}{*}{$(Re[C_S^L=4C_T],\text{Im}[C_S^L=4C_T])|_{60\%,30\%}$ } & (-0.04,$\pm0.28$)&2.64 &26.7 &3.72 & \multirow{2}{*}{$0.337$} & \multirow{2}{*}{$0.285$} &\multirow{2}{*}{ $0.288$} & \multirow{2}{*}{$0.455$} & \multirow{2}{*}{$-0.438$} & \multirow{2}{*}{$0.953$} &\multirow{2}{*}{$0.362$}
\tabularnewline
& (-0.04,$\pm0.28$) &5.47 & 14 &3.55 & $-0.26\sigma$ & $0\sigma$ & $-1.70\sigma$ & $-1.62\sigma$ & $-0.10\sigma$  & & $1.58\sigma$
 \tabularnewline
 &(-0.04,$\pm0.27$)  & 8.01& 9.1&3.59 &  &  & &  &  &  &  
  \tabularnewline
\hline
\multirow{3}{*}{$(Re[C_S^L=4C_T],\text{Im}[C_S^L=4C_T])|_{10\%}$ } &\multirow{3}{*}{(-0.02,$\pm0.23$)}&3.91 &14.1 &3.55 &\multirow{2}{*}{$0.333$} & \multirow{2}{*}{$0.274$} & \multirow{2}{*}{$0.276$} & \multirow{2}{*}{$0.457$} & \multirow{2}{*}{$-0.457$} & \multirow{2}{*}{$0.786$} &\multirow{2}{*}{$0.351$}
\tabularnewline
&  &6.92& 7.44&3.58 &$-0.42\sigma$ & $-0.91\sigma$ & $-1.75\sigma$ & $-1.59\sigma$ & $-0.14\sigma$ & & $1.43\sigma$
 \tabularnewline
 &  & 9.0&6.1 &3.45 &  &  & &  &  &  &
  \tabularnewline
\hline
\hline
\end{tabular}}
\caption{Fit results for 2D scenarios at 2TeV, including all available data: Best fit points, $\chi^2_{min}$,  p-value, $Pull_{SM}$ are given. The columns (2-5) represent the results for different parameters: First, second and third rows represent Fit A, B and C, respectively. The last eight columns show the predictions of different observables at the best fit point with the $\sigma$ deviation.}
\label{2D}
\end{table} 
The $1\sigma$ and $2\sigma$ intervals of these NP 1D and 2D scenarios are given in TABLE. (\ref{2sigma}).

\begin{table}[H]
\centering{}%
\scalebox{0.96}{
\begin{tabular}{|c|c|c|c|c|c|}
\toprule
\hline
\hline
1D Scenarios & 1$\sigma$ interval &2$\sigma$ interval & 2D Scenarios & 1$\sigma$ interval & 2$\sigma$ interval 
\tabularnewline
\midrule
\midrule 
\hline
\multirow{2}{*}{$C^L_S$} & \multirow{2}{*}{(0.03,0.11)} & \multirow{2}{*}{(-0.004,0.13)} &\multirow{2}{*}{$(C^L_V,C^L_S=-4C_{T})$} &\multirow{1}{*}{(0.04,0.07)$\epsilon$ $C_{V}^L$ }&(0.01,0.09)$\epsilon$ $C_{V}^L$ 
\tabularnewline
&  & & & (-0.02,0.04)$\epsilon$ $C_{S}^L$ &(-0.05,0.05)$\epsilon$ $C_{S}^L$ 
  \tabularnewline
\hline
\multirow{2}{*}{$C^R_S$} & \multirow{2}{*}{(0.04,0.10)} & \multirow{2}{*}{(0.01,0.12)} &\multirow{2}{*}{$(C^L_S,C^R_S)$} &(-0.25,-0.16)$\epsilon$ $C_{S}^L$ &(-0.27,-0.14)$\epsilon$ $C_{S}^L$ 
\tabularnewline
&  & & & (0.20,0.27)$\epsilon$ $C_{S}^R$ &(0.18,0.28)$\epsilon$ $C_{S}^R$ 
  \tabularnewline
\hline
\multirow{2}{*}{$C^L_V$} & \multirow{2}{*}{(0.03,0.08)} & \multirow{2}{*}{(0.02,0.09)} &\multirow{2}{*}{$(C^L_V,C^R_S)$} &(0.03,0.07)$\epsilon$ $C_{V}^L$ &(0.01,0.08)$\epsilon$ $C_{V}^L$ 
\tabularnewline
&  & & & (-0.02,0.05)$\epsilon$ $C_{S}^R$ &(-0.05,0.07)$\epsilon$ $C_{S}^R$ 
  \tabularnewline
\hline
\multirow{2}{*}{$C^L_S=4C_{T}$} & \multirow{2}{*}{(-0.04,0.05)} & \multirow{2}{*}{(-0.08,0.08)} &\multirow{2}{*}{$(Re[C_S^L=4C_T],\text{Im}[C_S^L=4C_T])$} &(-0.08,0.02)$\epsilon$ $RC_{S}^L$ &(-0.11,0.04)$\epsilon$ $RC_{S}^L$ 
\tabularnewline
&  & & & (-0.32,0.32)$\epsilon$ $ImC_{S}^L$ &(-0.35,0.35)$\epsilon$ $ImC_{S}^L$ 
  \tabularnewline
\hline
\hline
\end{tabular}}
\caption{ $1\sigma$ and $2\sigma$ allowed parametric space by using the $60\%$ branching ratio for 1D and 2D scenarios.}
\label{2sigma}
\end{table}  
\subsubsection{Discussion about $C_{V}^R$ related scenarios}
\begin{figure}[H]

 \centering{}
     \begin{subfigure}[b]{0.35\textwidth}
         \centering
         \includegraphics[width=7cm,height=6cm]{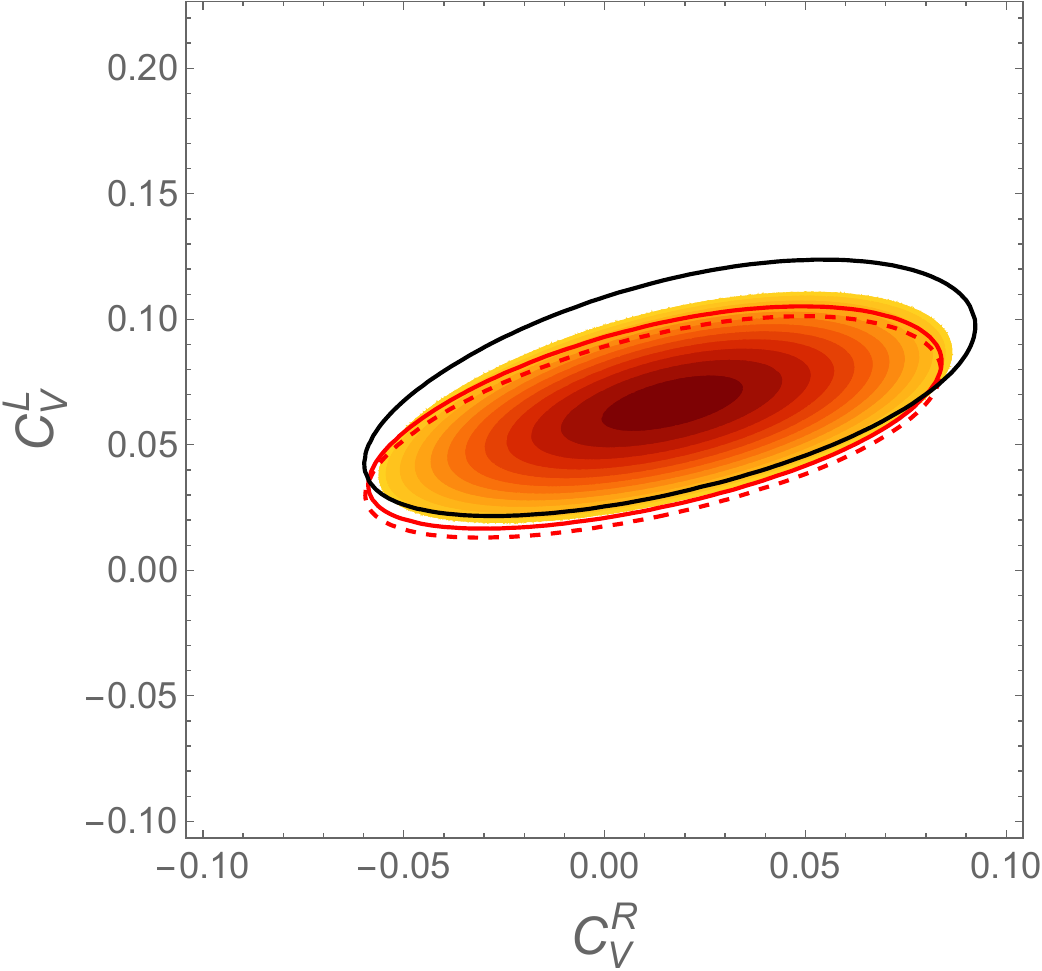}
         \caption{}
     \end{subfigure}\hspace{1.5cm}
\centering{}
     \begin{subfigure}[b]{0.35\textwidth}
         \centering{}
         \includegraphics[width=7cm,height=6cm]{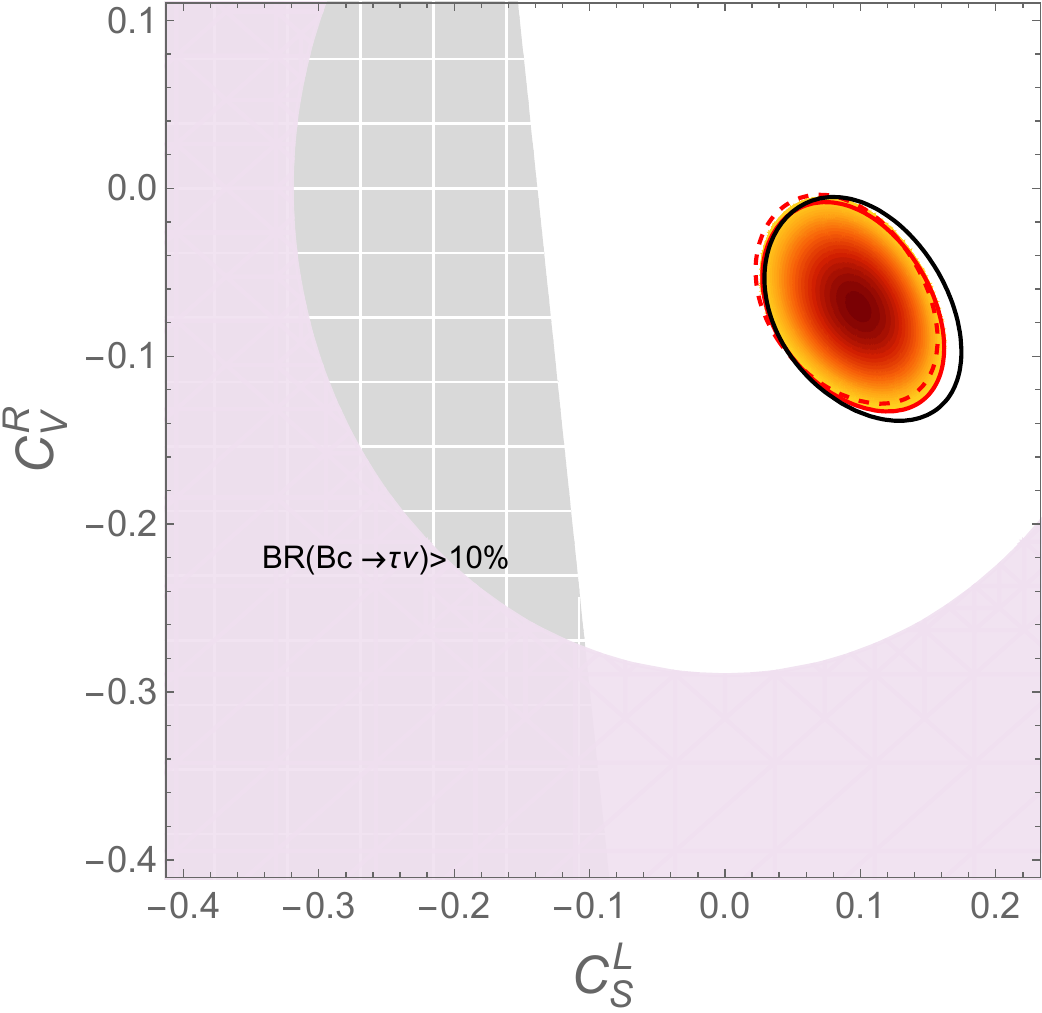}
        \caption{}
     \end{subfigure}\\

 \centering{}
     \begin{subfigure}[b]{0.35\textwidth}
         \centering{}
         \includegraphics[width=7cm,height=6cm]{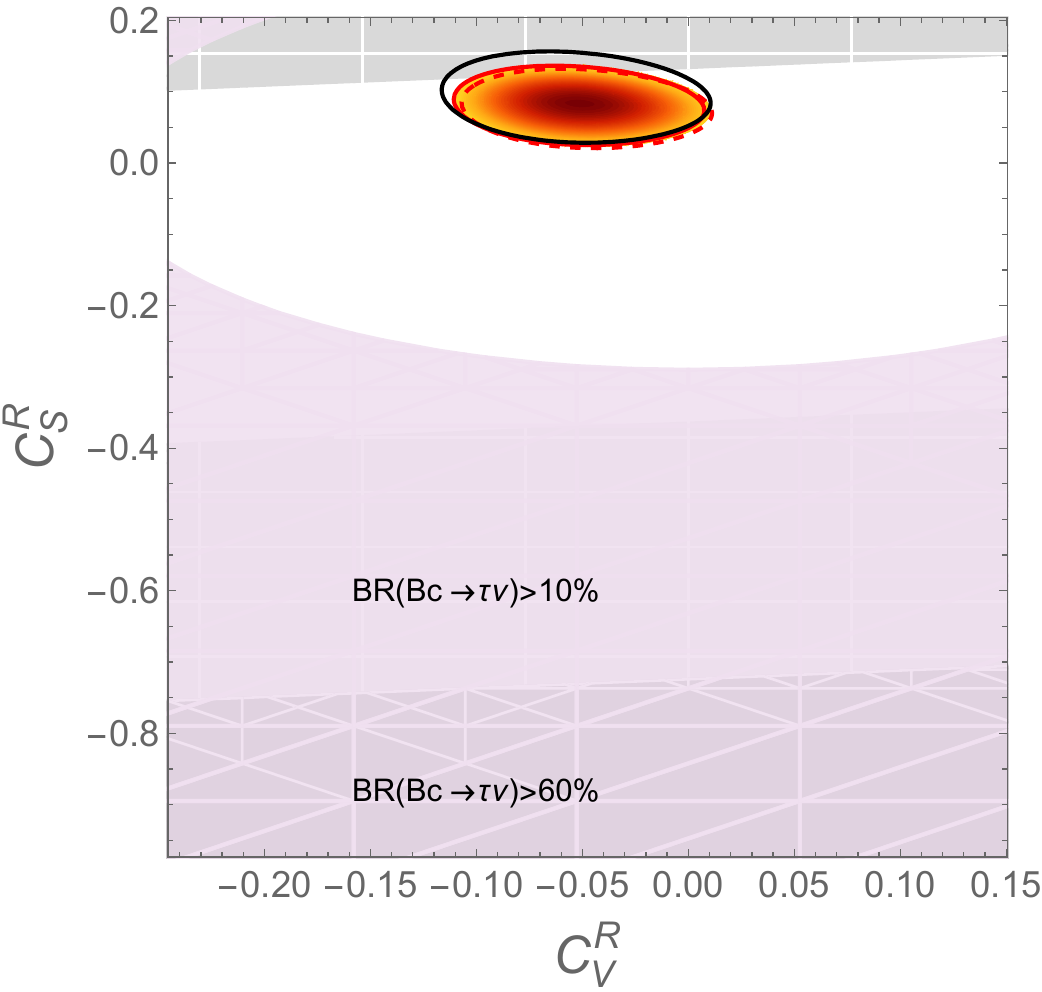}
         \caption{}
     \end{subfigure}\hspace{1.5cm}
\centering{}
     \begin{subfigure}[b]{0.35\textwidth}
         \centering{}
         \includegraphics[width=7cm,height=6cm]{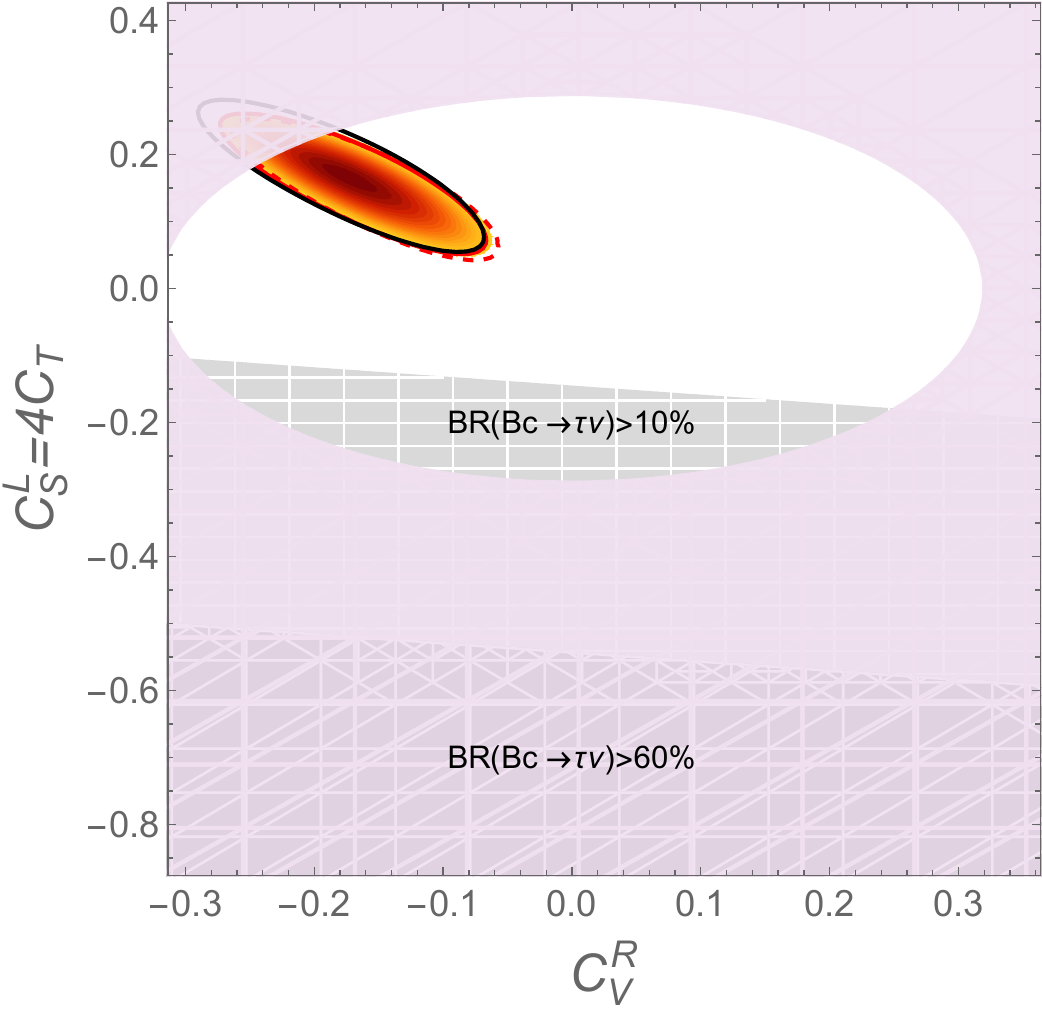}\includegraphics[width=0.7cm,height=6cm]{LEGENDS.png}
         \caption{}
     \end{subfigure}
\caption{Results of the fits for the 2D $C_{V}^{R}$ scenarios (a) $(C_{V}^L,C_{V}^R)$, (b) $(C_{S}^L,C_{V}^R)$, (c) $(C_{V}^R,C_{S}^R)$ and (d) $(C_{V}^R,C_{S}^L=4C_{T})$. The shaded contours represent Fit A, solid contours represent Fit B and the dashed contours represent Fit C. 
The gray portion represents the excluded regions by the $60\%$ and $10\%$ upper limits on branching ratio $B_c \rightarrow \tau \nu$. The black contours show the allowed parametric space at 1 TeV. The purple region shows the collider bounds at 2$\sigma$ level.}
\label{Fig3}
\end{figure}

In this section, we will discuss the possible scenarios associated with $C_{V}^R$, i.e., ($C_{V}^L,C_{V}^R$), ($C_{S}^L,C_{V}^R$), ($C_{S}^R,C_{V}^R$) and $(C^R_V,C^L_S=4C_T)$, where the last one is related with the $R_2$ model \cite{Iguro:2022yzr}. The corresponding plots are drawn in FIG. \ref{Fig3}, with shaded brown, solid red and dashed red contours representing the Fits A, B and C, respectively, at 2 TeV while the black contours are at 1 TeV. The purple region represents the LHC bounds at 2$\sigma$ level. It can be seen that these scenarios are independent of branching ratio constraints with the exception that the scenario ($C_{S}^R,C_{V}^R$) is slightly effected $\left(\approx 10\%\right)$ by the branching ratio constraint. Moreover, the $1\sigma$, $2\sigma$ allowed parametric space, and the best fit points of these scenarios are not significantly effected after including $R(J/\psi)$ (Fit B, solid red contour) and  $R({\Lambda_c})$ (Fit C, dashed red contour) in the analysis. It is also noted that the parametric space for the scenario $(C^R_V,C^L_S=4C_T)$ is impacted by the collider bounds, as seen by the purple region in FIG. (d). The  p-values and the other parameters of these scenarios are reported in the TABLE \ref{2DPSAB1}. The trend and the variation in the values of $\chi^2$, pull and p-value with respect to the $N_{obs}$ show that the scenario ($C_{S}^R,C_{V}^R$) is most favorable among the other scenarios for Fit A while for Fit B the scenarios ($C_{S}^R,C_{V}^R$) and $(C^R_V,C^L_S=4C_T)$ both are equally likely. However, to explore more about the NP scenarios, as mentioned in previous section, the theoretical prediction of $R(J/\psi)$ and the experimental measurement of $R(\Lambda_c)$ require further precision.

\begin{table}[H]
\centering{}%
\scalebox{0.87}{
\begin{tabular}{|c|c|c|c|c|c|c|c|c|c|c|c|}
\toprule
\multicolumn{12}{c}{$\chi^{2}_{SM}=(16.50,19.76,20.92)_{A,B,C}$, $p-value=(2.41,1.38,1.89)_{A,B,C}\times10^{-3}$, 
}\tabularnewline
\hline
\hline
Scenarios & Best fit & $\chi^2_{min}$ & $p-value\%$ & pull$_{SM}$ & $R(D)$  & $R(D^{*})$  & $R(J/\psi)$   &  $F_L (D^{*})$  & $P_{\tau}(D^{*})$ & $P_{\tau}(D)$ &$R(\Lambda_c)$
\tabularnewline
\midrule
\midrule 
\hline
\multirow{3}{*}{$(C^L_V,C^R_V)$} & \multirow{3}{*}{(-0.89,-0.99.)}&2.37 &30.5 &3.76 & \multirow{2}{*}{$0.332$} & \multirow{2}{*}{$0.286$} &\multirow{2}{*}{ $0.291$} & \multirow{2}{*}{$0.464$} & \multirow{2}{*}{$-0.496$} & \multirow{2}{*}{$0.331 $}& \multirow{2}{*}{$0.364$}
\tabularnewline
 &  &5.17 &15.9 &3.82 & $-0.46\sigma$ & $0.08\sigma$ & $-1.69\sigma$ & $-1.52\sigma$ & $-0.21\sigma$ & & $1.61\sigma$
 \tabularnewline
  & & 7.84&9.7 &3.62 &  &  & &  &    &  & 
\tabularnewline
\hline
\multirow{3}{*}{$(C^L_S,C^R_V)$} & \multirow{3}{*}{(0.09,-0.07)} &2.58 &27.5 &3.73 &\multirow{2}{*}{$0.334$} & \multirow{2}{*}{$0.285$} &\multirow{2}{*}{ $0.290$} & \multirow{2}{*}{$0.459$} & \multirow{2}{*}{$-0.522$} & \multirow{2}{*}{$0.553 $}& \multirow{2}{*}{$0.360$}
\tabularnewline
&  & 5.38&14.5 &3.79 &$-0.38\sigma$ & $0.\sigma$ & $-1.69\sigma$ & $-1.57\sigma$ & $-0.26\sigma$ &   & $1.55\sigma$
 \tabularnewline
 &   & 7.90&9.5 &3.61 &  &  & &  &    &  &
\tabularnewline
\hline
\multirow{3}{*}{$(C^R_S,C^R_V)$} & \multirow{3}{*}{(0.08,-0.05)} &1.87 &39.2 &3.82 & \multirow{2}{*}{$0.341$} &\multirow{2}{*}{ $0.284$} & \multirow{2}{*}{$0.288$} & \multirow{2}{*}{$0.477$} &\multirow{2}{*}{ $-0.469$} &\multirow{2}{*}{0.532 }  &\multirow{2}{*}{$0.363$}
\tabularnewline
& & 4.68&19.6 &3.88 &$-0.11\sigma$ & $-0.08\sigma$ & $-1.70\sigma$ & $-1.38\sigma$ & $-0.16\sigma$ &  & $1.59\sigma$
 \tabularnewline
  &  & 7.29&12.1 &3.69 &  &  & &  &    &  &
  \tabularnewline
\hline
\multirow{3}{*}{$(C^R_V,C^L_S=4C_{T})$} & \multirow{3}{*}{(-0.17,0.17)} &2.14 &34.5 &3.79 & \multirow{2}{*}{$0.350$} & \multirow{2}{*}{$0.282$} & \multirow{2}{*}{$0.284$} & \multirow{2}{*}{$0.472$} & \multirow{2}{*}{$-0.568$} & \multirow{2}{*}{$0.763 $}& \multirow{2}{*}{$0.364$}
\tabularnewline
&  &4.97 & 17.4&3.85&$0.23\sigma$ & $-0.25\sigma$ & $-1.72\sigma$ & $-1.43\sigma$ & $-0.34\sigma$ & &   $1.61\sigma$
 \tabularnewline
 &  & 8.58&10.7 &3.65 &  &  & &  &  &   &
  \tabularnewline

\hline
\hline
\end{tabular}}
\caption{The Best fit point, $\chi^2_{min}$,  p-value and $Pull_{SM}$ of 2D scenarios related to $C_V^{R}$ at 2 TeV are given in the columns (2-5). First row represents Fit A, second row represents Fit B and last row shows Fit C. Last seven columns shows the predictions of different observables at the best fit point with the sigma deviation.}
\label{2DPSAB1}
\end{table}  
\subsection{Impact of collider (LHC) bounds on 2D NP Scenarios}\label{CB}
We would like to mention here that at the LHC, the analysis of the $\tau$+ missing searches has given the upper limits on the NP WC(\cite{Greljo:2018tzh,Endo:2021lhi}) at the scale  $\mu = m_b$, which are listed in TABLE \ref{CB}. By using Eq. (\ref{eq3}), the values of these NP WC at  $\mu = 2$ TeV are also given in TABLE \ref{CB} where the values in the parenthesis correspond to HL-LHC 3000fb$^{-1}$ limit.

\begin{table}[H]
\centering{}
\begin{tabular}{|cccccc|}
\hline
& $C_V^{L}$ &$C_V^{R}$ & $C_S^{L}$ & $C_S^{R}$ & $C_{T}$ 
\\ \hline
\hline
 \ $\mu = m_b$  &  \ 0.30(0.14)  \ & \ 0.32(0.15) \ &   \ 0.55(0.25)   \ &   \ 0.55(0.25)  \ &   \ 0.15(0.07)   \  \\
  \hline 
\ $\mu = 2$ TeV  \  &  \  0.27(0.13)  \ & \ 0.29(0.14) \ &   \ 0.32(0.15)   \ &   \ 0.28(0.13) \ &   \ 0.17(0.08)  \  \\
\hline
   \end{tabular}
\caption{The current collider bounds of the NP WC's with 139fb$^{-1}$ based on the $\tau^\pm\nu$ search at $\mu = m_{b} $\cite{Endo:2021lhi}.
The 2nd row represents the values of these NP WC's at  $\mu = 2$ TeV where the values in the parenthesis correspond to HL-LHC 3000fb$^{-1}$ limit.}
\label{CB}
\end{table}

We have imposed these upper limits on the considered 2D NP scenarios. We have found that all the NP scenarios are within the limits coming from the analysis of the $\tau$+ missing search blue shown by purple region. However, the future prospects imposed the severe constraints on the scenarios: $(Re[C_S^L=4C_T],\text{Im}[C_S^L=4C_T])$, $(C^L_S,C^R_S)$. We have shown these constraints by the blue shaded region in the plots of FIG. \ref{Monika fig2} (c,d). One can see the lower portion of the scenario $(Re[C_S^L=4C_T],\text{Im}[C_S^L=4C_T])$ is  completely excluded while the best fit point of upper portion is excluded, and the allowed region is drastically squeezed. Similarly, for the scenario $(C^L_S,C^R_S)$, the best fit points and the large areas of both the upper and lower portions are also excluded.
\section{Sum Rule and Correlation of Observables}\label{correlation}
In this section, we have calculated and analyzed the correlation among different observables in 2D scenarios under consideration - but before that we want to validate the sum rule among the observables $R({D^{(*)}})$ and $R(\Lambda_c)$ given in refs. \cite{Blanke:2018yud,Blanke:2019qrx,Fedele:2022xyz}. We can see that by using the Eqs. (\ref{eq4}, \ref{eq5}) and Eq. (\ref{eq10}), the sum rule reads as
\begin{equation}
\frac{R\left(\Lambda_{c}\right)}{R_{SM}\left(\Lambda_{c}\right)}=0.276\frac{R\left(D\right)}{R_{SM}\left(D\right)}+0.724\frac{R\left(D^{*}\right)}{R_{SM}\left(D^{*}\right)}+x_1,\label{sUMRULE}
\end{equation}
where one can see that even with the different analytical expressions of $R({D^{(*)}})$, the coefficients of first two terms on the right hand side in the above equation are almost similar to the ref. \cite{Fedele:2022xyz}. The only change appears in the remainder $x_1$, which in our case becomes
\begin{equation}
x_{1}=-Re\left[\left(1+C_{V}^{L}\right)\left(0.122\left(C_{T}\right)^{*}+0.019\left(C_{S}^{L}\right)^{*}+0.132\left(C_{V}^{R}\right)^{*}\right)\right]+ Re[(C_{V}^{R})^{*}](0.018C_{S}^{R}+0.351C_{T}).
\end{equation}
The values of $x_1$ for 1D [2D] scenarios are $10^{-5}$ [$10^{-3}$] and the updated predicted value of $R(\Lambda_c)$ by using the latest data of $R({D^{(*)}})$ \cite{HFLAV:2023link} is 
\begin{eqnarray}
 \begin{aligned}
 R(\Lambda_c) \,=\,& R_{\rm SM}(\Lambda_c) \left( 1.140 \pm 0.041\right) 
  \label{eq:predlc1} \\
  \,=\,&  0.369 \pm 0.013 \pm 0.005      \label{eq:predlc2} ,
\end{aligned} 
\end{eqnarray}
 which is not significantly different from the numbers reported in \cite{Fedele:2022xyz}. This shows that the latest data of $R({D^{(*)}})$ again confirms the validity of the above sum rule. In Eq. (\ref{eq:predlc2}), the first and the second errors come from the experimental and form factors uncertainties, respectively. However, the current experimentally measured value of $R(\Lambda_c)$ is larger than the predicted value by sum rule as well as inconsistent with the $R({D^{(*)}})$ data pattern and need further experimental confirmation as discussed in ref. \cite{Fedele:2022xyz}.\\ 
 In addition, the latest predicted SM value of the observable $R(J/\psi)=0.258\pm0.038$ ~\cite{RJSHI:12}, is smaller than its experimentally measured value: $0.71\pm0.17\pm0.18$ \cite{Aaij:2017tyk,Fedele:2022xyz}, which shows the same behavior as $R(\Lambda_c)$, i.e. deficit in taus in contrast of $R({D^{(*)}})$ data. Therefore, it is also interesting to find out the sum rule of  $R(J/\psi)$ in terms of $R({D^{(*)}})$, which we have derived by using the Eqs. (\ref{eq4}, \ref{eq5}) and Eq. (\ref{eq9}) as follows:
 \begin{equation}
\frac{R\left(J/\psi\right)}{R_{SM}\left(J/\psi\right)} =0.006\frac{R\left(D\right)}{R_{SM}\left(D\right)}+0.994\frac{R\left(D^{*}\right)}{R_{SM}\left(D^{*}\right)}+x_{2},\label{RJSUM}
\end{equation}
where 
\begin{eqnarray}
x_{2}&=&-Re\left[\left(1+C_{V}^{L}\right)\left(0.019C_{S}^{R^*}+0.257C_{T}^{*}+0.013C_{V}^{R^*}-0.0004C_{S}^{L^{*}}\right)\right]+0.006\left(\left|C_{R}^{S}\right|^{2}+\left|C_{L}^{S}\right|^{2}\right)\notag \\
&&+0.013Re[C_{S}^{L}C_{S}^{R^*}]-1.205\left|C_{T}\right|^{2}-Re[C_{V}^{R^*}]\left[0.018C_{S}^{L}-0.0031C_{T}-0.004C_{S}^{R}\right].\notag
\end{eqnarray}
and the remainder $x_2$ for this observable in 1D [2D] scenarios are $10^{-5}$ [$10^{-3}$]. The predicted value of $R(J/\psi)$ by using the above sum rule is
\begin{eqnarray}
 \begin{aligned}
 R(J/\psi) \,=\,& R_{\rm SM}(J/\psi) \left( 1.119\pm 0.046 \right) 
  \label{eq:predlc12} \\
  \,=\,&  0.289 \pm 0.013 \pm 0.043.      \label{eq:predlc22}
\end{aligned} 
\end{eqnarray}
One can see that both the SM value of $R(J/\psi)$ as well as its predicted value obtained by sum rule using updated data are smaller than its experimental value and follow the coherent pattern as in the $R({D^{(*)}})$ case, i.e. abundance of taus in comparison of light leptons. But in case of $R(J/\psi)$, even though its tensor form factors are not precisely calculated yet, the theoretical predicted values are quite smaller compared to its experimental value with large uncertainties, $0.71\pm0.18\pm0.17$. We hope that in future, this value will be further scrutinized at some ongoing and planned experiments.

\begin{figure}[H]
\centering
\includegraphics[width=5cm, height=3.9cm]{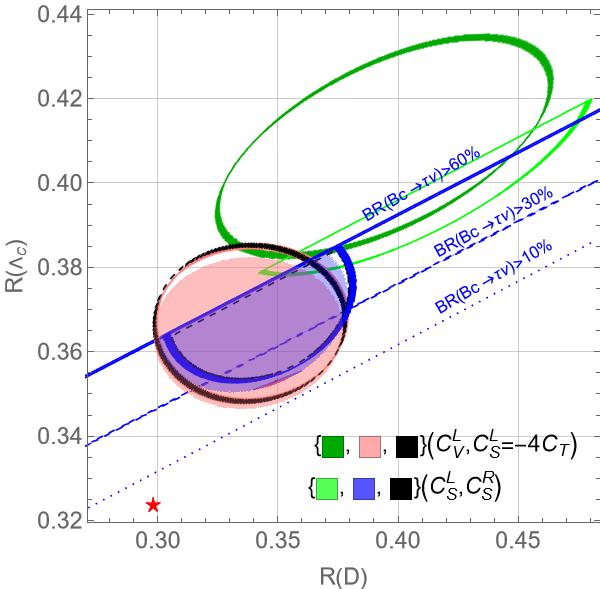}
\includegraphics[width=5cm, height=3.9cm]{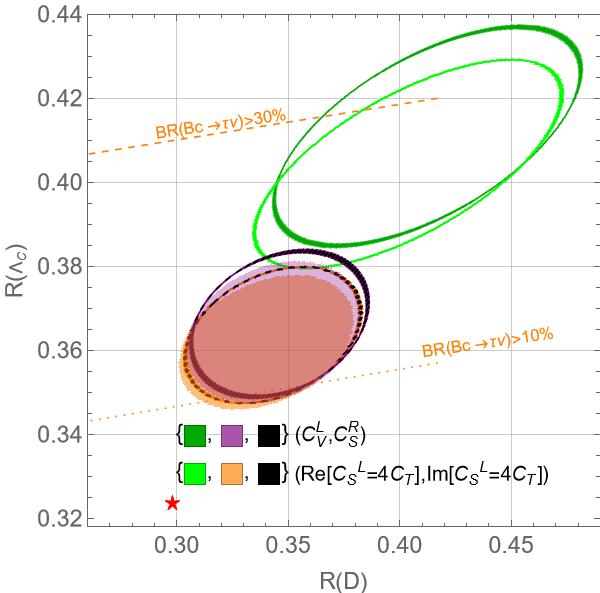}
\includegraphics[width=5cm, height=3.9cm]{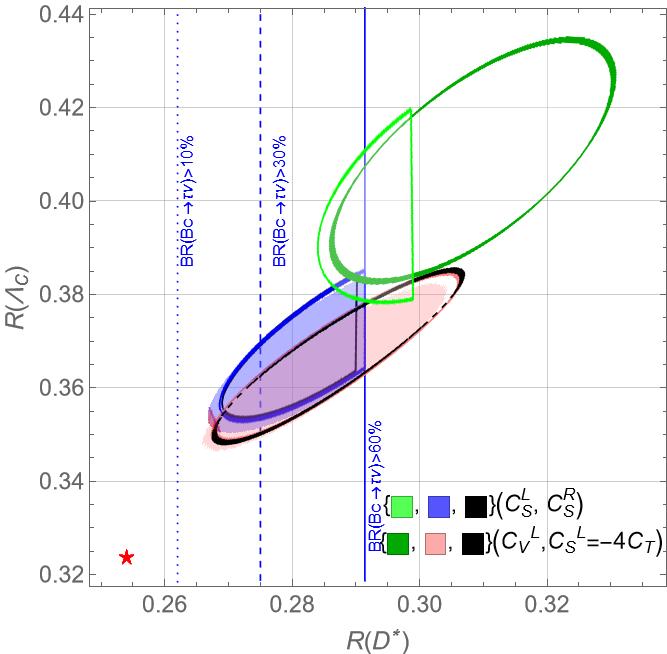}
\includegraphics[width=5cm, height=3.9cm]{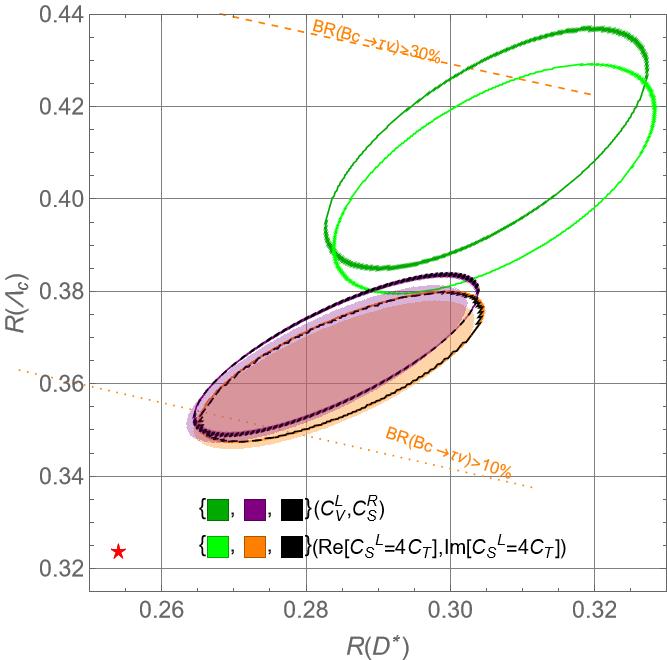}
\includegraphics[width=5cm, height=3.9cm]{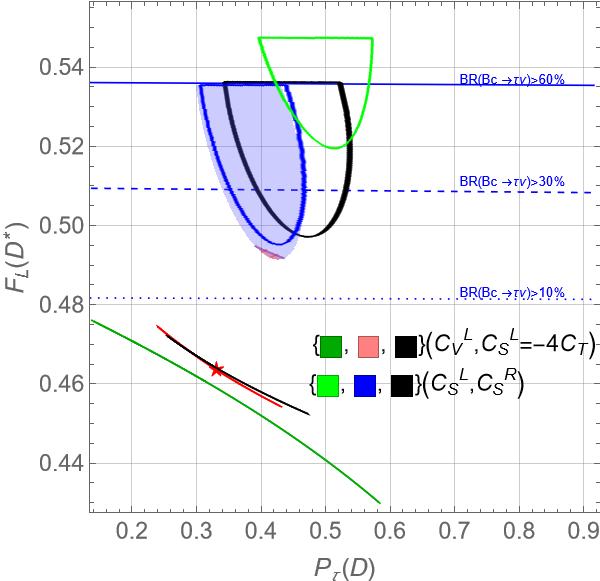}
\includegraphics[width=5cm, height=3.9cm]{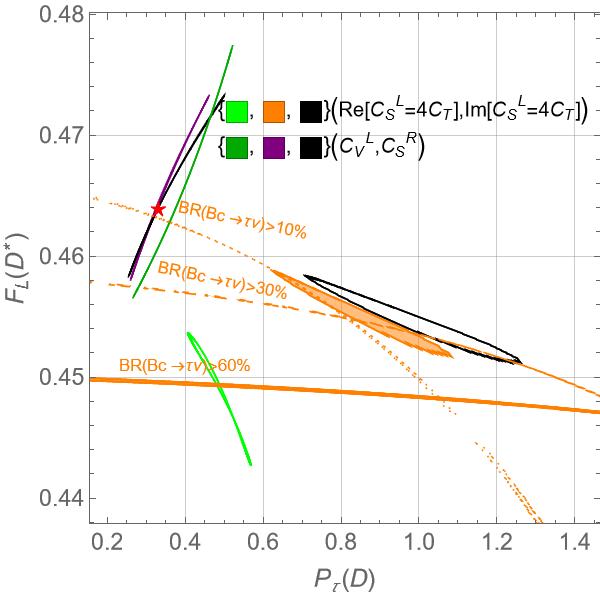}
\includegraphics[width=5cm, height=3.9cm]{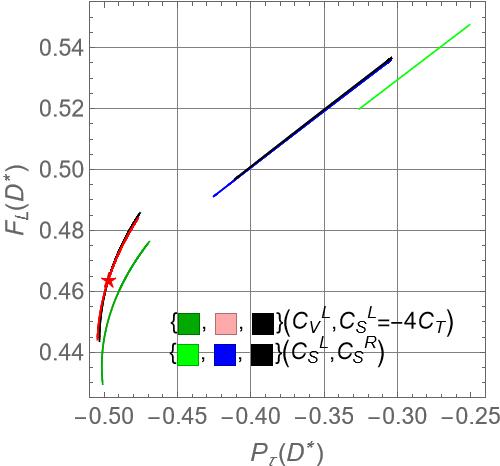}
\includegraphics[width=5cm, height=3.9cm]{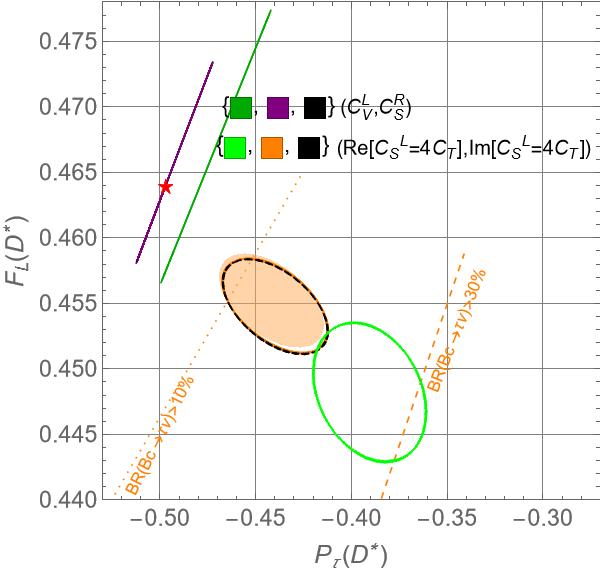}
\includegraphics[width=5cm, height=3.9cm]{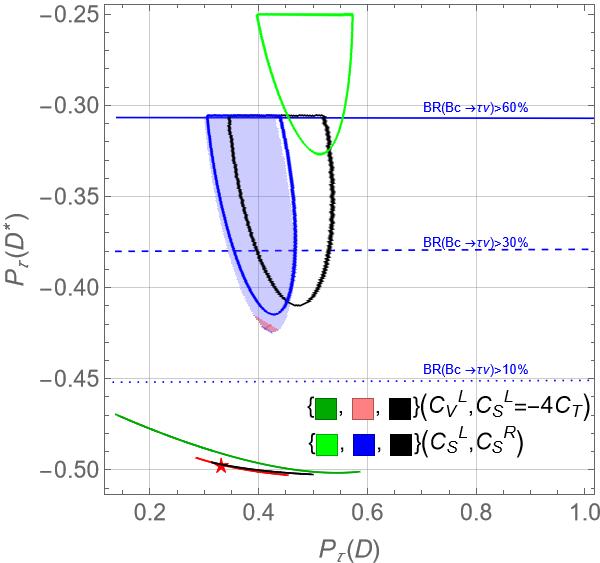}
\includegraphics[width=5cm, height=3.9cm]{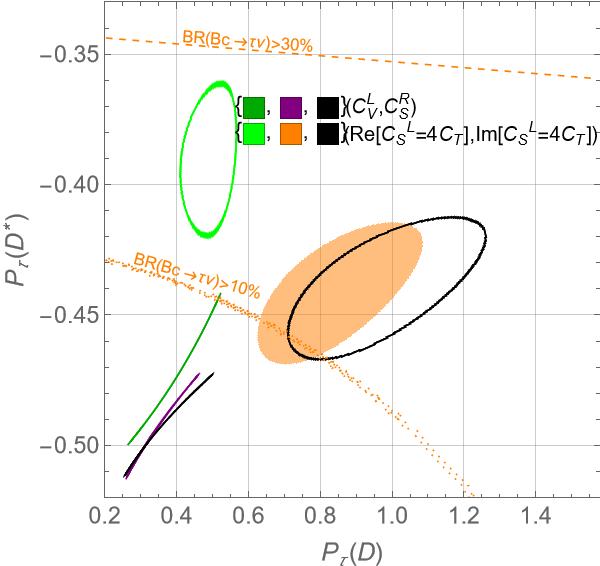}
\includegraphics[width=5cm, height=3.9cm]{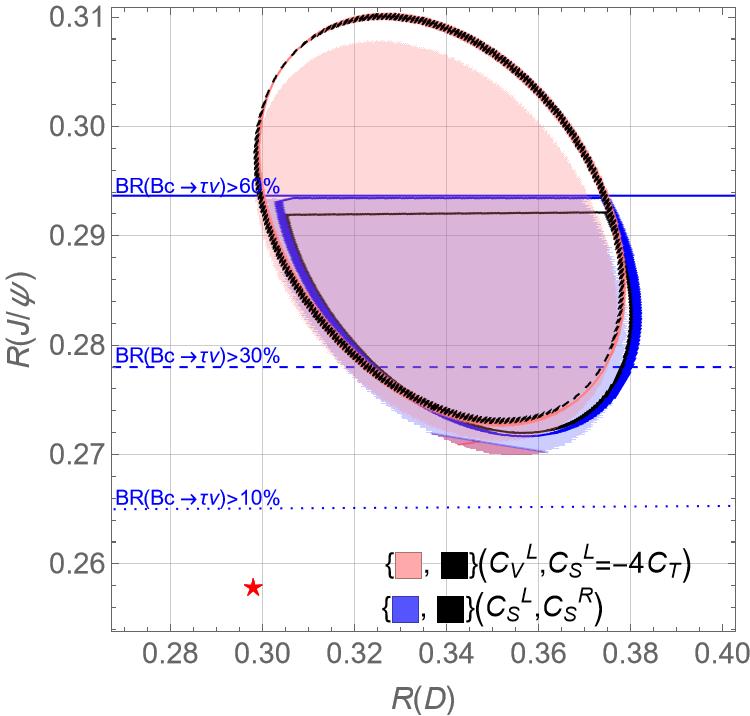}
\includegraphics[width=5cm, height=3.9cm]{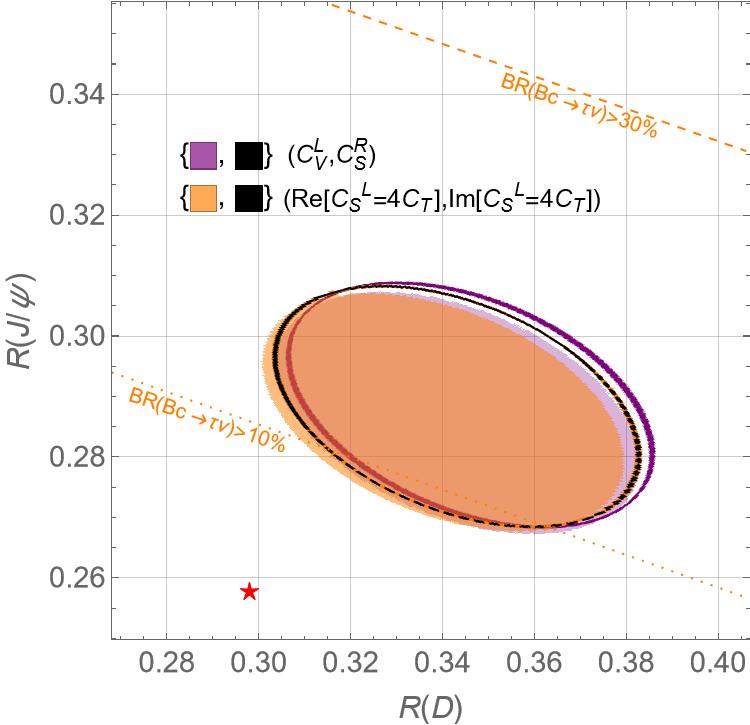}
\includegraphics[width=5cm, height=3.9cm]{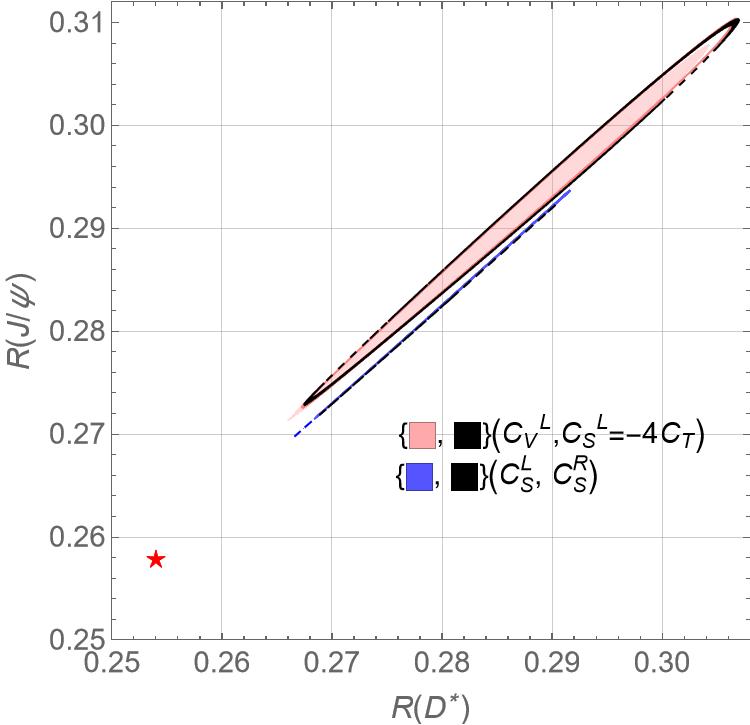}
\includegraphics[width=5cm, height=3.9cm]{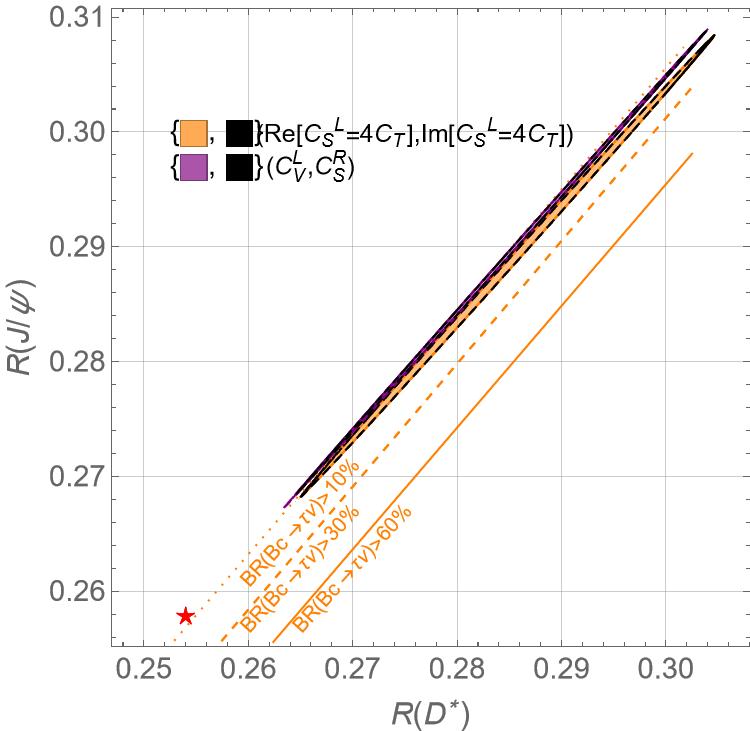}
\caption{\label{fig4cor}Correlations plots for $1\sigma$ with the $BR<60\%$. The green (black) contours represent the plots of ref. \cite{Blanke:2018yud} while filled (unfilled) contours for Fit C (A). The solid, dashed and dotted lines for $BR(60,30,10)\%$, respectively, while red star represents SM. The meshed regions corresponds the future constraints of $\tau\nu$ analysis. }
\label{correlationcv}
\end{figure}

\begin{figure}[H]
\centering
\includegraphics[width=5cm, height=3.9cm]{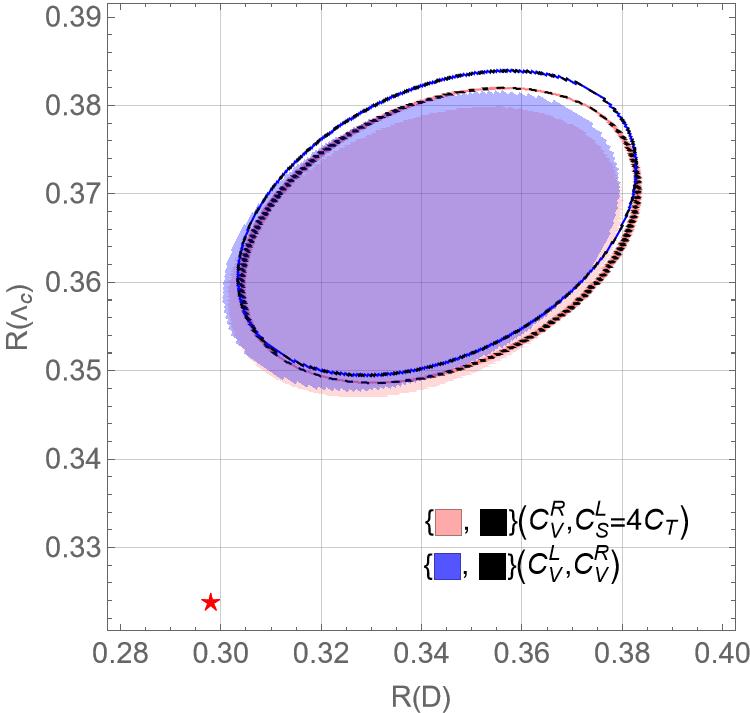}
\includegraphics[width=5cm, height=3.9cm]{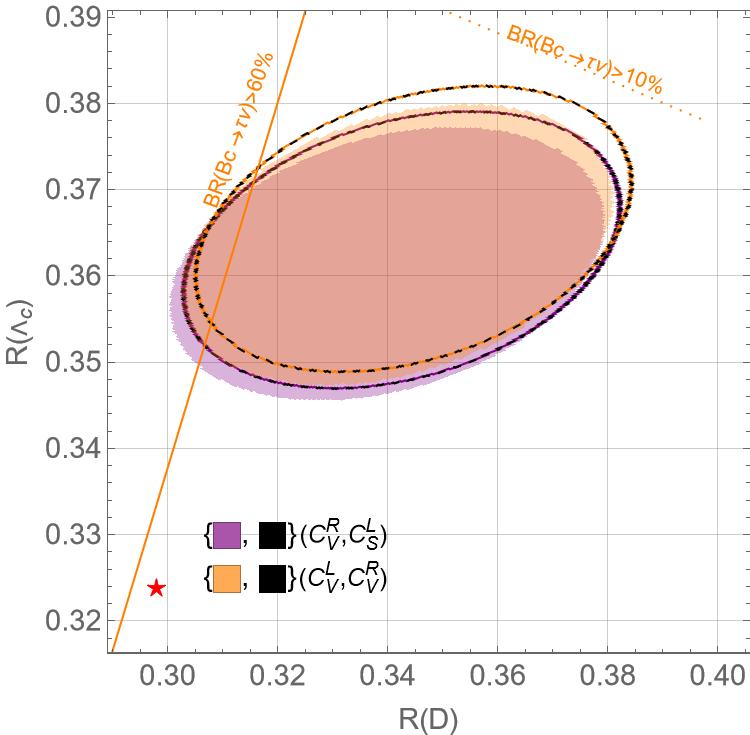}
\includegraphics[width=5cm, height=3.9cm]{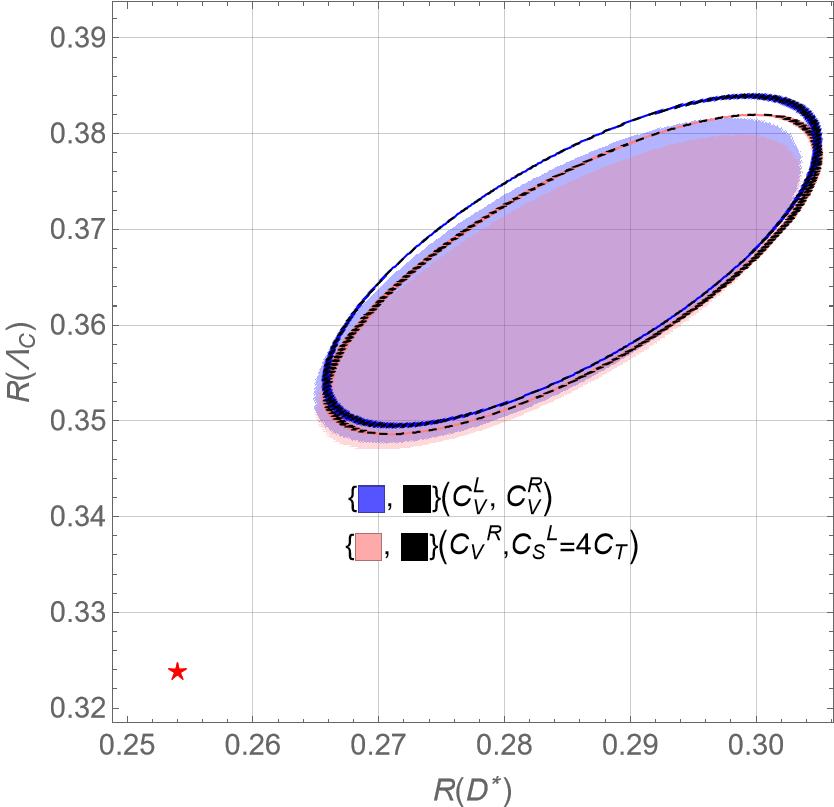}
\includegraphics[width=5cm, height=3.9cm]{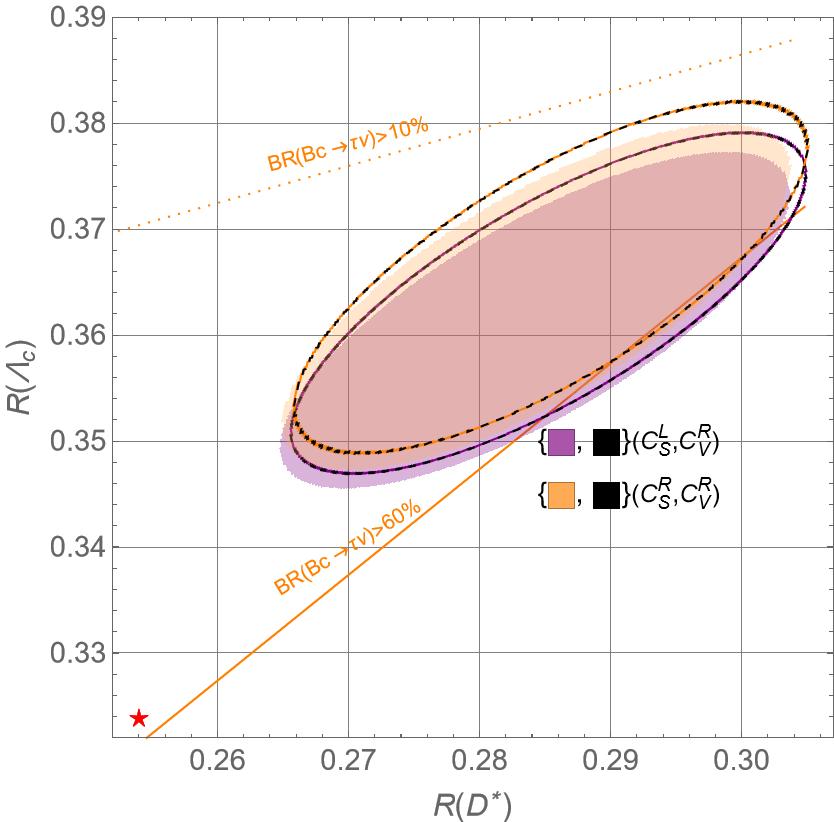}
\includegraphics[width=5cm, height=3.9cm]{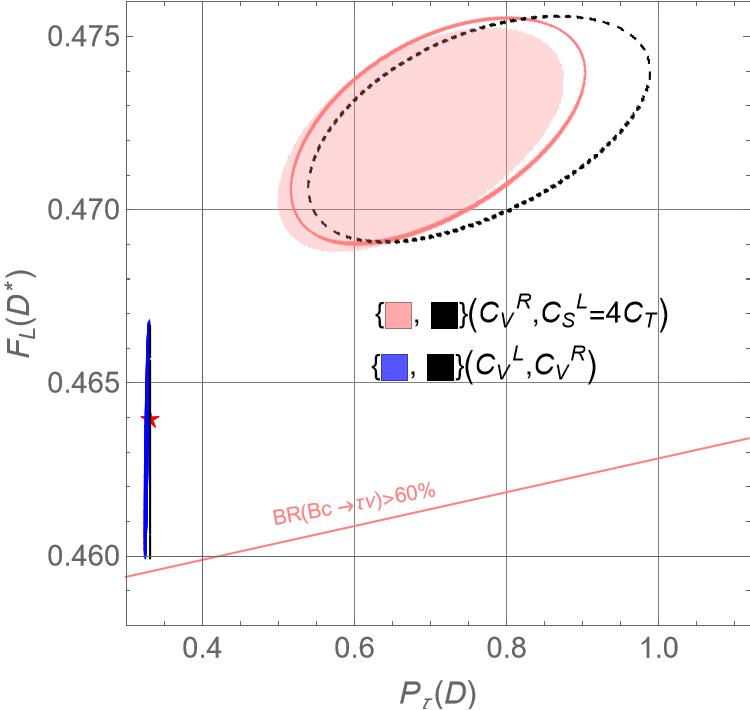}
\includegraphics[width=5cm, height=3.9cm]{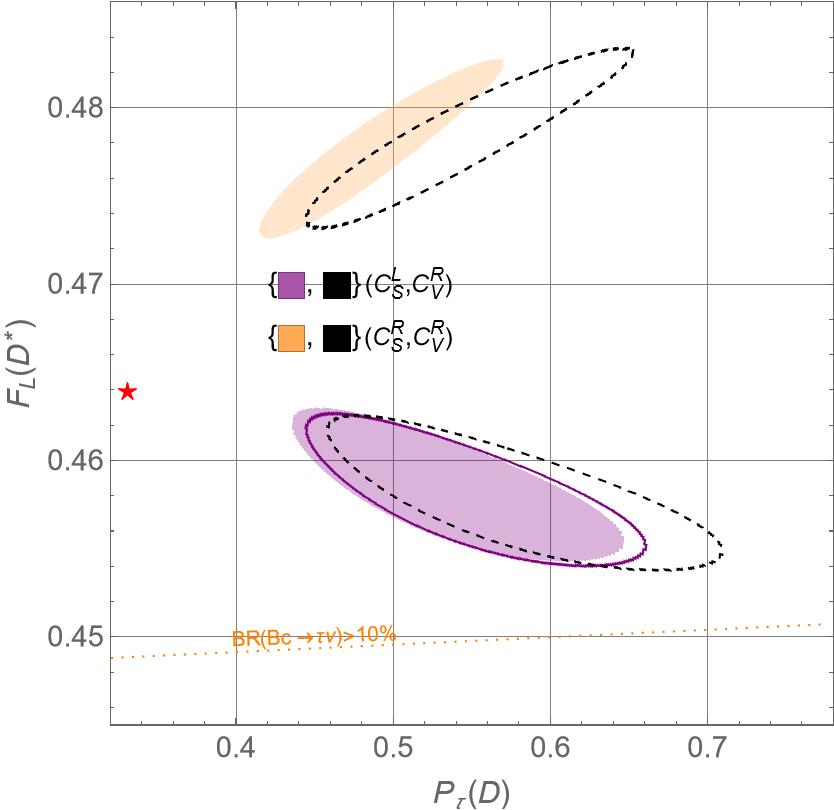}
\includegraphics[width=5cm, height=3.9cm]{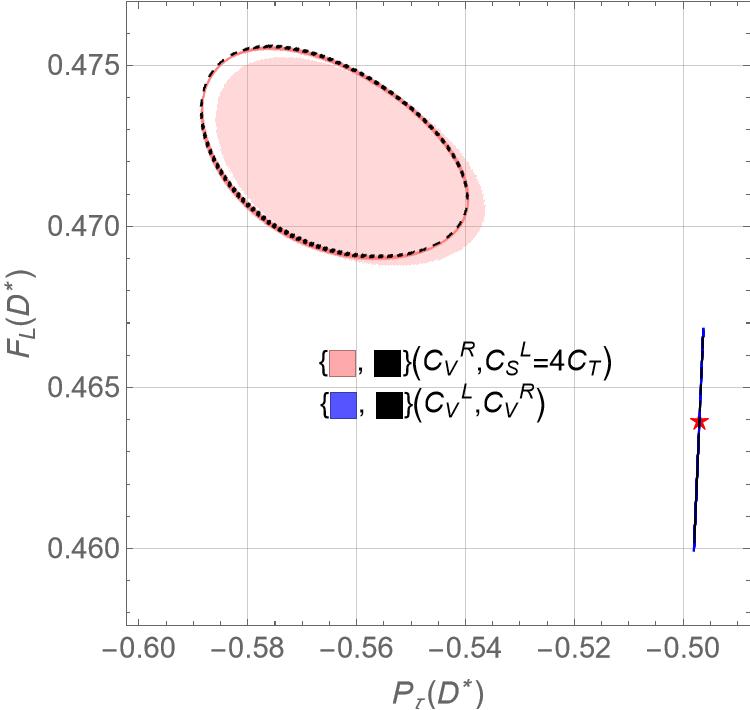}
\includegraphics[width=5cm, height=3.9cm]{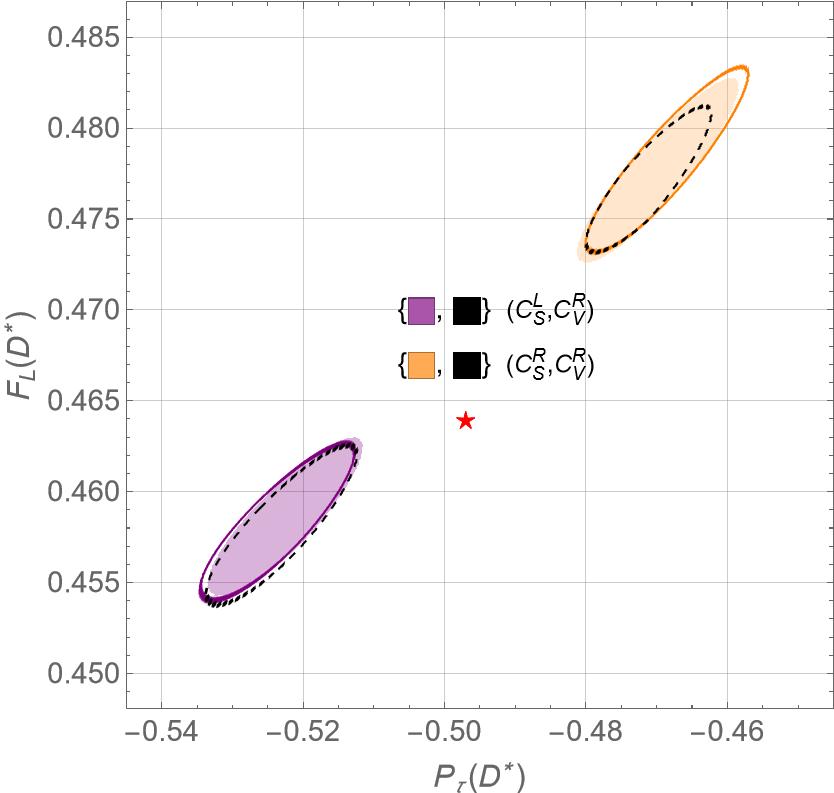}
\includegraphics[width=5cm, height=3.9cm]{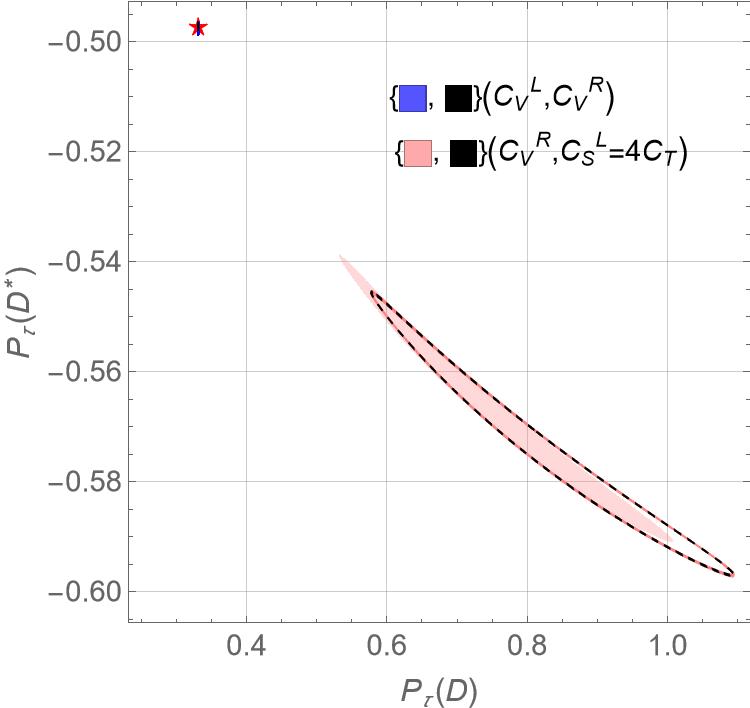}
\includegraphics[width=5cm, height=3.9cm]{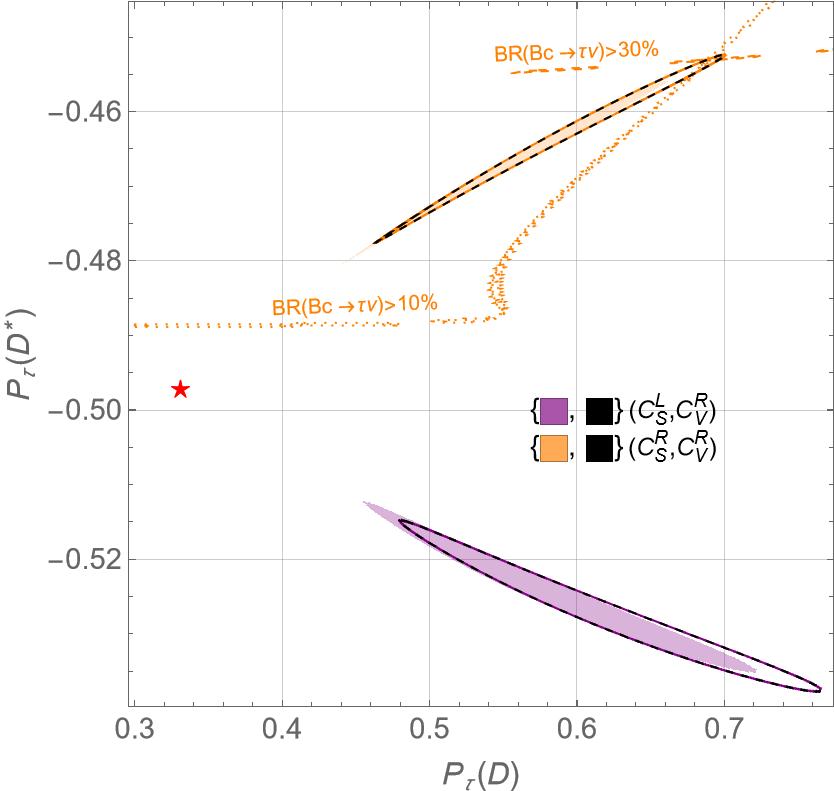}
\includegraphics[width=5cm, height=3.9cm]{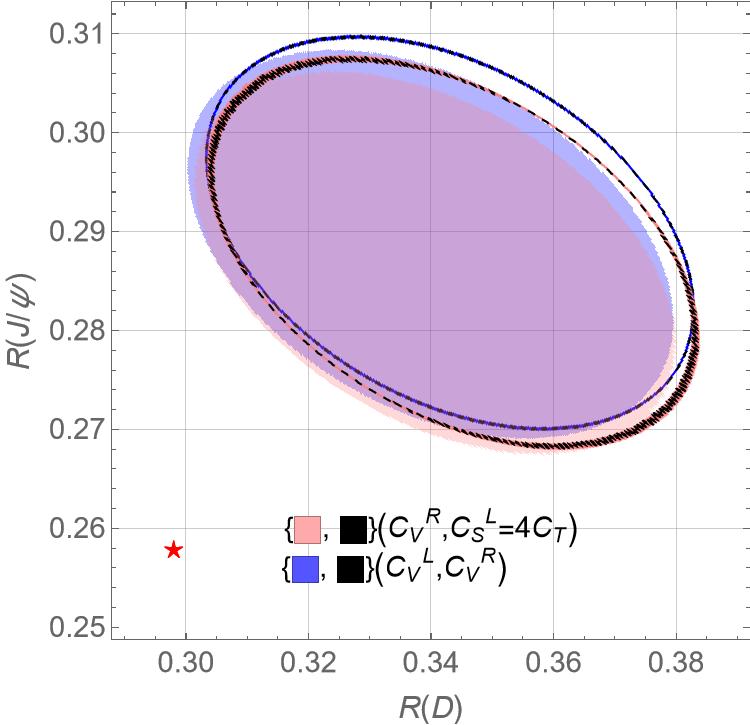}
\includegraphics[width=5cm, height=3.9cm]{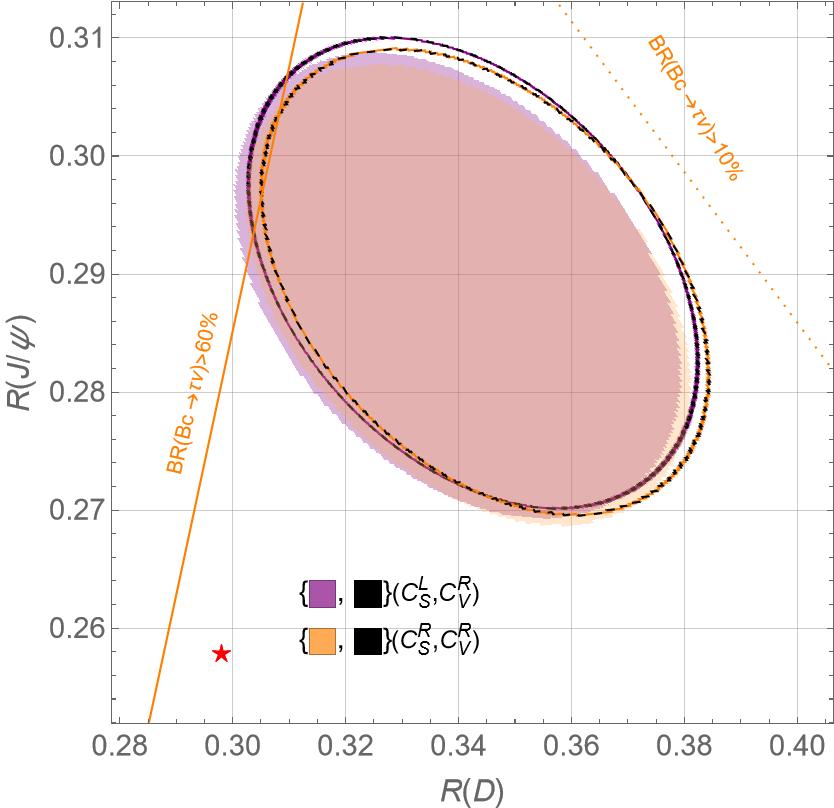}
\includegraphics[width=5cm, height=3.9cm]{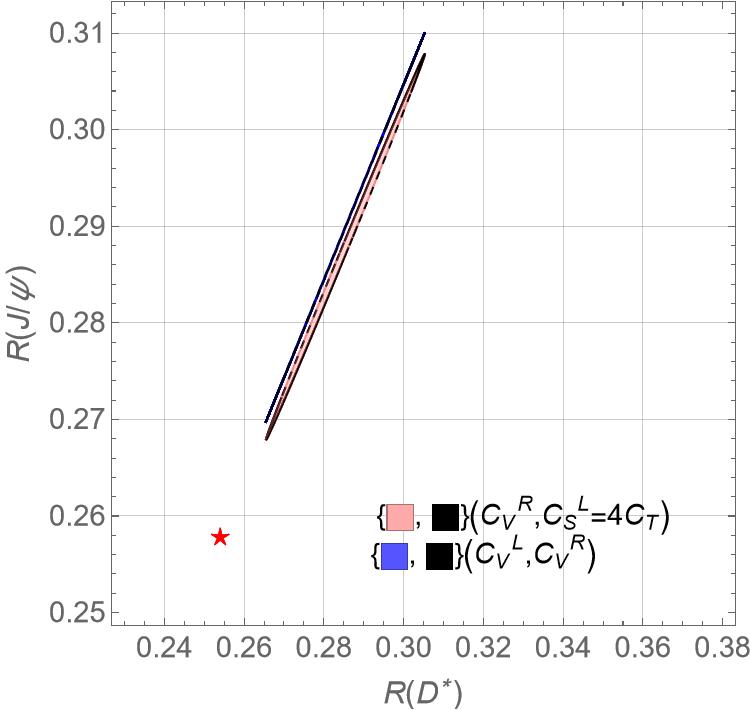}
\includegraphics[width=5cm, height=3.9cm]{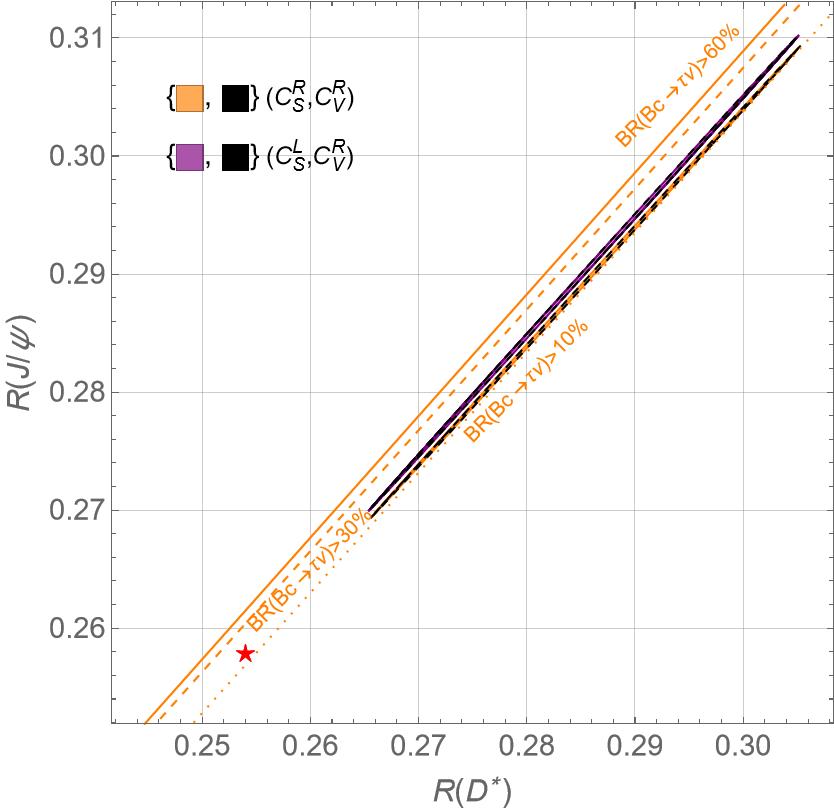}
\caption{\label{fig4cor1}The legends are same as described in FIG. \ref{fig4cor} but for $C_V^{R}$ related scenarios.}
\label{correlationcvr}
\end{figure}

It will be interesting to see how the the updated data change the correlation among the $R({D^{(*)}})$, $R(\Lambda_c)$, $F_L(D^*)$ and $P_\tau(D^{(*)})$ observables which are calculated in \cite{Blanke:2018yud}, and for that the updated and the previous results are shown in FIG. \ref{correlationcv} by using $1\sigma$ parametric range of 2D scenarios. It is worth mentioning here that these correlations are significantly effected by considering the updated values of $R({D^{(*)}})$ while the inclusion of $R(J/\psi)$ and  the recently measured $R(\Lambda_c)$ data  mildly effect the behavior of  correlations. However, to see their effects explicitly, we have also shown the correlations of $1\sigma$ parametric space of Fit-A (un-filled region) and Fit-C (filled region) in FIG. \ref{correlationcv}. In addition, to see the effects by the scale of WC's on the correlations, we have also calculated these correlations at 1 TeV for Fit A and shown by the black curves.  As mentioned above in section III-B that the updated values of $R({D^{(*)}})$ squeezed the parametric space of 2D scenarios significantly (see FIG. 2). Consequently, it shrink and lower the values of correlation regions among the observables $R({D^{(*)}})$ and $R(\Lambda_c)$, which is more close to the SM values of these observables as can be seen from the first four plots of FIG. 4. Similarly, the changes by the updated data of $R({D^{(*)}})$ in the correlation regions among the observables $P_{\tau}(D^{(*)})$ and $F_L(D^*)$ are depicted in plots five to ten of FIG. 4. In addition, the correlations among $R({D^{(*)}})$ and $R(J/\psi)$ are also shown in last four plots of FIG. \ref{correlationcv}. On the other hand, for the scenarios related to $C_{V}^R$, the correlation among the observables by using the $1\sigma$ parametric space are plotted in FIG. \ref{correlationcvr}.

Finally, from Eq. (\ref{discrepancy}), the predicted values of observables used in fitting analysis are also calculated by using the $1\sigma$ parametric space of the 2D NP scenarios and listed them in TABLEs \ref{2D} and \ref{2DPSAB1}. It can be seen that the predicted values of  $R({D^{(*)}})$ and $P_{\tau}(D^{*})$ show less than $1\sigma$ deviations except for the scenarios $(Re[C_S^L=4C_T],\text{Im}[C_S^L=4C_T])$ and $(C_{S}^L,C_{S}^R)$ for $10\%$ branching ratio. However, the observables $ R(J /\psi)$ and $F_{L}(D^{*})$ exhibit $2\sigma$ deviation except the scenario $(C_{S}^L,C_{S}^R)$ for $60\%$ branching ratio for $ R(J /\psi)$ . Similarly, the predicted value of $R(\Lambda_c)$ is showing approximately $2\sigma$ deviation with respect to its experimentally measured numbers. 

\section{Sensitivity of angular observables to New Physics (NP)}\label{Senseang}
For the NP point of view, it is important to mention here that the form factors are the main source of hadronic uncertainties, consequently, generate the errors in the theoretical predictions which may preclude the effects of NP. Therefore, we need to select those observables which are not only sensitive to NP but also the variation in their values in the presence of NP may provide a discriminatory tool among the different NP scenarios.

To accomplish this purpose, we have considered the lepton forward-backward asymmetry ($A^\ell_{FB}$), the forward-backward
asymmetry of transversely polarized $D^*$ meson ($A^{T,L}_{FB}$), the longitudinal polarization fraction of the $D^*$ meson ($F_L(D^{*})$), the ratios ($R_{A,B}$, $R_{L,T}$), and the angular asymmetries: $A_3$, $A_4$, $A_5$, $A_{6s}$ for the decay channel $B\to D^{\ast}\tau\nu_\tau$ which are relatively clean and also sensitive to the NP. Therefore, the variation in their values in the presence of 1D and 2D NP scenarios (under consideration) could be used to discriminate these NP scenarios. These CP-even angular observables are discussed in detail in the literature and their analytical expressions in terms of angular coefficients, $I'$s, can be found in \cite{Alok:2016qyh,Mandal:2020htr,Becirevic:2019tpx}. Furthermore, in the current study, we rely on the form factors which are calculated in ref. \cite{Alok:2016qyh,Mandal:2020htr} and observables mentioned above have been presented with their theoretical uncertainties.

To see the sensitivity of these angular observables to NP, we have plotted them against the square of transverse momentum, $q^2$, in FIGs. \ref{1Dang} and \ref{2Dang} for 1D and 2D NP scenarios, respectively. In these figures, the black(gray) band shows the SM values of these observables where the width corresponds to the uncertainty in the values due to the form factors. The color bands represent their values in the presence of NP. For the NP dependence of these observables, we have used the central values of the form factors and the width of light and dark color bands show the uncertainty due to $1\sigma$ and $2\sigma$ intervals in the NP WCs at 2 TeV, respectively. The effects of different 1D and 2D NP scenarios on the above mentioned angular observables are discussed in the following sections. In addition, to see the direct influence of the scenarios on observables, we have also found the expressions $I's$ in terms of NP WCs, $C^{L(R)}_i$, after integrating $q^2$ and these are given in Appendix.

One can immediately see that the expressions of coefficients $I'$s in terms of NP WCs make the study of NP in different angular observables quite trivial. Therefore, by using these expressions of $I'$s, we have calculated the variation in the amplitude of different angular observables in the presence of NP and shown by the bar plots in FIGs. \ref{1Dbar} and \ref{2Dbar} for 1D and 2D NP scenarios, respectively. The corresponding SM predictions and values at different scenarios are also listed in TABLEs \ref{SMang1} and \ref{SMang2}, respectively.

As we have mentioned above in Section \ref{Fit} that for the scenarios: $(C^L_V,C^L_S=-4C_{T})$, $(C^L_V,C^R_S)$, the allowed parametric space of NP does not significantly change by the number of observables and the branching ratio constraints, while the allowed parametric space of the scenarios: $(Re[C_S^L=4C_T],\text{Im}[C_S^L=4C_T])$, $(C^L_S,C^R_S)$ are effected only by the branching ratio constraints. It is worth mentioning here that we have found the allowed $1\sigma$ and $2\sigma$ parametric space for the Fit A, B and C, are almost same. However, to see the impact of NP effects on the numerical values of angular observables, we use the $1\sigma$ and $2\sigma$ parametric space with the $60\%$ branching ratio that are given in TABLE  \ref{2sigma} with the constraints coming from LHC bounds on the 2D scenarios: $(C^L_V,C^L_S=-4C_{T})$, $(C^L_V,C^R_S)$ which are discussed in section (\ref{CB}).
\subsection{Effects of 1D scenarios on observables}
\begin{figure}[H]
\centering
\includegraphics[width=5cm,height=5cm]{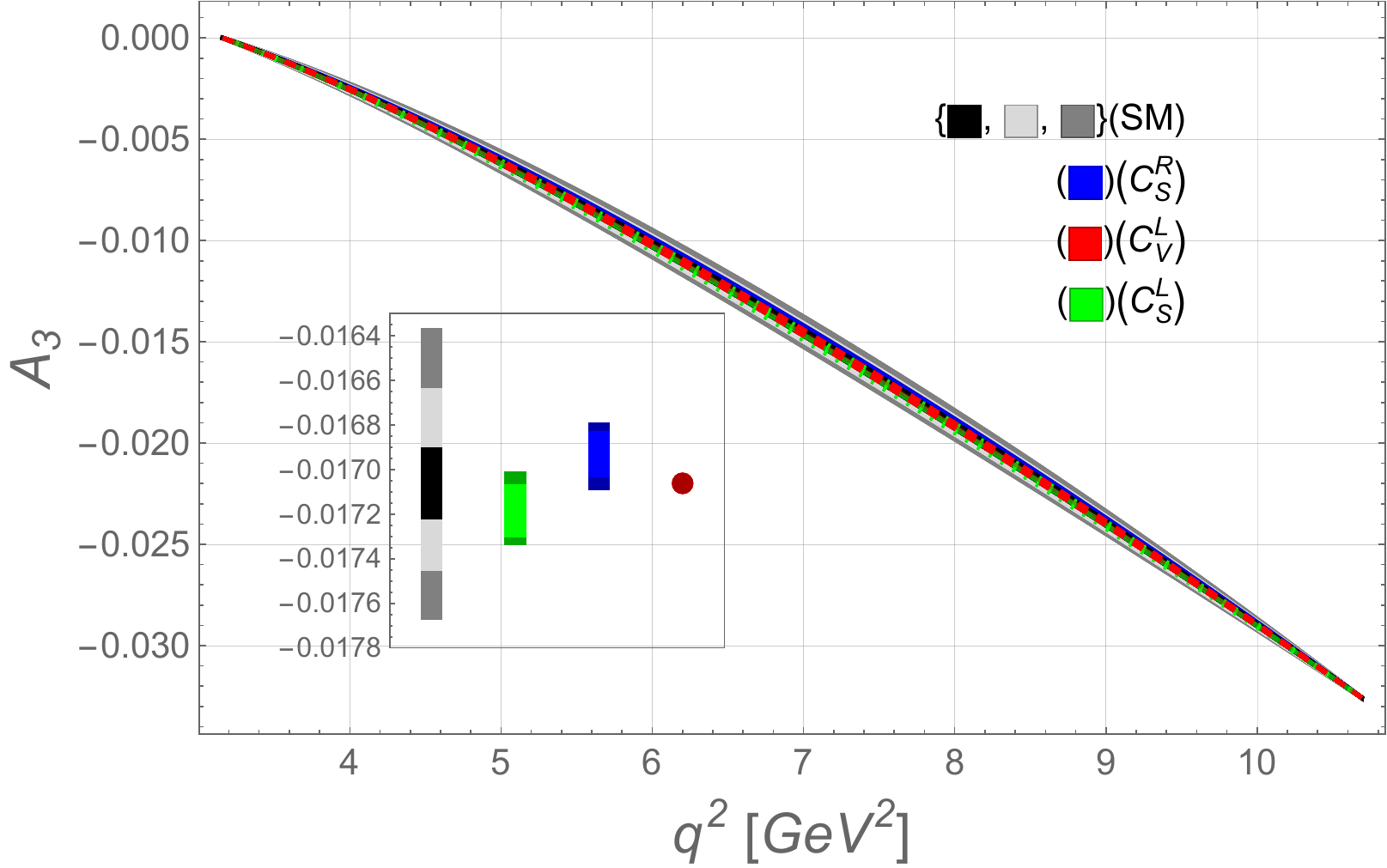} 
\includegraphics[width=5cm,height=5cm]{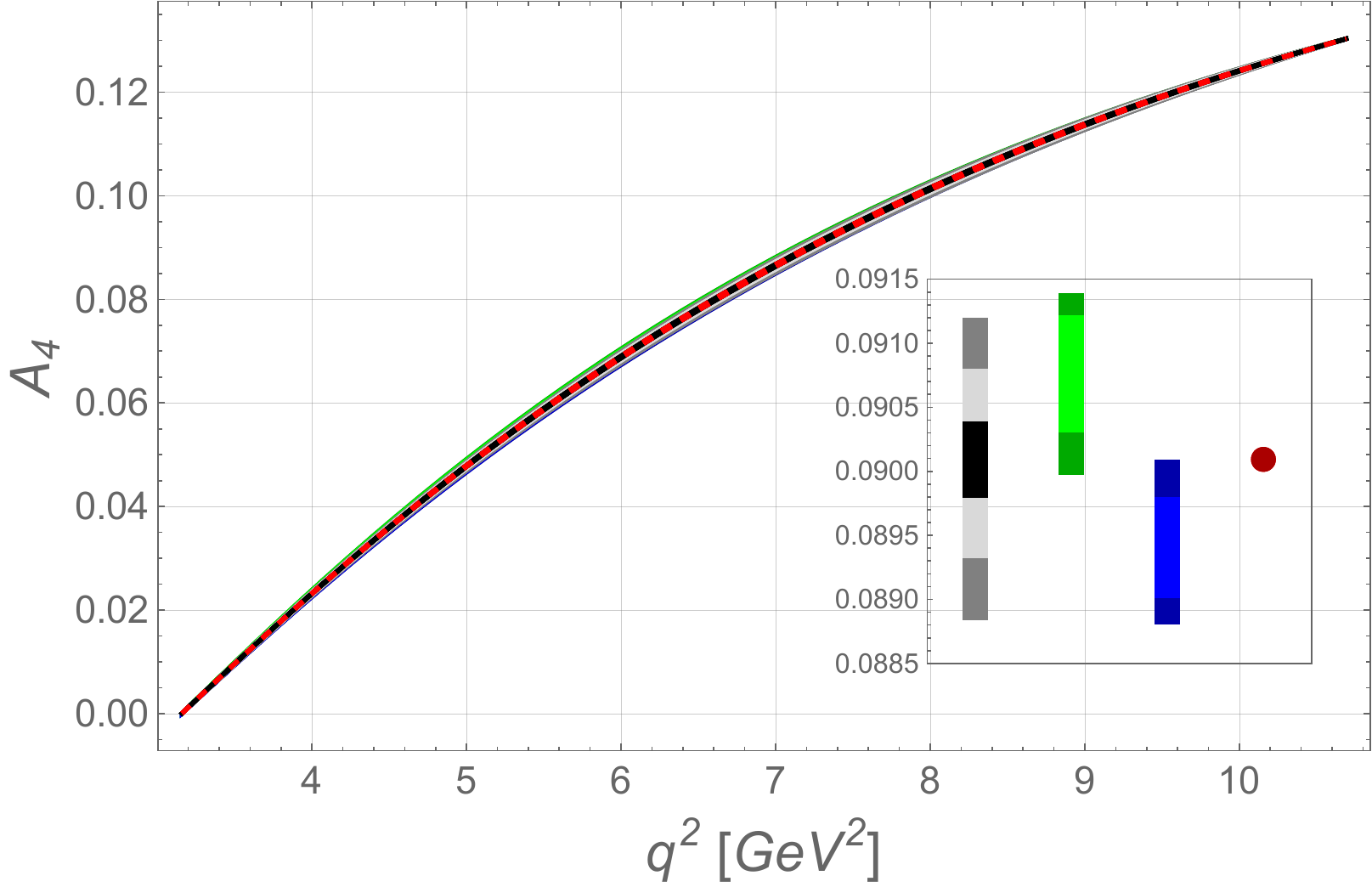}
\includegraphics[width=5cm,height=5cm]{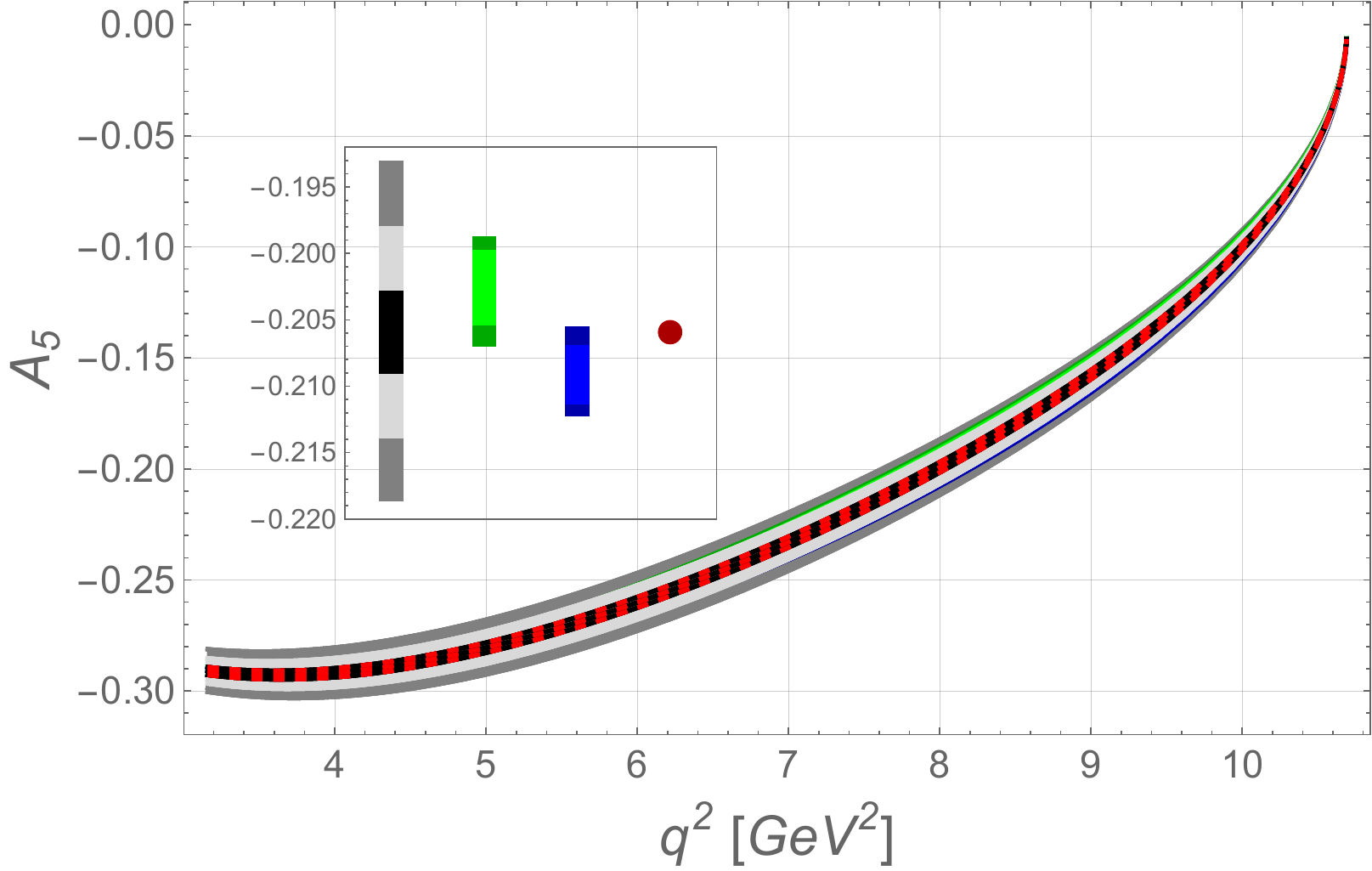}
\includegraphics[width=5cm,height=5cm]{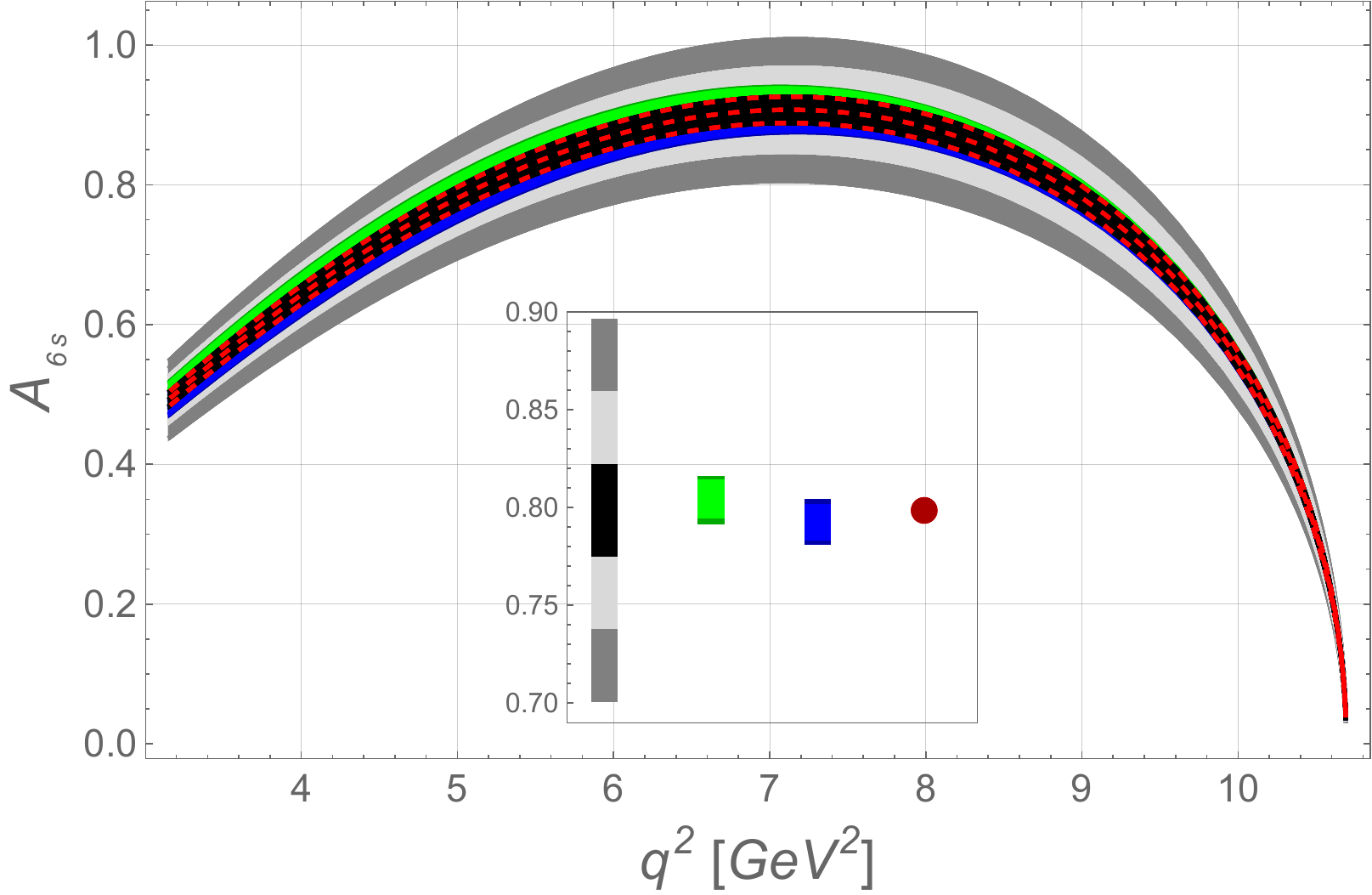}
\includegraphics[width=5cm,height=5cm]{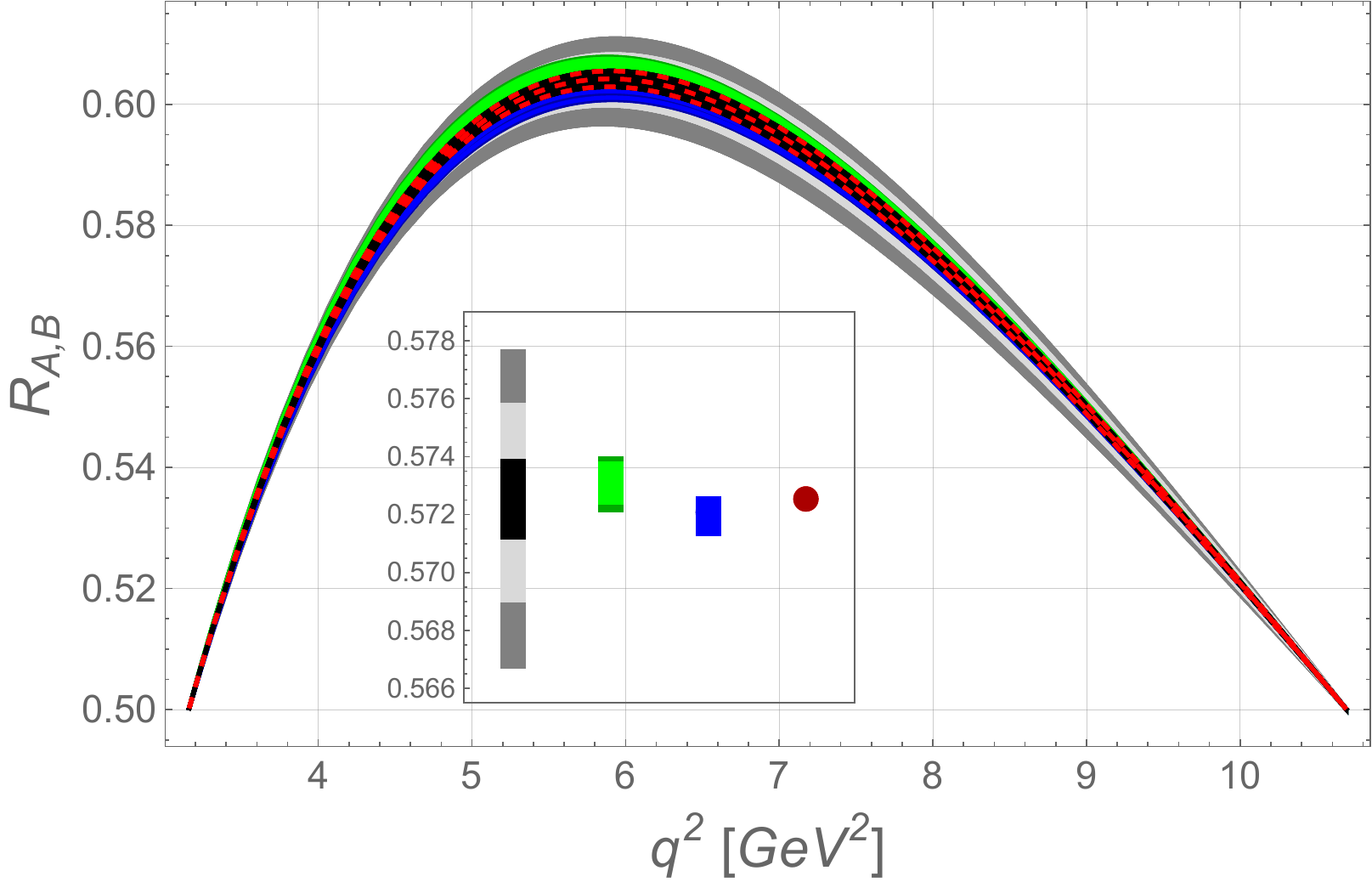}
\includegraphics[width=5cm,height=5cm]{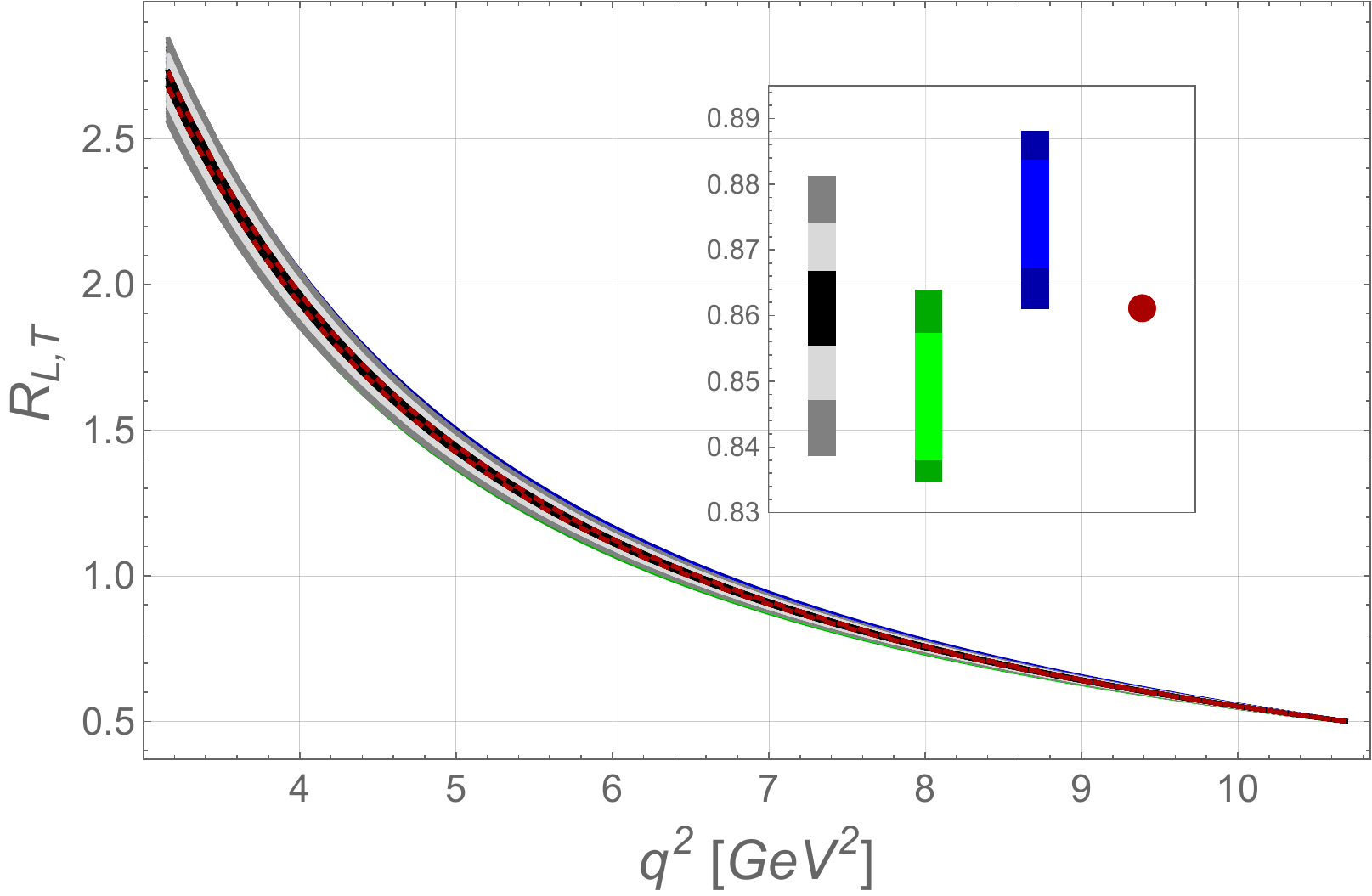}
\includegraphics[width=5cm,height=5cm]{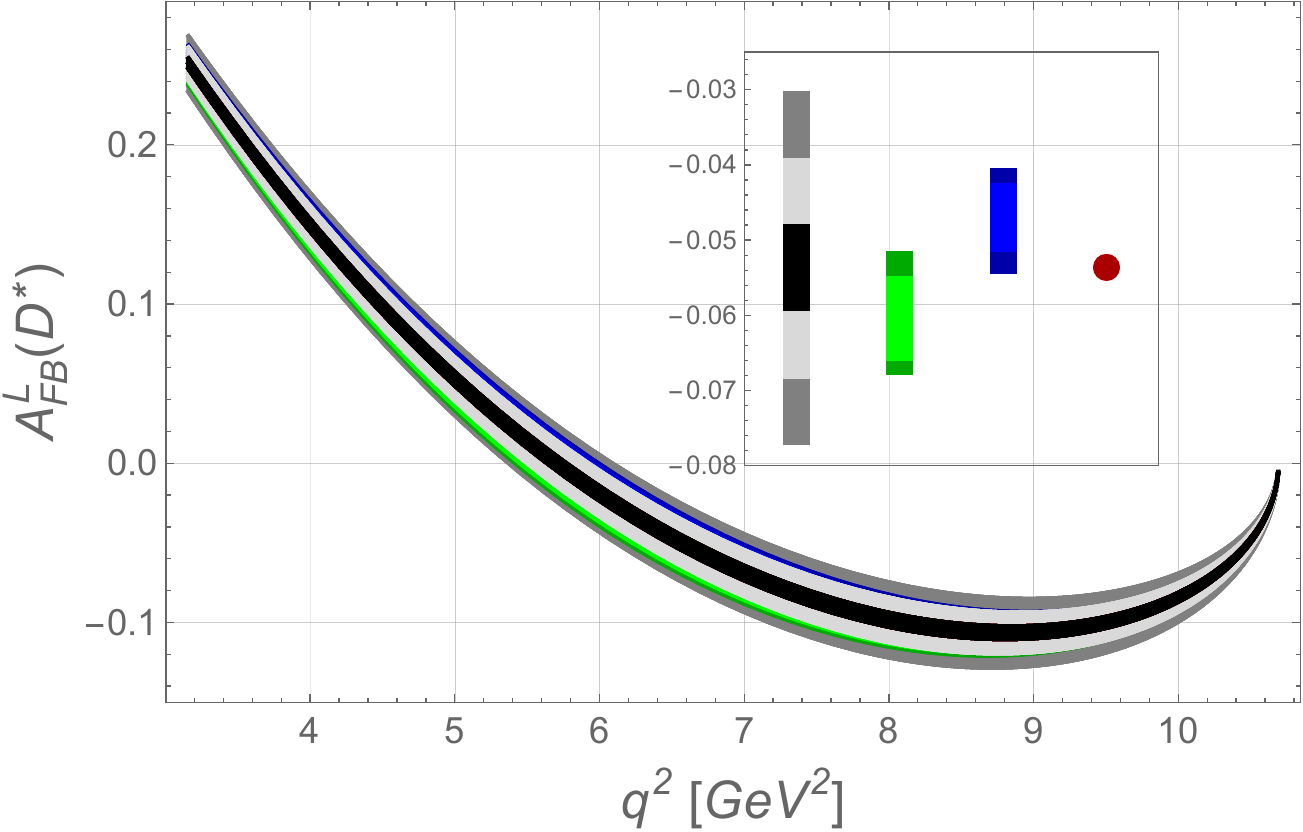}
\includegraphics[width=5cm,height=5cm]{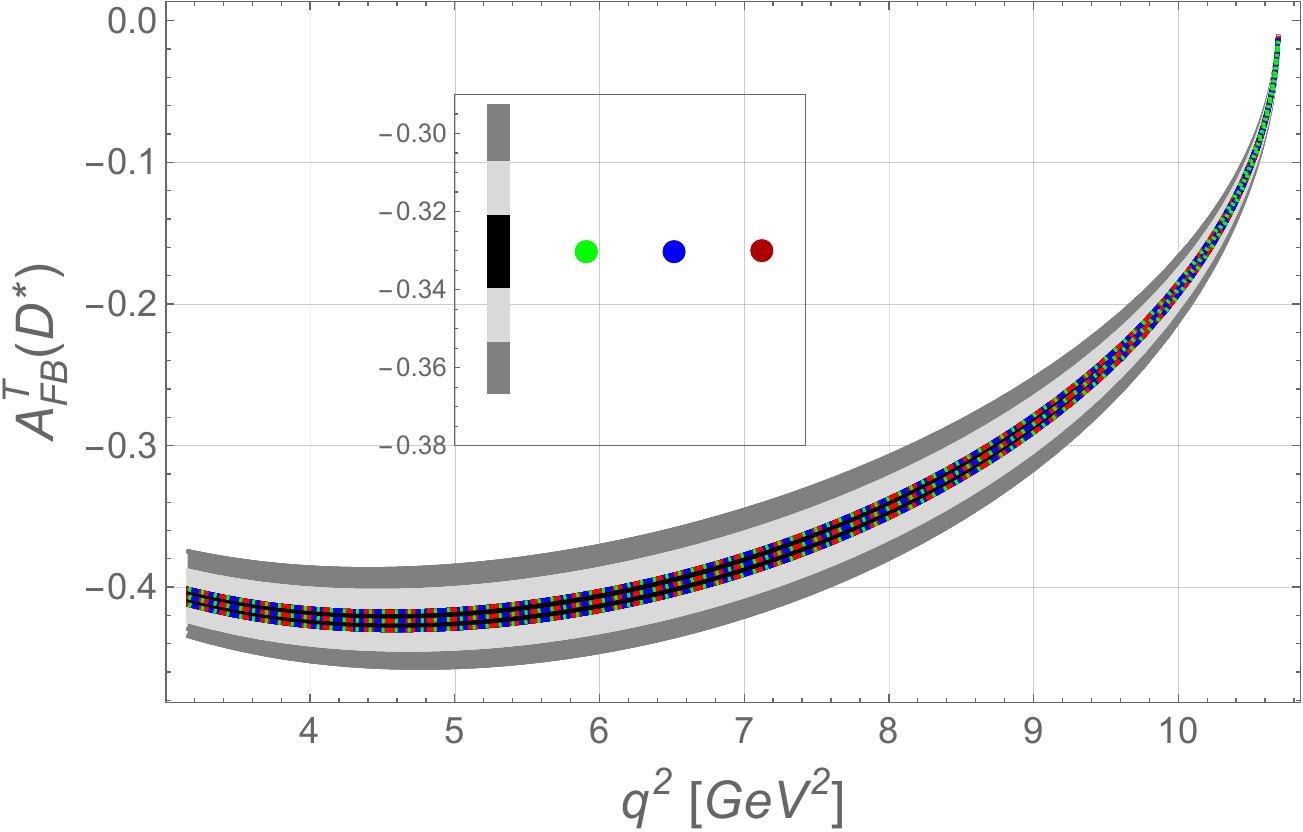}
\includegraphics[width=5cm,height=5cm]{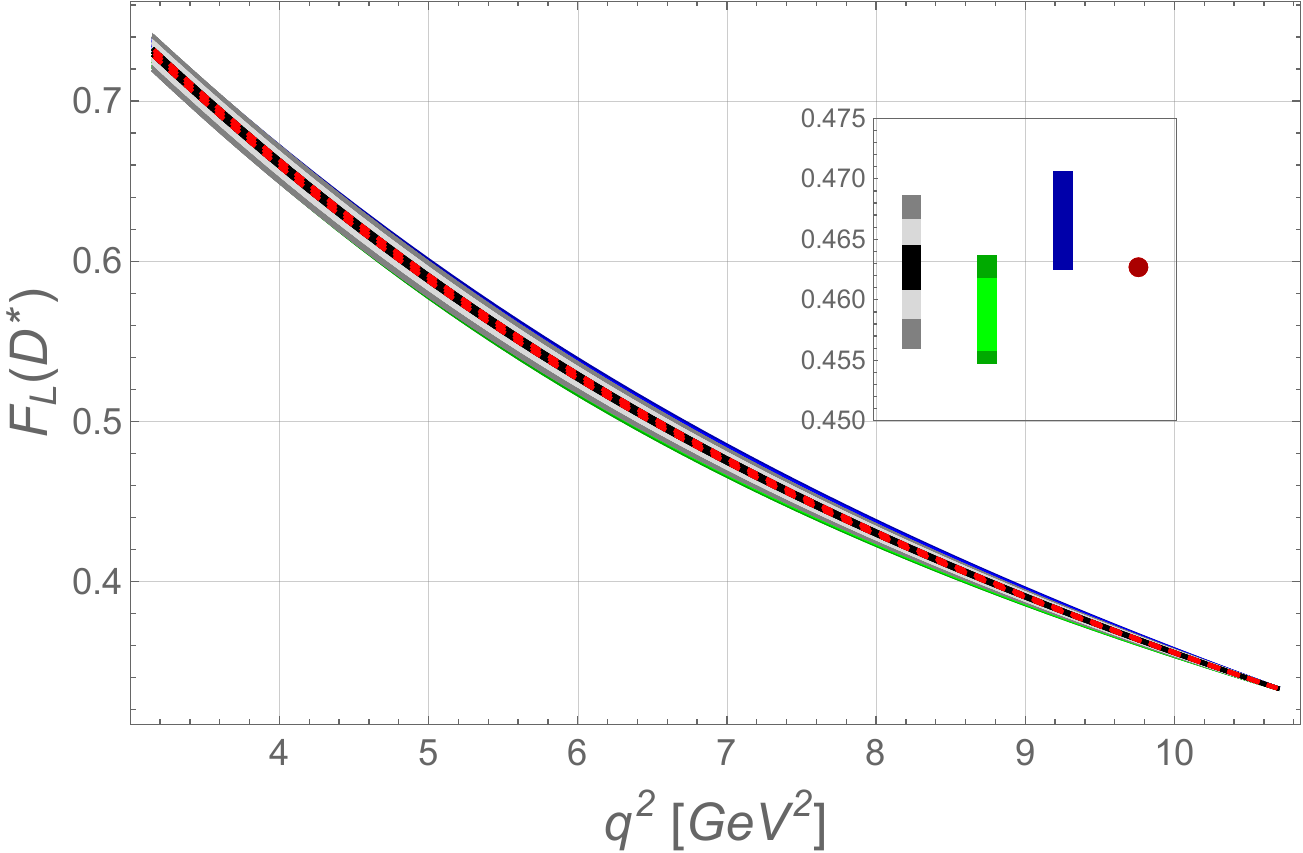}
\includegraphics[width=5cm,height=5cm]{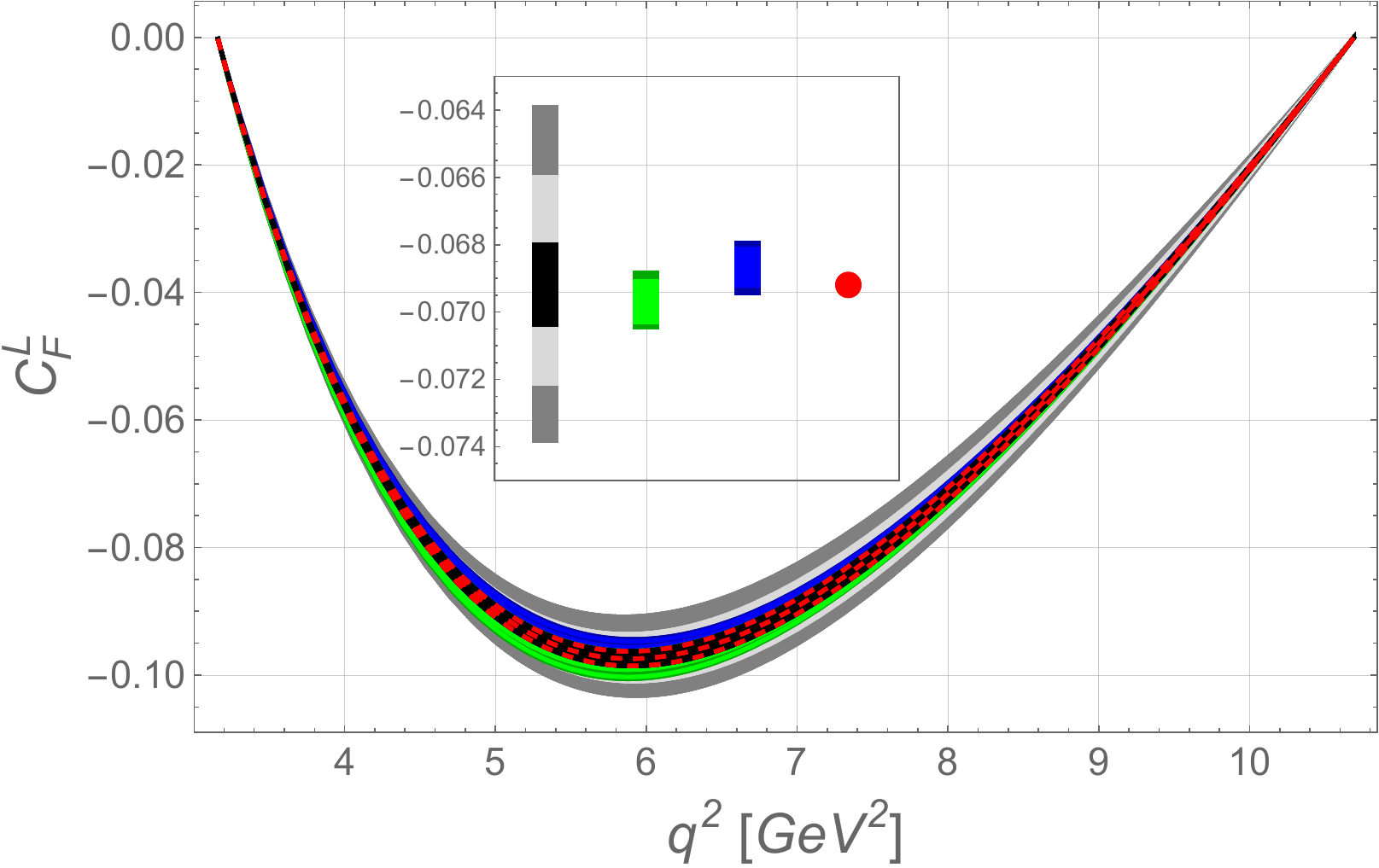}
\caption{ $A_{3-5}$ , $A_{6s}$, $R_{A,B}$, $R_{L,T}$, $A_{FB}^{L,T}(D^*)$, $F_{L}(D^*)$ and $C_{F}^{L}$  for $\bar{B}\rightarrow D^* \tau \bar{\nu}$ are shown for allowed values of NP couplings for 1D scenarios as function of $q^2$. The black, light gray and gray portions represent the variation in the SM value  due to the uncertainties of the form factors, and 5$\%$ and 10$\%$ statistical uncertainties of future experiments, respectively. While the light and dark color bands reflect the 1$\sigma$ and 2 $\sigma$ranges of NP scenarios.}
\label{1Dang}
\end{figure}

\begin{figure}[H]
     \begin{subfigure}
     [b]{0.35\textwidth}
         \centering
\centering{}
\includegraphics[width=16cm,height=5cm]{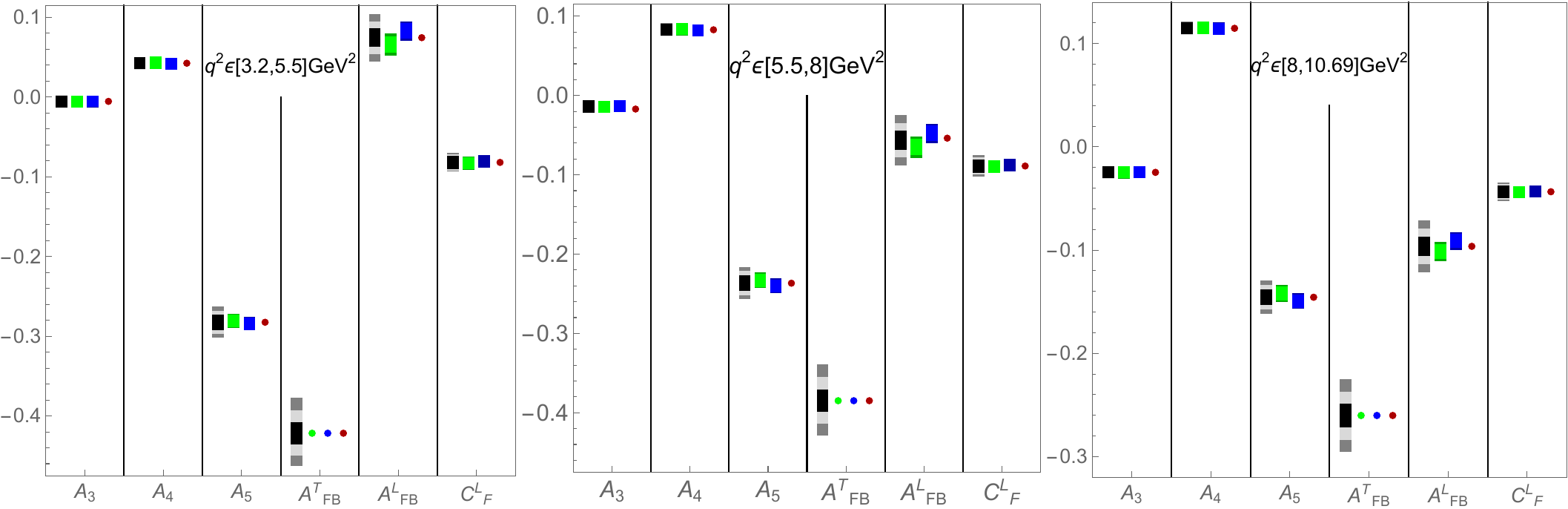}
    \hspace{1.5cm}
    \newline
    \includegraphics[width=16cm,height=5cm]{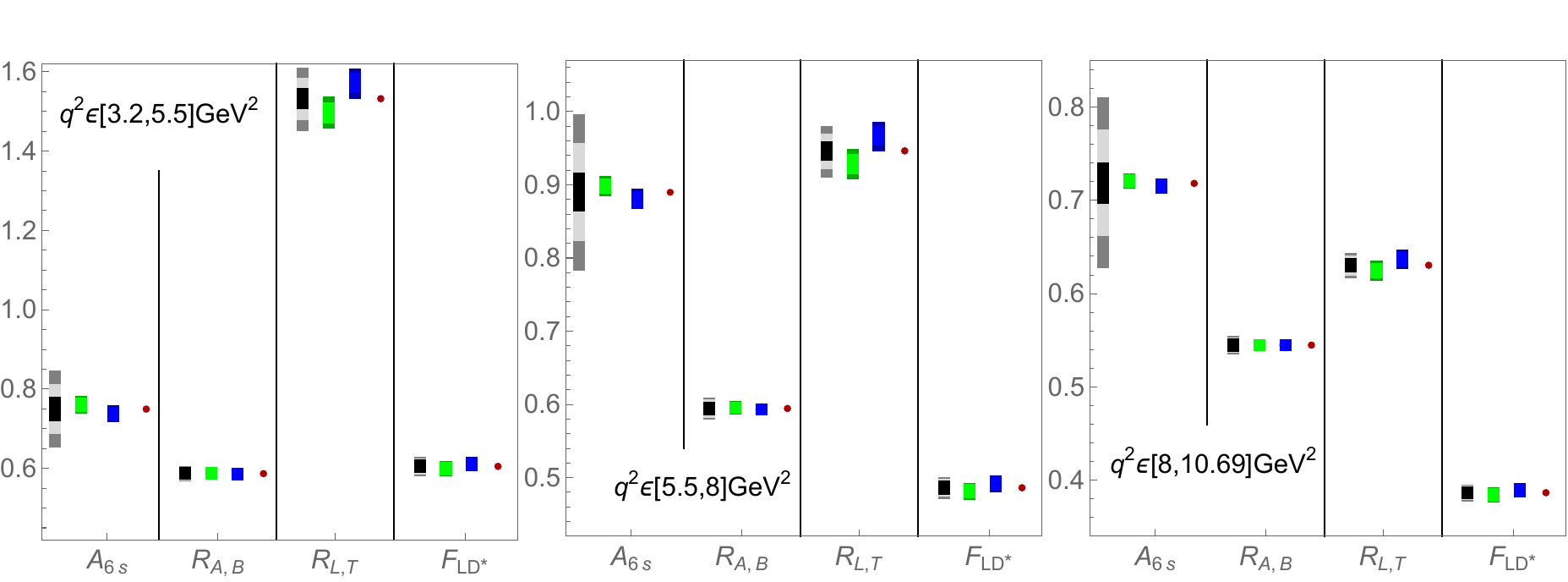}

\end{subfigure}
\caption {The bar plots of  $A_{3-5}$, $A_{6s}$, 
 $R_{A, B}$, $R_{L, T}$, $A_{FB}^{L, T}(D^*)$, $F_{L}(D^*)$ and $C_{F}^{L}$  for $\bar{B}\rightarrow D^* \tau \bar{\nu}$  are shown. The bar plots in first and second row show the numerical values in different $q^2$ bins. The black bar shows the SM variation of angular observables and the color bars show the variation in the numerical values in the different NP scenarios.}
\label{1Dbar}
\end{figure}

The sensitivity of different angular observables by taking one of the NP WCs, $C^{(L,R)}_{i}$, is set to be non zero and are plotted in FIG. \ref{1Dang}. In these plots, we took the numerical values of the NP WCs after imposing the current prospects of the the collider bounds. As for scenario $C_{S}^L=4C_T$ ,$Pull_ {SM}$ value is too small, therefore, this scenario is not included in the $q^2$ analysis. In this figure, the black, light gray and gray portions represent the variation in the SM value  due to the uncertainties of the form factors, and 5$\%$ and 10$\%$ statistical uncertainties of future experiments, respectively. One can notice that the 1D scenarios are almost preclude by these uncertainties. However, if the future experimental uncertainties are limited to 5$\%$ then the values of the observables $A_4,R_{L,T}, F_L^D$ are exceeded to their SM values which may help to distinguish the 1D NP scenarios. Therefore, the bin wise precise measurements of the observables are also important to probe the NP and to distinguish among different NP scenarios. For this purpose, we have also calculated the full variation in the values of observables by using the $1\sigma$ to $2\sigma$ ranges of WC's of 1D scenarios after integrating the full and the different $q^2$ bins and are shown by the bar plots in FIG. \ref{1Dbar}. The SM values and the values in the different 1D NP scenarios of the angular observables in the full $q^2$ region are also given in TABLE. \ref{SMang1}. \\

\begin{table}[H]
\centering{}%
\scalebox{0.95}{
\begin{tabular}{|c|c|c|c|c|}
\toprule  
\hline
\hline
\vspace{0.05cm}
Observables  & SM   &$C_{S}^{L}$ & $C_{S}^R$& $C_{V}^L$   
\tabularnewline
\midrule
\hline
$A_{3}$  & $-0.0170\pm0.0001$  & $-0.0171\pm0.0001$ & $-0.0169\pm0.0001$&$-0.0170\pm0.0001$  
\tabularnewline
\midrule 
\hline
 $A_{4}$  & $0.0900\pm0.0002$  & $0.0907\pm0.0006 $ & $0.0894\pm0.0005$&$0.0900\pm0.0002$ 
 \tabularnewline
\midrule 
\hline
$A_{5}$  & $-0.2059\pm0.002$  & $-0.2025\pm0.003 $ &  $-0.2091\pm0.003$&  $-0.2059\pm0.002$
\tabularnewline
\midrule 
\hline
$A_{6s}$  & $0.7985\pm0.017$  & $0.8043\pm0.005 $ &  $0.7924\pm0.005$&$0.7985\pm0.017$ 
\tabularnewline
\midrule 
\hline
$R_{A,B}$  & $0.5725\pm0.0009$  & $0.5730\pm0.0005$  & $0.5719\pm0.0001$ &  $0.5725\pm0.0009$
\tabularnewline
\midrule
\hline
$R_{L,T}$  &$0.8611\pm0.0036$ & $0.8475\pm0.011$ & $0.8754\pm0.011$& $0.8611\pm0.0036$
\tabularnewline
\midrule 
\hline
$A_{FB}^{T} $  &$-0.3302\pm0.006$  & $-0.3302\pm0.006$ &$-0.3302\pm0.006$ &$-0.3302\pm0.006$
 \tabularnewline
\midrule 
\hline
$A_{FB}^{L} $  & $-0.0536\pm0.004$  &  $-0.0603\pm0.006$ &  $-0.0469\pm0.005$ & $-0.0536\pm0.004$
\tabularnewline
\midrule 
\hline
 $F_{L}(D^{*}) $  & $0.4626\pm0.001$  & $0.4587\pm0.004$ &$0.4667\pm0.003$    &$0.4649\pm0.001$ 
 \tabularnewline
\midrule 
\hline
$C_{F}^{L}$   &  $-0.0691\pm0.0009$  & $-0.0697\pm0.0006$  & $-0.0686\pm0.0005$& $-0.0691\pm0.0009$
\tabularnewline
\midrule
\hline
\bottomrule
\hline
\hline
\end{tabular}}
\caption{ The SM and the NP values of angular observables at $2\sigma$ allowed parametric space in the full $q^2$ region for different 1D scenarios.}
\label{SMang1} 
\end{table}

\subsection{Effects of 2D scenarios on observables}
\begin{figure}[H]
\centering
\includegraphics[width=5cm,height=5cm]{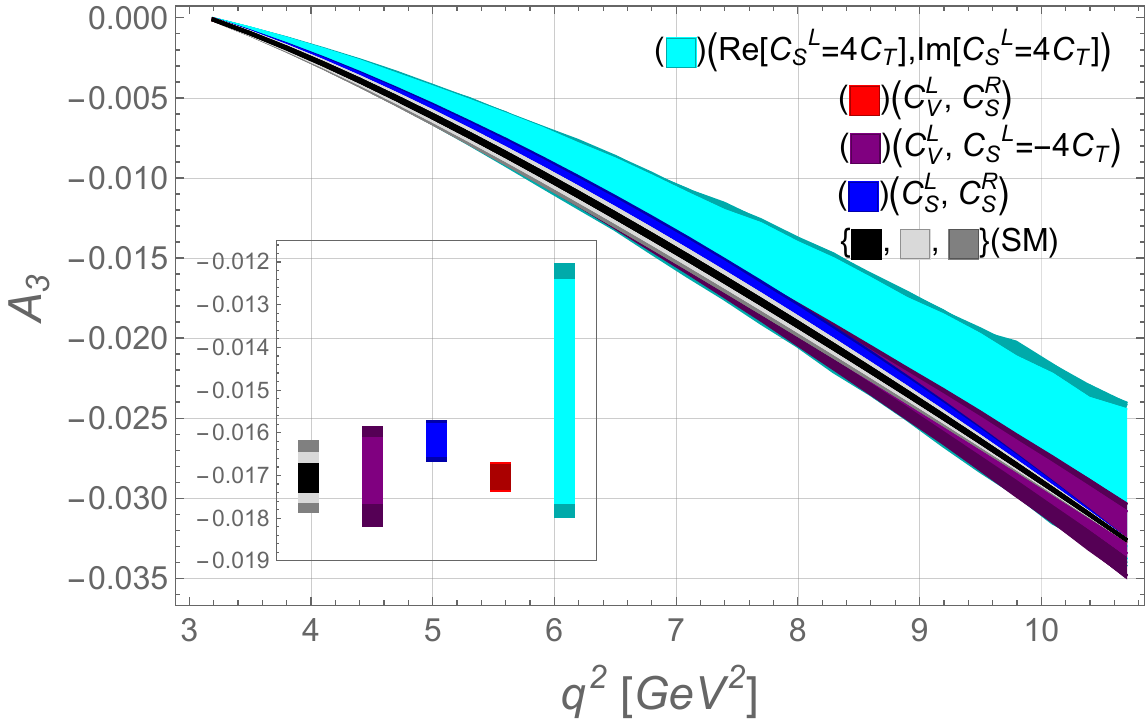 } 
\includegraphics[width=5cm,height=5cm]{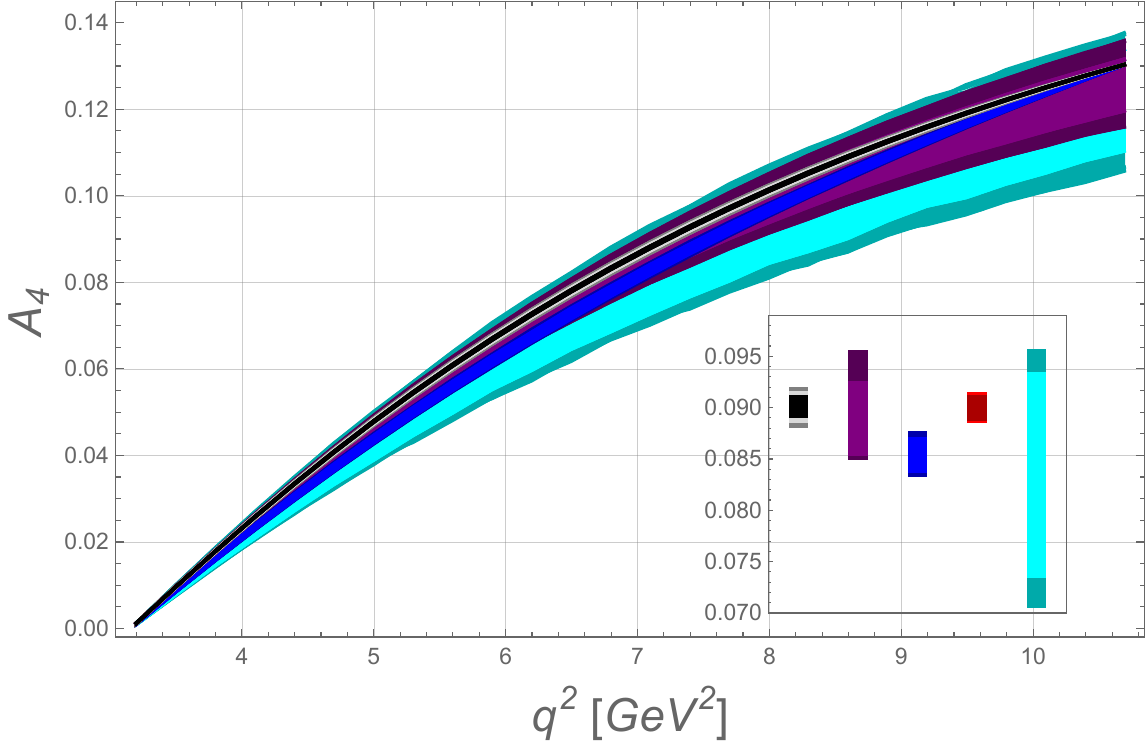 }
\includegraphics[width=5cm,height=5cm]{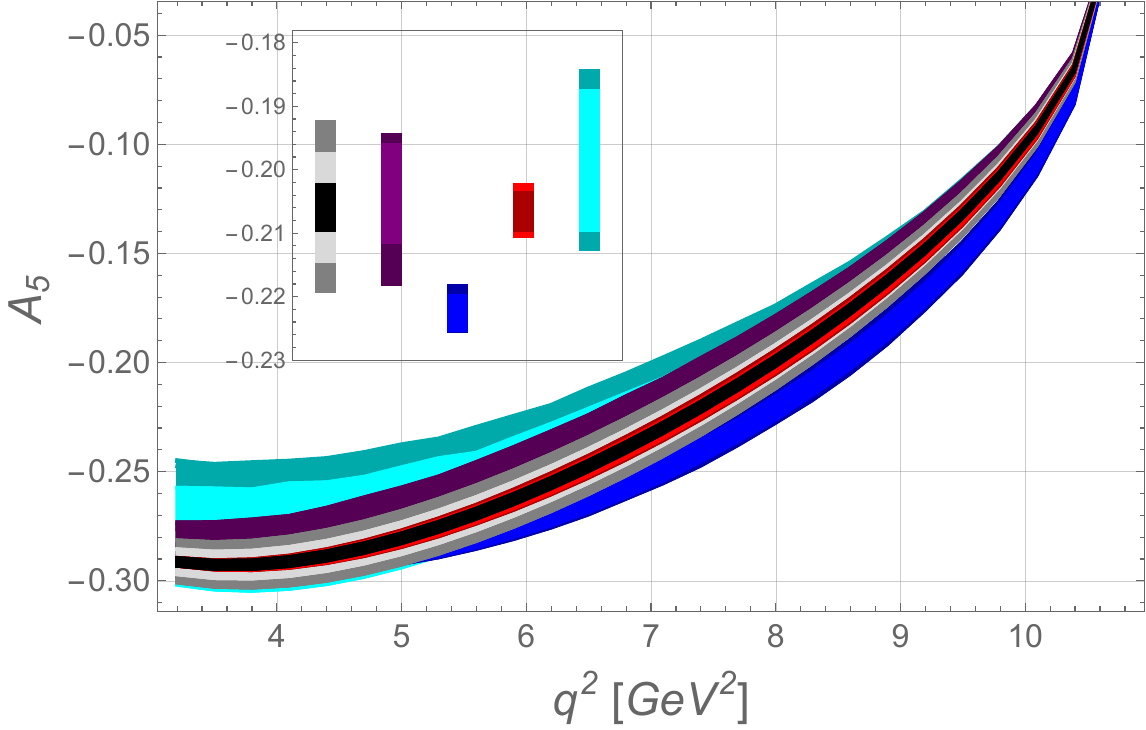}
\includegraphics[width=5cm,height=5cm]{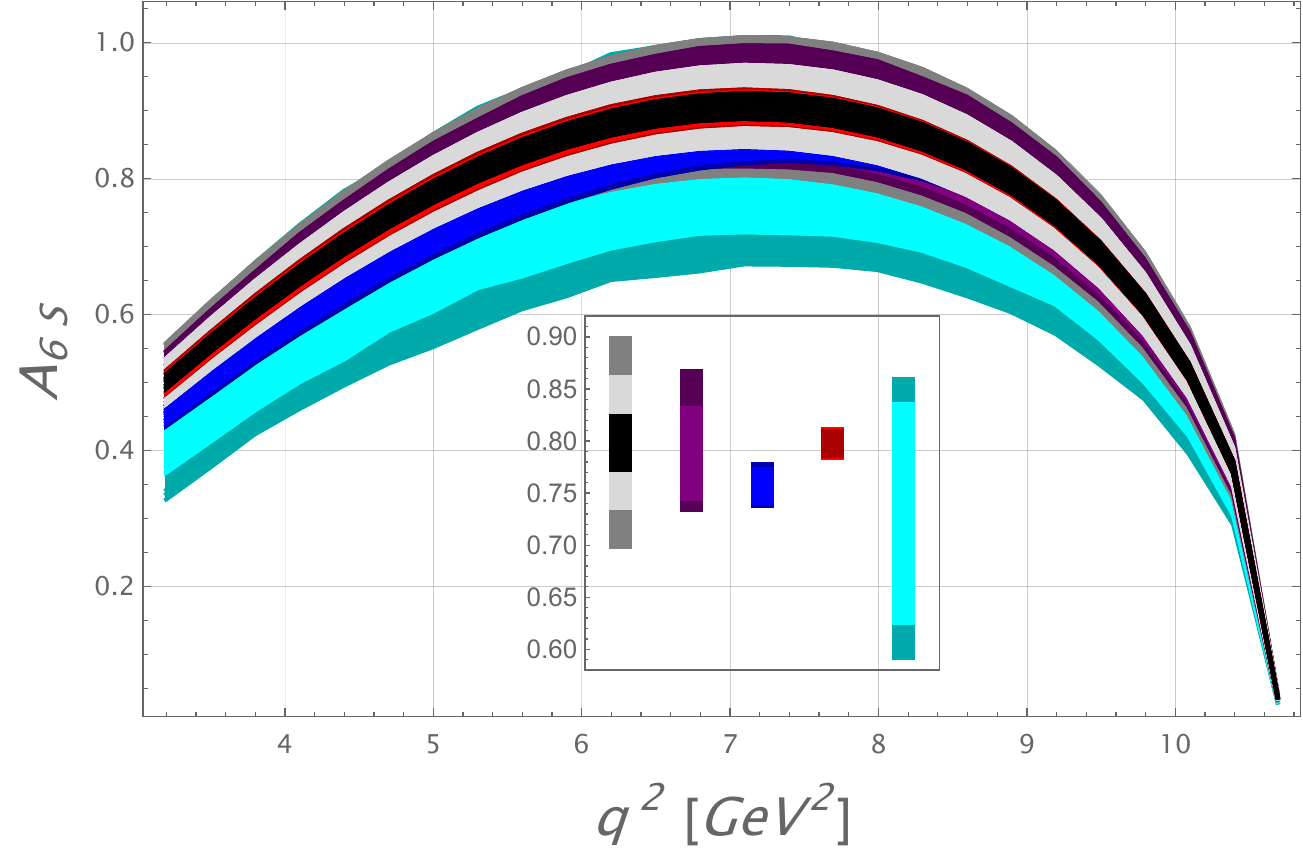}
\includegraphics[width=5cm,height=5cm]{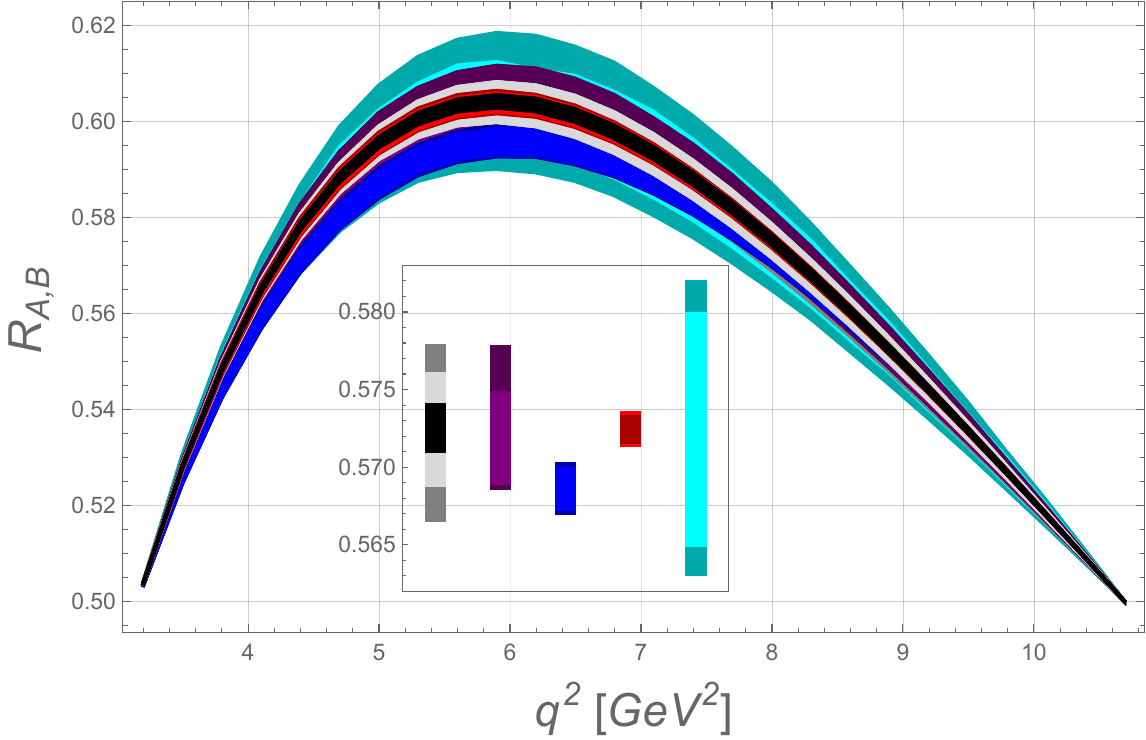}
\includegraphics[width=5cm,height=5cm]{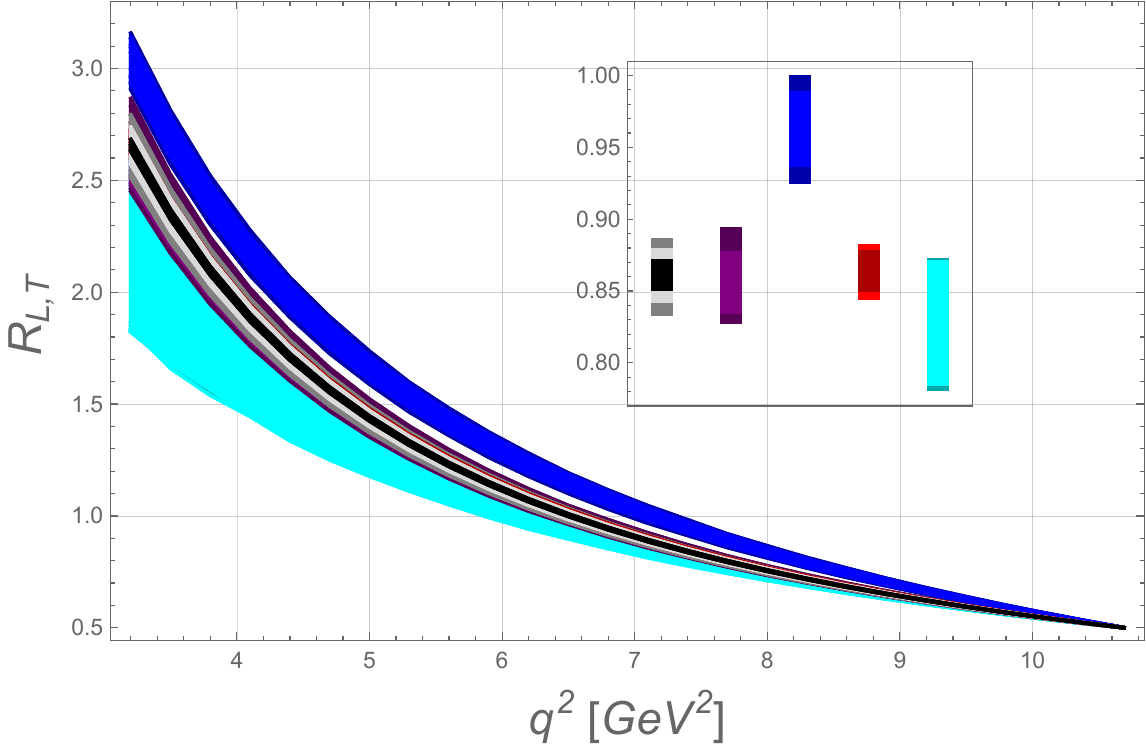}
\includegraphics[width=5cm,height=5cm]{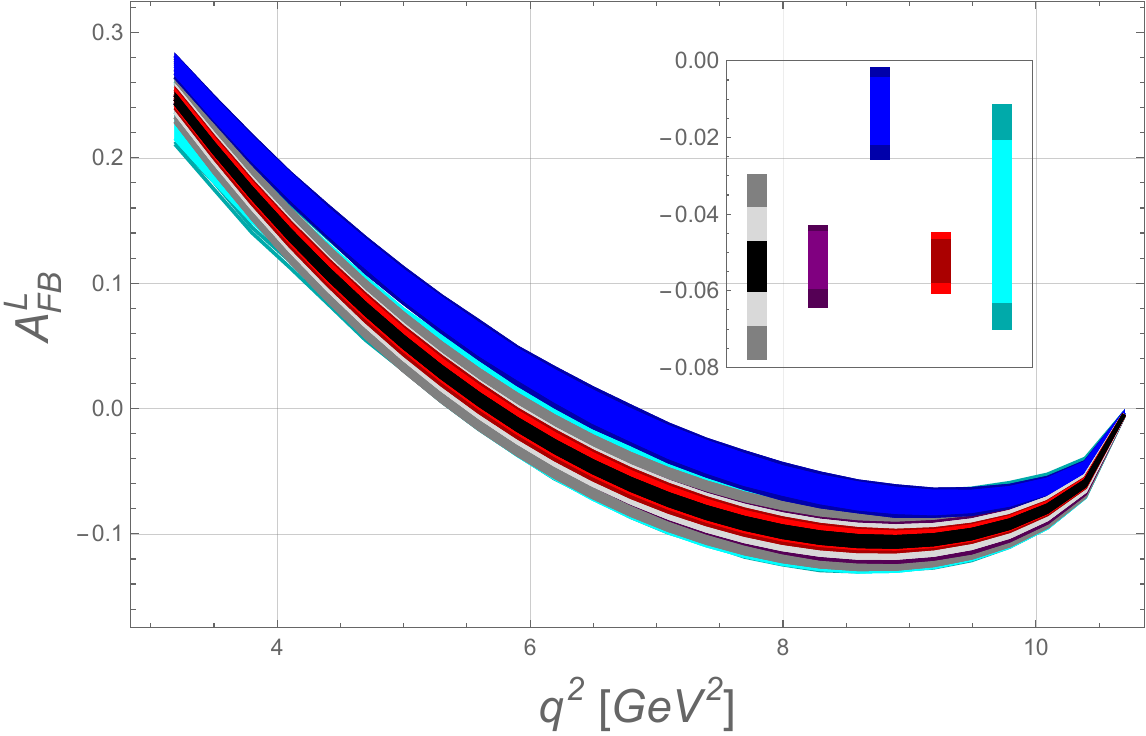 }
\includegraphics[width=5cm,height=5cm]{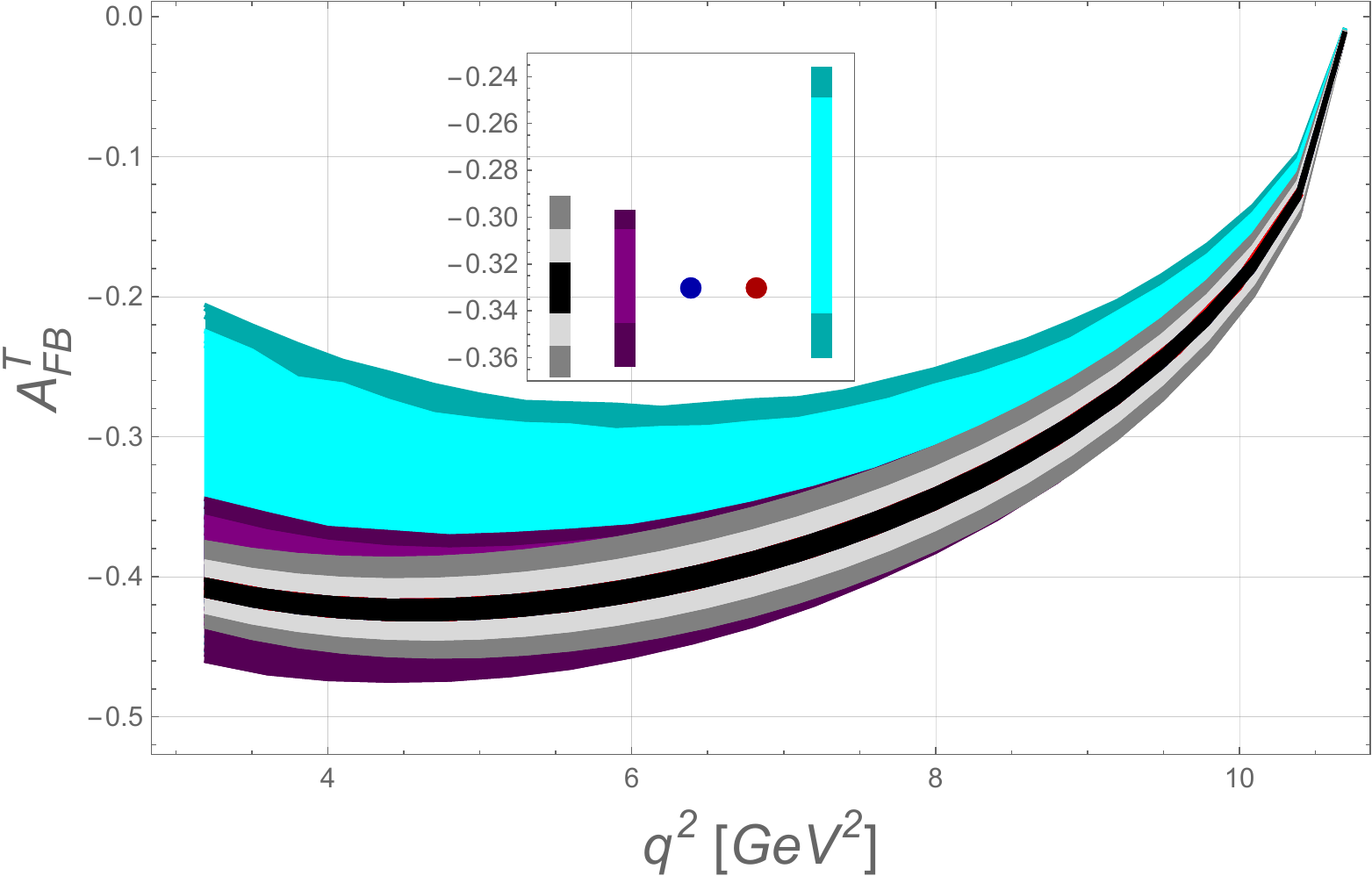}
\includegraphics[width=5cm,height=5cm]{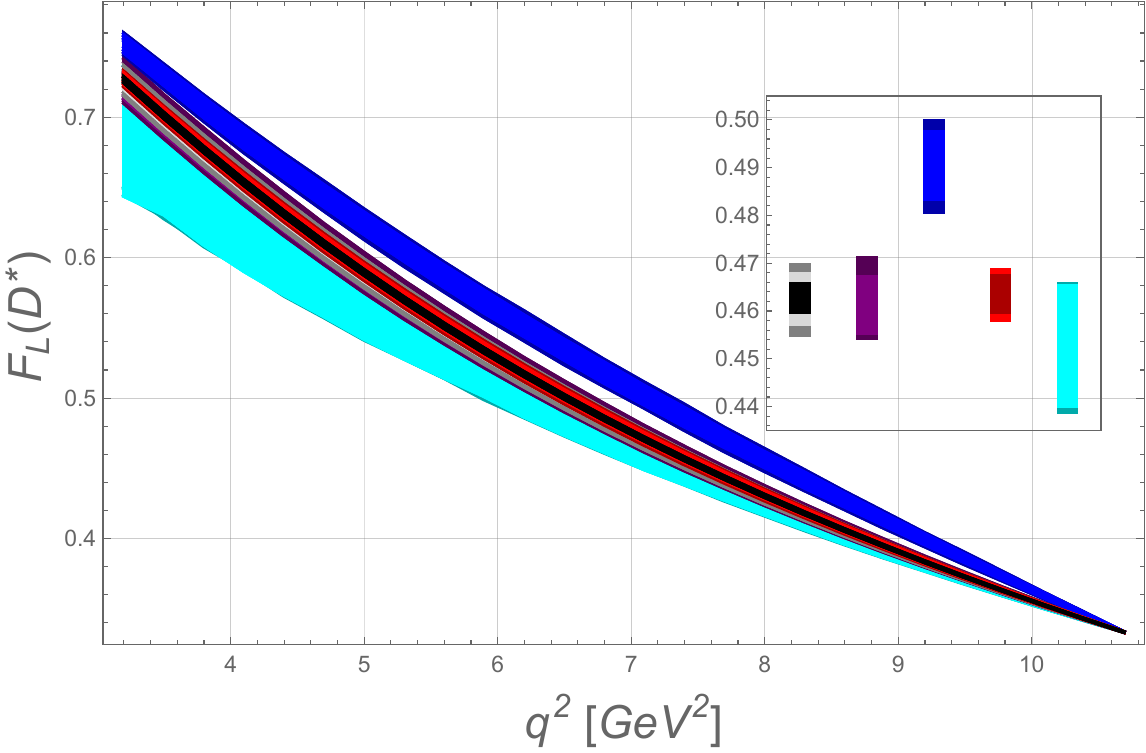}
\includegraphics[width=5cm,height=5cm]{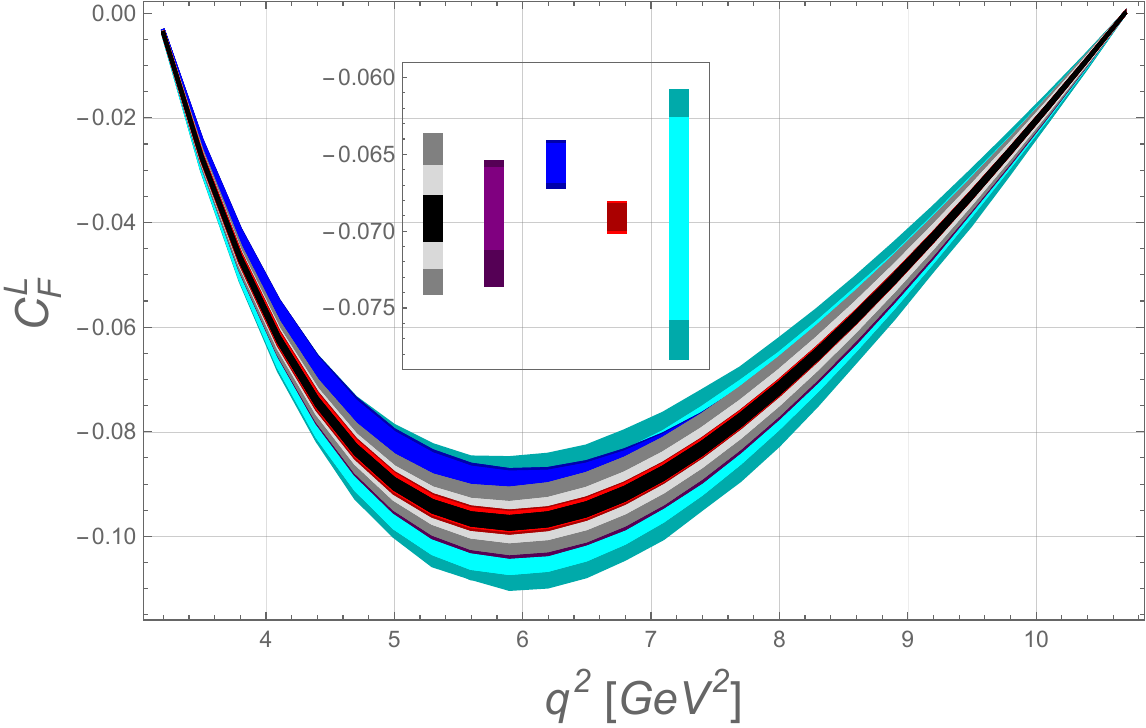}
\caption {The legends are same as described in FIG. \ref{1Dang} but for 2D scenarios.}
\label{2Dang}
\end{figure}

\begin{figure}[H]
 \centering{}
 \begin{subfigure}[b]{0.8\textwidth}
         \centering{}
\includegraphics[width=16cm,height=5cm]{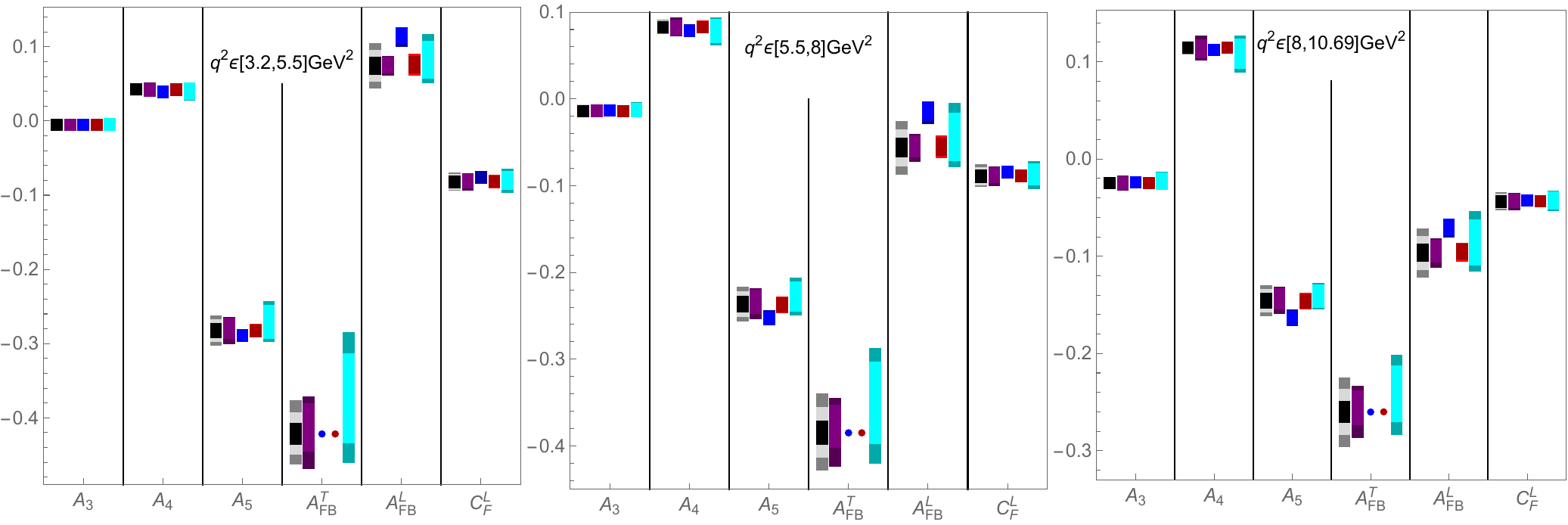}
    \hspace{1.5cm}
    \newline
    \includegraphics[width=16cm,height=5cm]{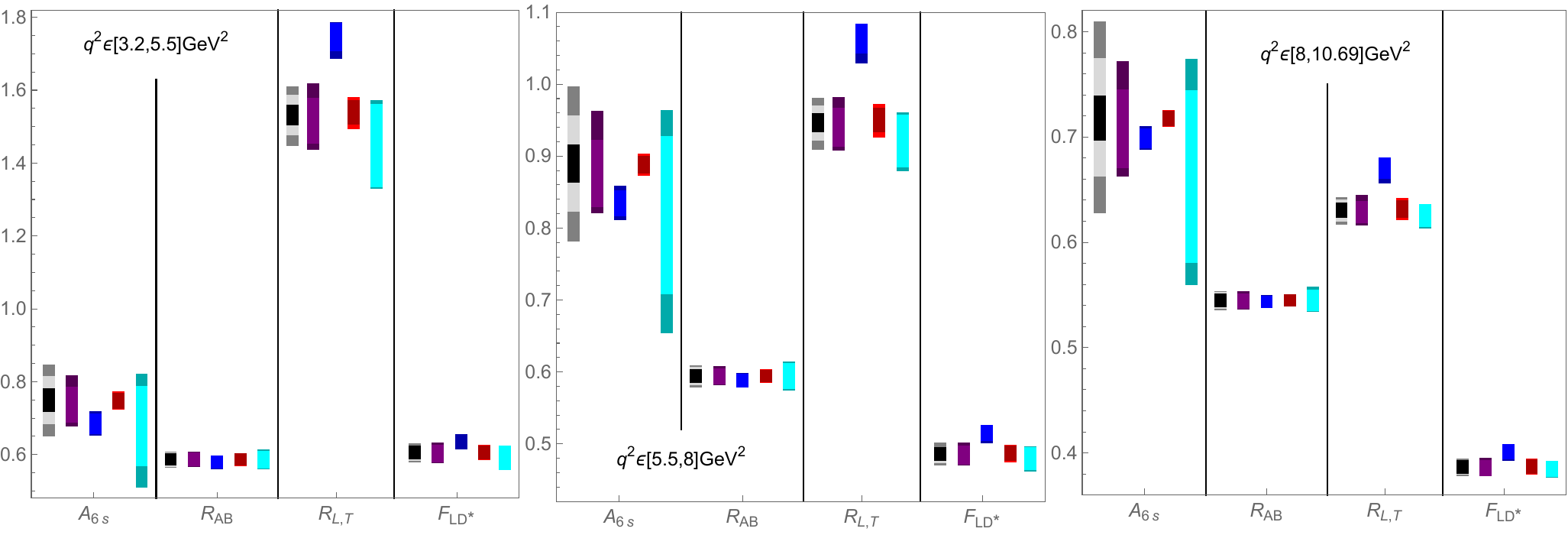}
     \end{subfigure}
\caption {The legends are same as described in FIG. \ref{1Dbar} but for 2D scenarios.}
\label{2Dbar}
\end{figure}

In order to keep the analysis consistent, we imposed the current collider bounds, as seen in Fig.\ref{2Dang}, to depict the angular observable. It is clear from the figure that the observables are not sensitive to the scenario $(C_{V}^L,C_{S}^R)$ (red band), whereas they are influenced by the other 2D scenarios, particularly $(Re[C_{S}^L=4C_T],\text{Im}[C_{S}^L=4C_T])$ has large effect (cyan band) on the values of the angular observables. One can also see that the effects of NP scenarios, $(Re[C_{S}^L=4C_T],\text{Im}[C_{S}^L=4C_T])$ (cyan band), $(C_{V}^L,C_S^L=-4C_T)$ (purple band) are increased (decreased) when  the $q^2$ value increases for $A_3$, $A_4$ $(A_5$, $R_{L,T}$, $F_L (D^*)$ and $A^T_{FB})$, whereas the observables $A_{6S}$, $R_{A,B}$, $A^L_{FB}$ and $C^L_F$ are largely effected in the middle of $q^2$ region. Therefore, we have not only plotted the variation in the amplitude due to the 2D NP scenarios when integrated over whole $q^2$ region but also in  different $q^2$ bins and shown in Fig. \ref{2Dbar}.  Moreover, the scalar coupling scenario $(C_S^L,C_S^R)$ (blue band), only increase (decrease) the values of $A_3$, $R_{L,T}$, $A^L_{FB}$, $F_L(D^*)$, $C_L^F$ ($A_{4}$, $A_{5}$, $A_{6S}$, $R_{A,B}$, $A^T_{FB}$) with respect to their SM values throughout the $q^2$ region. On the other hand, the scenarios $(Re[C_{S}^L=4C_T]$, $\text{Im}[C_{S}^L=4C_T])$, $(C_{V}^L,C_S^L=-4C_T)$, raise and lower the values of these observables values from their SM predictions throughout the $q^2$ region. In addition, to see the total variation in the magnitude of numerical values of different angular observables, we have also calculated their numerical values by using the 2$\sigma$ parametric space of 2D scenarios and listed them in Table \ref{SMang2} along with their SM results. 

\begin{table}[H]
\centering{}%
\scalebox{0.95}{
\begin{tabular}{|c|c|c|c|c|c|}
\toprule  
\hline
\hline
\vspace{0.05cm}
Observables  & SM   &$(C_{V}^{L},C_{S}^L=-4C_T)$ & $(C_{S}^L,C_{S}^R)$& $(C_{V}^L,C_{S}^L)$ &$(Re[C_S^L=4C_T],\text{Im}[C_S^L=4C_T])$  
\tabularnewline
\midrule
\hline
$A_{3}$  & $-0.0170\pm0.0001$  & $-0.0170\pm0.0001$ & $-0.0162\pm0.0001$&$-0.0001\pm0.0001$ &$-0.0148\pm0.003$ 
\tabularnewline
\midrule 
\hline
 $A_{4}$  & $0.0900\pm0.0002$  & $0.0890\pm0.007$ & $0.0856\pm0.0002$&$0.900\pm0.0005$ & $0.0828\pm0.009$
 \tabularnewline
\midrule 
\hline
$A_{5}$  & $-0.2059\pm0.002$  & $-0.2037\pm0.009 $ &  $-0.2233\pm0.0006$&  $-0.2063\pm0.003$&$-0.1935\pm0.007$
\tabularnewline
\midrule 
\hline
$A_{6s}$  & $0.7985\pm0.017$  & $0.7861\pm0.04 $ &  $0.7591\pm0.002$&$0.7977\pm0.006$ & $0.7240\pm0.011$
\tabularnewline
\midrule 
\hline
$R_{A,B}$  & $0.5725\pm0.001$  & $0.5717\pm0.008$  & $0.5688\pm0.0009$ &  $0.5724\pm0.0005$ & $0.5726\pm0.0009$
\tabularnewline
\midrule
\hline
$R_{L,T}$  &$0.8611\pm0.0036$ & $0.8559\pm0.002$ & $0.9760\pm0.003$& $0.8630\pm0.011$&$0.8257\pm0.011$
\tabularnewline
\midrule 
\hline
$A_{FB}^{T} $  &$-0.3302\pm0.006$  & $-0.3242\pm0.033$ &$-0.3302\pm0.006$ &$-0.3302\pm0.006$&$-0.2855\pm0.044$
 \tabularnewline
\midrule 
\hline
$A_{FB}^{L} $  & $-0.0536\pm0.004$  &  $-0.0518\pm0.008$ &  $-0.0149\pm0.008$ & $-0.0527\pm0.006$&$-0.0407\pm0.03$
\tabularnewline
\midrule 
\hline
 $F_{L}(D^{*}) $  & $0.4626\pm0.001$  & $0.4611\pm0.009$ &$0.4891\pm0.0001$    &$0.4632\pm0.003$ &$0.4522\pm0.002$
 \tabularnewline
\midrule 
\hline
$C_{F}^{L}$   &  $-0.0691\pm0.0009$  & $-0.0684\pm0.005$  & $-0.0657\pm0.0001$& $-0.0691\pm0.0004$ &$-0.0692\pm0.008$
\tabularnewline
\midrule
\hline
\bottomrule
\hline
\hline
\end{tabular}}
\caption{The SM and the NP values of angular observables at 2$\sigma$ allowed parametric space in the full
$q^2$ region for different 2D scenarios.}
\label{SMang2} 
\end{table}
\section{Summary and Conclusions}
In the current study, we have first checked the impact of recently measured data of $R({D^{(*)}})$ and $R(\Lambda_c)$ \cite{HFLAV:2023link} on different 1D and 2D NP scenarios that are considered in refs. \cite{Fedele:2022xyz, Blanke:2018yud,Blanke:2019qrx}. In addition, we have also included the $R(J/\psi)$ data in the analysis which is not considered in the previous studies and found that their influence on the best fit point and the parametric space are not significantly large. We have also validated (in light of recent data) the robustness of the sum rule of $R(\Lambda_c)$ and, similarly, found the sum rule for $R(J/\psi)$ in terms of $R({D^{(*)}})$. From this sum rule, we have also predicted the value of $R(J/\psi)$ which is smaller than its experimental value. Instead, the form factors of $R(J/\psi)$ are not precisely calculated but the difference between its experimental and theoretical values is quite large, therefore, to see the agility of the sum rule of $R(J/\psi)$ and the NP, it is mandatory to confirm the $R(J/\psi)$ value from further experiments. Furthermore, we have also modified the correlation among the different observables given in ref. \cite{Blanke:2018yud} according to the recent development in the data and also shown some more interesting correlations among observables. Finally, to discriminate the NP scenarios from each other, we have plotted the different angular observables against $q^2$, by using the $1\sigma$ and $2\sigma$ parametric space of NP scenarios with the 
 constraints of collider bounds. Further to see the influence of NP on the amplitude of the angular observables, we have also calculated their numerical values in different $q^2$ bins and shown them through the bar plots. We have found that angular observables are not only sensitive to  NP but also very fertile to find out the precise values of possible NP couplings, consequently, helping us in discriminating various NP scenarios. 
\section*{Acknowledgements}
The authors would like to express their sincere gratitude to Prof. Monika Blanke $et. al.$, and Dr. Nestor Quintero for valuable feedback on our queries during the course of this study which enable us to understand their work. 
\section{Appendix}
\subsection{Expressions of $I's$}
As mentioned in section \ref{Senseang}, the expressions of $I_i's$ can be found in \cite{Alok:2016qyh,Mandal:2020htr,Becirevic:2019tpx}. However, $I_i's$ after integrating over $q^2$ can be expressed in terms of $C^{L(R)}_i$ as follows. These expression are given as
\begin{equation}
\begin{aligned}
I_{1}^{s}=&6.02\times10^{-15} \left\{ |1+C_{V}^{L}|^{2}+|C_{V}^{R}|^{2}-1.63 \text{Re}\left[\left(1+C_{V}^{L}\right)\left(C_{V}^{R}\right)^{*}\right]-4.68 \text{Re}\left[\left(1+C_{V}^{L}\right)\left(C_{T}\right)^{*}\right] \right. \notag \\
& \left. +16.2|C_{T}|^{2} +7.6 \text{Re}\left[\left(C_{V}^{R}\right)\left(C_{T}\right)^{*}\right]\right\}, \notag \\
I_{1}^{c}=&9.05\times10^{-15}\left\{ \left|1+C_{V}^{L}-C_{V}^{R}\right|^{2}+0.08\left|1+C_{S}^{R}-C_{S}^{L}\right|^{2}-0.242 \text{Re}\left[\left(1+C_{V}^{L}-C_{V}^{R}\right)\left(C_{S}^{L}\right)^{*}\right] \right. \notag \\
&  \left. +0.242 \text{Re}\left[\left(1+C_{V}^{L}-C_{V}^{R}\right)\left(C_{S}^{R}\right)^{*}\right]-4.07 \text{Re}\left[\left(1+C_{V}^{L}(C_{V}^{R}\right)\left(C_{S}^{R}\right)^{*}\right]\right\} +6.24\left|C_{T}\right|^{2}\notag \\ 
I_{2}^{c}=&-5.43\times10^{-15}\left\{ \left|1+C_{V}^{L}-C_{V}^{R}\right|^{2}+8.31\left|C_{T}\right|^{2}\right\} ,\nonumber \\
I_{2}^{s}=&9.99\times10^{-16}\left\{ \left|1+C_{V}^{L}-C_{V}^{R}\right|^{2}-21.7\left|C_{T}\right|^{2}\right\} ,\nonumber \\
I_{3}=&1.67\times10^{-15}\left\{ -\left|1+C_{V}^{L}-C_{V}^{R}\right|^{2}-12.5\left|C_{T}\right|^{2}\right\} ,\nonumber \\
I_{4}=&2.23\times10^{-15}\left\{ -\left|1+C_{V}^{L}-C_{V}^{R}\right|^{2}-12\left|C_{T}\right|^{2}\right\} ,\nonumber \\
I_{5}=& 4.49\times10^{-15}\left\{ \left|1+C_{V}^{L}-C_{V}^{R}\right|^{2}-5.75\left|C_{T}\right|^{2}+2.04\left[\left|C_{S}^{L}-C_{S}^{R}\right|C_{T}^{*}\right]-1.26 \text{Re}\left[\left(C_{V}^{R}\right)\left(C_{T}\right)^{*}\right]\right\} ,\nonumber \\
I_{6}^{c}=&5.44\times10^{-15}\left\{ \left|1+C_{V}^{L}-C_{V}^{R}\right|^{2}-0.67 \text{Re}\left[\left(1+C_{V}^{L}-C_{V}^{R}\right)\left(C_{S}^{L*}\right)\right]+0.67 \text{Re}\left[\left(1+C_{V}^{L}-C_{V}^{R}\right)\left(C_{S}^{R*}\right)\right] \right. \nonumber\\
& \left. -3.76 \text{Re}\left[\left(1+C_{V}^{L}\right)\left(C_{T}\right)^{*}\right]+\left[\left|C_{S}^{L}-C_{S}^{R}\right|C_{T}^{*}\right]\right\}, \notag \\
I_{6}^{s}=&-3.85\times10^{-15}\left\{
\left|1+C_{V}^{L}\right|^{2}+\left|C_{V}^{R}\right|^{2}-17.5\left|C_{T}\right|^{2}-2.59 \text{Re} \left[ \left(1+C_{V}^{L}\right)\left(C_{T}\right)^{*}\right]
-8.54 \text{Re}\left[\left(C_{V}^{R}\right)\left(C_{T}\right)\right]\right\}. \notag
\label{IntIs}
\end{aligned}
\end{equation}


\begin{thebibliography}{99}
\bibitem{Blanke:2018yud} 
M.~Blanke, A.~Crivellin, S.~de Boer, T.~Kitahara, M.~Moscati, U.~Nierste and I.~Nisandzic,
Phys. Rev. D \textbf{99}, no.7, 075006 (2019)
arXiv:1811.09603 [hep-ph].
\bibitem{Fedele:2022xyz}
M. Fedele, M. Blanke, A. Crivellin, S. Iguro, T. Kitahara, U. Nierste and R. Watanabe(2022), arXiv:2211.14172 [hep-ph].
\bibitem{Blanke:2019qrx}
M.~Blanke, A.~Crivellin, T.~Kitahara, M.~Moscati, U.~Nierste and I.~Nisandzic,
Phys. Rev. D \textbf{100}, 035035 (2019)
arXiv:1905.08253 [hep-ph].

\bibitem{Dutta:2013qaa}
R.~Dutta, A.~Bhol and A.~K.~Giri,
Phys. Rev. D \textbf{88}, no.11, 114023 (2013)
arXiv:1307.6653 [hep-ph].
\bibitem{Dutta:2017xmj}
R.~Dutta and A.~Bhol,
Phys. Rev. D \textbf{96}, no.7, 076001 (2017)
arXiv:1701.08598 [hep-ph].
\bibitem{Dutta:2017wpq}
R.~Dutta,
arXiv:1710.00351 [hep-ph].
\bibitem{Dutta:2018jxz}
R.~Dutta and N.~Rajeev,
Phys. Rev. D \textbf{97}, no.9, 095045 (2018)
arXiv:1803.03038 [hep-ph].
\bibitem{Azatov:2018kzb}
A.~Azatov, D.~Barducci, D.~Ghosh, D.~Marzocca and L.~Ubaldi,
JHEP {\bf 1810} 092, (2018) 
arXiv:1807.10745 [hep-ph].
\bibitem{Heeck:2018ntp}
J.~Heeck and D.~Teresi,
JHEP {\bf 1812} 103, (2018) 
arXiv:1808.07492 [hep-ph].
\bibitem{Li:2016vvp} 
X.~Qiang~Li, Y.~D.~Yang and X.~Zhang,
{JHEP {\bf 1608}, 054 (2016)}
arXiv:1605.09308 [hep-ph].
\bibitem{Aaij:2014ora} 
R.~Aaij {\it et al.} [LHCb Collaboration],
{Phys.\ Rev.\ Lett.\  {\bf 113}, 151601 (2014)}
arXiv:1406.6482 [hep-ex].
\bibitem{Sakaki:2013bfa} 
Y.~Sakaki, M.~Tanaka, A.~Tayduganov and R.~Watanabe,
Phys.\ Rev.\ D {\bf 88}, no. 9, 094012 (2013)
arXiv:1309.0301 [hep-ph].
\bibitem{Caprini:1997mu} 
I.~Caprini, L.~Lellouch and M.~Neubert,
Nucl.\ Phys.\ B {\bf 530}, 153 (1998)
 [hep-ph/9712417].
\bibitem{Bailey:2014tva} 
J.~A.~Bailey {\it et al.} [Fermilab Lattice and MILC Collaborations],
Phys.\ Rev.\ D {\bf 89}, no. 11, 114504 (2014)
arXiv:1403.0635 [hep-lat].
\bibitem{Ligeti:2016npd}
Z.~Ligeti, M.~Papucci and D.~J.~Robinson,
JHEP {\bf 1701}  083, (2017)
arXiv:1610.02045 [hep-ph]. 

\bibitem{Abdesselam:2019wbt} 
A.~Abdesselam {\it et al.} [Belle Collaboration],
arXiv:1903.03102 [hep-ex].
\bibitem{Abdesselam:2019dgh}
A.~Abdesselam \textit{et al.}
arXiv:1904.08794 [hep-ex].
\bibitem{Aaij:2017tyk}
R.~Aaij  {\it et al.} [LHCb Collaboration], {Phys.\ Rev.\ Lett.\  {\bf 120}, 121801 (2018)}
arXiv:1711.05623 [hep-ex].
\bibitem{Murphy:2018sqg}
C.~W.~Murphay and A.~Soni, (2018)
arXiv:1808.05392 [hep-ph].

\bibitem{Kamali:2018fhr}
S.~Kamali, A.~Rashed, and A.~Datta, 
{Phys.\ Rev.\ D. {\bf 97},
 no. 9, 095034 (2018)},
arXiv:1801.08259 [hep-ph].
\bibitem{Grossman:1994ax}
Y.~Grossman and Z.~Ligeti,
{{Phys. Lett.} {\bf B332} 373--380 (1994) },
arXiv:hep-ph/9403376 [hep-ph].
\bibitem{Colangelo:2016ymy}
P.~Colangelo and F.~De~Fazio,
{Phys.\ Rev.\ D. {\bf 95} no.~1, 011701 (2017)}
arXiv:1611.07387 [hep-ph].
\bibitem{Aaij:2022abc}  
  R.~Aaij {\it et al.} [LHCb Collaboration],
  {Phys.\ Rev.\ Lett.\  {\bf128},no. 19, 191803 (2022)} 
 \bibitem{Detmold:2015aaa}
W. Detmold, C. Lehner, and S. Meinel, Phys. Rev.
D92, 034503 (2015), arXiv:1503.01421 [hep-lat]. 

\bibitem{Lees:2012xj} J.~P.~Lees {\it et al.} [BaBar Collaboration],
{\ Phys. \ Rev. \ Lett. {\bf 109}, 101802 (2012)}
arXiv:1205.5442 [hep-ex].
\bibitem{Lees:2013uzd} J.~P.~Lees {\it et al.} [BaBar Collaboration],
{\ Phys.\ Rev.\ D {\bf 88},no. 7, 072012 (2013)}
 arXiv:1303.0571 [hep-ex].
\bibitem{Huschle:2015rga} 
M.~Huschle {\it et al.} [Belle Collaboration],
{Phys.\ Rev.\ D {\bf 92}, no. 7, 072014 (2015)}
arXiv:1507.03233 [hep-ex].
\bibitem{Sato:2016svk} 
Y.~Sato {\it et al.} [Belle Collaboration],
 {Phys.\ Rev.\ D {\bf 94}, no. 7, 072007 (2016)}
 arXiv:1607.07923 [hep-ex].
\bibitem{Abdesselam:2016xqt} 
A.~Abdesselam {\it et al.},
arXiv:1608.06391 [hep-ex].
\bibitem{G.Caria:2019ff}
G. Caria et al.,[Belle collaboration] 
Phys. Rev. Lett. 124 (2020) 161803, arXiv:1910.05864.
\bibitem{Belle:2019rba}
G. Caria {\it et al.} [Belle Collaboration],
{\ Phys. \ Rev. \ Lett. {\bf 124}, n0. 16, 161803 (2020)}
arXiv:1910.05864 [hep-ex].  
\bibitem{Aaij:2015yra} 
R.~Aaij {\it et al.} [LHCb Collaboration],
{Phys.\ Rev.\ Lett.\  {\bf 115}, no. 11, 111803 (2015)} 
arXiv:1506.08614 [hep-ex].
\bibitem{Aaij:2017uff} 
R.~Aaij {\it et al.} [LHCb Collaboration],
arXiv:1708.08856 [hep-ex].
\bibitem{HFLAV:2023link} Preliminary average of $R(D)$ and $R(D^{*})$
for Moriond 2024, \url{https://hflav-eos.web.cern.ch/hflav-eos/semi/moriond24/html/RDsDsstar/RDRDs.html}.
%
\bibitem{Fajfer:2012vx} 
S.~Fajfer, J.~F.~Kamenik and I.~Nisandzic,
{Phys.\ Rev.\ D {\bf 85}, 094025 (2012)}
arXiv:1203.2654 [hep-ph].
\bibitem{Kamenik:2008tj} 
J.~F.~Kamenik and F.~Mescia,
Phys.\ Rev.\ D {\bf 78}, 014003 (2008)
arXiv:0802.3790 [hep-ph].
 \bibitem{Amhis:2016xyh}
 Y.~Amhis {\it et al.} [Heavy Flavor Averaging Collaboration],  Eur.\ Phys.\ J.\ C {\bf 77},no.12, 895 (2017) 
  arXiv:1612.07233 [hep-ph].
\bibitem{Bailey:2012jg}
J. A.~Bailey {\it et al.}, {Phys.\ Rev.\ Lett. {\bf 109}, 071802 (2012)}
arXiv:1206.4992 [hep-ph].
\bibitem{Bailey:2015xy}
 J. A.~Bailey {\it et al.} [MILC Collaboration] {Phys.\ Rev.\ D. {\bf 92}, no.3, 034506 (2015)}
 arXiv:1503.07237 [hep-ph].
\bibitem{Aoki:2016frl}
S.~Aoki {\it et al.}  Eur.\ Phys.\ J.\ C {\bf 77}, no. 2, 112 (2017) 
arXiv:1607.00299 [hep-lat].
\bibitem{Hirose:2016wfn}
S.~Hirose {\it et al.}[BELLE], {Phys.\ Rev.\ Lett. {\bf 118}, 211801 (2017)}
arXiv:1612.00529 [hep-ex].
\bibitem{Hirose:2017dxl} 
 S.~Hirose {\it et al.}[BELLE], {Phys.\ Rev.\ D. {\bf 97}, 012004 (2018)}
arXiv:1709.00529 [hep-ex]
\bibitem{Tanaka:2012nw} 
M.~Tanaka and R.~Watanabe,
{Phys.\ Rev.\ D {\bf 87}, no. 3, 034028 (2013)}
arXiv:1212.1878 [hep-ph].
  
\bibitem{Asadi:2018sym} 
P.~Asadi, M.~R.~Buckley and D.~Shih,
arXiv:1810.06597 [hep-ph].
\bibitem{Alok:2016qyh}
A.~K.~Alok, D.~Kumar, S.~Kumbhakar, and S.~U.~Sankar, {Phys.\ Rev.\ D. {\bf 95}, 115038 (2017)}
arXiv:1606.03164 [hep-ph].
\bibitem{Tanaka:2010se} 
M.~Tanaka and R.~Watanabe,
{Phys.\ Rev.\ D {\bf 82}, 034027 (2010)}
arXiv:1005.4306 [hep-ph].

\bibitem{LHCbRJpsi}
LHCb-PAPER-2017-035. 
\bibitem{LHCbStatus}
 Presentation by M.~Fontana, on behalf of LHCb Collaboration, 
 [talk slide].
\bibitem{Watanabe:2017mip}
R.~Watanabe,  {Phys.\ Rev.\ Lett. {\bf B 776}, no. 5 (2018)}
arXiv:1709.08644 [hep-ph].
\bibitem{Chauhan:2017uil}
B.~Chauhan and B.~Kindra, (2017)
arXiv:1709.09989 [hep-ph].
\bibitem{Cohen:2018dgz}
T.~D.~Cohen,H.~Lamm, and R.~F.~Lebed,JHEP {\bf 09}, 168 (2018)
arXiv:1807.02730 [hep-ph].
\bibitem{Tran:2018kuv}
C.~T.~Tran, M.~A.~Ivanov, J.~G.~Korner, and P.~Santorelli, {Phys.\ Rev.\ D. {\bf 97}, 054014 (2018)}
 arXiv:1801.06927 [hep-ph]. 

\bibitem{Issadykov:2023afe}
A.~Issadykov and M.~A.~Ivanov,
Phys. Part. Nucl. Lett. \textbf{20}, no.3, 355-359 (2023)
[arXiv:2307.05013 [hep-ph]].
\bibitem{Azizi:2019aaf}
K.~Azizi, Y.~Sarac and H.~Sundu,
Phys. Rev. D \textbf{99}, no.11, 113004 (2019)
doi:10.1103/PhysRevD.99.113004
[arXiv:1904.08267 [hep-ph]].






\bibitem{Adamczyk}
K.~Adamczyk,
(2018) talk at 10th International Workshop on the CKM unitarity Triangle, Heidelberg, 17-21 Sep. 2018.
\bibitem{Celis:2016azn} 
A.~Celis, M.~Jung, X.~Q.~Li and A.~Pich,
{Phys.\ Lett.\ B {\bf 771}, 168 (2017)}
arXiv:1612.07757 [hep-ph].
  
\bibitem{Asadi:2018wea} 
  P.~Asadi, M.~R.~Buckley and D.~Shih,
  JHEP {\bf 1809}, 010 (2018)
  [arXiv:1804.04135 [hep-ph]].
\bibitem{Iguro:2018vqb} 
  S.~Iguro, T.~Kitahara, Y.~Omura, R.~Watanabe and K.~Yamamoto,
JHEP {\bf 1902}, 194 (2019)
arXiv:1811.08899 [hep-ph].
\bibitem{Alonso:2016oyd} 
R.~Alonso, B.~Grinstein and J.~Martin Camalich,
{Phys.\ Rev.\ Lett.\  {\bf 118}, no. 8, 081802 (2017)}
arXiv:1611.06676 [hep-ph].
\bibitem{Akeroyd:2017mhr} 
A.~G.~Akeroyd and C.~H.~Chen,
{Phys.\ Rev.\ D {\bf 96}, no. 7, 075011 (2017)}
arXiv:1708.04072 [hep-ph].
\bibitem{Gershtein:1994jw}
S. S. Gershtein, V. V. Kiselev, A. K. Likhoded, and
A. V. Tkabladze, Phys. Usp. 38, 1 (1995),
\bibitem{Bigi:1995fs}
I. I. Y. Bigi, Phys. Lett. B371, 105 (1996), arXiv:hep-ph/9510325 [hep-ph].
\bibitem{Beneke:1996xe}
M. Beneke and G. Buchalla, Phys. Rev. D53, 4991
(1996), arXiv:hep-ph/9601249 [hep-ph].
\bibitem{Chang:2000ac}
C.-H. Chang, S.-L. Chen, T.-F. Feng, and X.-Q. Li,
Phys. Rev. D64, 014003 (2001), arXiv:hep-ph/0007162 [hep-ph].
\bibitem{Kiselev:2000pp}
V. V. Kiselev, A. E. Kovalsky, and A. K. Likhoded,
Nucl. Phys. B585, 353 (2000), arXiv:hep-ph/0002127
[hep-ph].

\bibitem{Kamali:2018bdp} 
S.~Kamali,
Int.\ J.\ Mod.\ Phys.\ A {\bf 34}, no. 06, 1950036 (2019)
arXiv:1811.07393 [hep-ph]. 

\bibitem{Shi:2019gxi} 
R.~X.~Shi, L.~S.~Geng, B.~Grinstein, S.~Jager and J.~Martin Camalich,
JHEP {\bf 1912}, 065 (2019)
arXiv:1905.08498 [hep-ph].
\bibitem{Bhattacharya:2011qm}
T.~Bhattacharya, V.~Cirigliano, S.~D. Cohen, A.~Filipuzzi,
M.~Gonzalez-Alonso, et~al.Phys. Rev. D \textbf{85}, 054512,(2012)
arXiv:1110.6448 [hep-ph].
\bibitem{Antonelli:2008jg}
FlaviaNet Working Group on Kaon Decays Collaboration, M.~Antonelli et~al.,
Italy, 7-10 April 2008, 2008.
arXiv:0801.1817  [hep-ph].
\bibitem{Becirevic:2019tpx}
B. Damir, M. Fedele, 
 I. Nisandzic and A. Tayduganov
arXiv:1907.02257 [hep-ph].
\bibitem{Zhang:2020dla}
L.~Zhang, X.~W.~Kang, X.~H.~Guo, L.~Y.~Dai, T.~Luo and C.~Wang,
JHEP \textbf{02}, 179 (2021)
doi:10.1007/JHEP02(2021)179
[arXiv:2012.04417 [hep-ph]].
70B
\bibitem{Faustov:2022ybm}
R.~N.~Faustov, V.~O.~Galkin and X.~W.~Kang,
Phys. Rev. D \textbf{106}, no.1, 013004 (2022)
doi:10.1103/PhysRevD.106.013004
[arXiv:2206.10277 [hep-ph]].
\bibitem{Cardozo:2020abc}
J. Cardozo, J. H. Munoz, N. Quintero, E.Rojas,
J. Phys. G: Nucl. Part. Phys. 48, 035001 (2021). arXiv:2006.07751 [hep-ph]
\bibitem{Gomez:2019xfw}
J.~D.~Gomez, N.~Quintero and E.~Rojas,
Phys.\ Rev.\ D {\bf 100} (2019) no.9,  093003
arXiv:1907.08357 [hep-ph].
\bibitem{Cui:2023bzr}
B.~Y.~Cui, Y.~K.~Huang, Y.~M.~Wang and X.~C.~Zhao,
[arXiv:2301.12391 [hep-ph]].

\bibitem{Iguro:2022yzr}
S.~Iguro, T.~Kitahara and R.~Watanabe,
[arXiv:2210.10751 [hep-ph]].

\bibitem{RJSHI:12}
LATTICE-HPQCD Collaboration, 
Phys. Rev. Lett. 125 (2020) 222003 [arXiv:2007.06956].

\bibitem{Harrison:2020gvo}
 HPQCD Collaboration,
Phys. Rev. D 102 (2020) 094518 [arXiv:2007.06957]

\bibitem{LHCb:2022piu}
 LHCb Collaboration, Phys. Rev. Lett. 128 (2022) 191803
[arXiv:2201.03497].


\bibitem{belle:2}
Belle II Collaboration: presented at Lepton Photon 2023 [Lepton Photon 2023]
\bibitem{LHCB:1}
LHCb Collaboration: accepted by PRL [arXiv:2302.02886]
\bibitem{LHCB:2}
LHCb Collaboration: accepted by PRD [arXiv:2305.01463]
\bibitem{LHCb:2017vlu}
 R. Aaij et al. (LHCb), Phys. Rev. Lett. 120, 121801
(2018), arXiv:1711.05623 [hep-ex].
\bibitem{Mandal:2020htr} R.~Mandal, C.~Murgui, A.~Penuelas and A.~Pich,
JHEP \textbf{08}, 022 (2020)
arXiv:2004.06726 [hep-ph].

\bibitem{Murgui:2019czp}
C.~Murgui, A.~Pe\~nuelas, M.~Jung and A.~Pich,
JHEP \textbf{09}, 103 (2019)
doi:10.1007/JHEP09(2019)103
[arXiv:1904.09311 [hep-ph]].

\bibitem{Jung:2018a}
M. Jung and D. M. Straub, 
JHEP 01 (2019) 009,
arXiv:1801.01112 [hep-ph].

\bibitem{Gonzalez-Alonso:2017iyc} M.~Gonzalez-Alonso, J.~Martin Camalich and K.~Mimouni,
Phys. Lett. B \textbf{772} (2017), 777-785
[arXiv:1706.00410 [hep-ph]].
\bibitem{Datta:2019abc} A. Datta, S. Kamali, S. Meinel, and A. Rashed,
JHEP 08 (2017) 131 [arXiv:1702.02243 [hep-ph]].



\bibitem {Alonso:2015sja}
R. Alonso, B. Grinstein,and J. Martin~Camalich, {JHEP}\textbf {10}, {184} (2015){arXiv:1505.05164 [hep-ph]} 
\bibitem{Calibbi:2015kma} L. Calibbi, A. Crivellin, and T. Ota, Phys. Rev. Lett.
115, 181801 (2015), arXiv:1506.02661 [hep-ph]. 
\bibitem{Fajfer:2015ycq} S. Fajfer and N. Koˇsnik, Phys. Lett. B755, 270 (2016),
arXiv:1511.06024 [hep-ph].
\bibitem{Barbieri:2015yvd} R. Barbieri, G. Isidori, A. Pattori, and F. Senia, Eur.
Phys. J. C76, 67 (2016), arXiv:1512.01560 [hep-ph].
\bibitem{Barbieri:2016las} R. Barbieri, C. W. Murphy, and F. Senia, Eur. Phys.
J. C77, 8 (2017), arXiv:1611.04930 [hep-ph]
\bibitem{Hiller:2016kry} G. Hiller, D. Loose, and K. Schonwald, JHEP 12, 027
(2016), arXiv:1609.08895 [hep-ph].
\bibitem{Bhattacharya:2016mcc} B. Bhattacharya, A. Datta, J.-P. Gu´evin, D. London, and R. Watanabe, JHEP 01, 015 (2017), arXiv:1609.09078 [hep-ph].
\bibitem{Buttazzo:2017ixm} D. Buttazzo, A. Greljo, G. Isidori, and D. Marzocca, JHEP 11, 044 (2017), arXiv:1706.07808 [hep-ph]
\bibitem{Kumar:2018kmr} J. Kumar, D. London, and R. Watanabe, (2018),
arXiv:1806.07403 [hep-ph]
\bibitem{Assad:2017iib} N. Assad, B. Fornal, and B. Grinstein, Phys. Lett.
B777, 324 (2018), arXiv:1708.06350 [hep-ph].
\bibitem{DiLuzio:2017vat} L. Di Luzio, A. Greljo, and M. Nardecchia, Phys. Rev.
D96, 115011 (2017), arXiv:1708.08450 [hep-ph].
\bibitem{Calibbi:2017qbu} L. Calibbi, A. Crivellin, and T. Li, (2017),
arXiv:1709.00692 [hep-ph].
\bibitem{Bordone:2017bld}
M. Bordone, C. Cornella, J. Fuentes-Martin,
and G. Isidori, Phys. Lett. B779, 317 (2018),
arXiv:1712.01368 [hep-ph]
\bibitem{Barbieri:2017tuq}
R. Barbieri and A. Tesi, Eur. Phys. J. C78, 193 (2018),
arXiv:1712.06844 [hep-ph]
\bibitem{Blanke:2018sro} M. Blanke and A. Crivellin, Phys. Rev. Lett. 121,
011801 (2018), arXiv:1801.07256 [hep-ph].
\bibitem{Greljo:2018tuh} A. Greljo and B. A. Stefanek, Phys. Lett. B782, 131
(2018), arXiv:1802.04274 [hep-ph].
\bibitem{Bordone:2018nbg} M. Bordone, C. Cornella, J. Fuentes-Mart´ın, and
G. Isidori, (2018), arXiv:1805.09328 [hep-ph]
\bibitem{Matsuzaki:2018jui}
 S. Matsuzaki, K. Nishiwaki, and K. Yamamoto, (2018),
arXiv:1806.02312 [hep-ph].
\bibitem{Crivellin:2018yvo}
A. Crivellin, C. Greub, F. Saturnino, and D. Muller,
(2018), arXiv:1807.02068 [hep-ph]
\bibitem{DiLuzio:2018zxy}
 L. Di Luzio, J. Fuentes-Martin, A. Greljo, M. Nardecchia, and S. Renner, (2018), arXiv:1808.00942 [hepph]
\bibitem{Biswas:2018snp}
 A. Biswas, D. Kumar Ghosh, N. Ghosh, A. Shaw, and
A. K. Swain, (2018), arXiv:1808.04169 [hep-ph]
\bibitem{Deshpande:2012rr} N. G. Deshpande and A. Menon, JHEP 01, 025 (2013),
arXiv:1208.4134 [hep-ph].
[hep-ph].
\bibitem{Bauer:2015knc} 
M. Bauer and M. Neubert, Phys. Rev. Lett. 116, 141802
(2016), arXiv:1511.01900 [hep-ph]
\bibitem{Cai:2017wry}
Y. Cai, J. Gargalionis, M. A. Schmidt, and R. R.
Volkas, JHEP 10, 047 (2017), arXiv:1704.05849 [hepph]
\bibitem{Crivellin:2017zlb} 
A. Crivellin, D. Muller, and T. Ota, JHEP 09, 040
(2017), arXiv:1703.09226 [hep-ph]
\bibitem{Altmannshofer:2017poe}
W. Altmannshofer, P. Bhupal Dev, and A. Soni, Phys.
Rev. D96, 095010 (2017), arXiv:1704.06659 [hep-ph].
\bibitem{Marzocca:2018wcf}
 D. Marzocca, JHEP 07, 121 (2018), arXiv:1803.10972
[hep-ph]



\bibitem{He:2012zp}
X.-G. He and G. Valencia, Phys. Rev. D87, 014014
(2013), arXiv:1211.0348 [hep-ph].
\bibitem{Greljo:2015mma}
A. Greljo, G. Isidori, and D. Marzocca, JHEP 07, 142
(2015), arXiv:1506.01705 [hep-ph].
\bibitem{Boucenna:2016wpr}
S. M. Boucenna, A. Celis, J. Fuentes-Martin, A. Vicente, and J. Virto, Phys. Lett. B760, 214 (2016),
arXiv:1604.03088 [hep-ph]
\bibitem{He:2017bft}
X.-G. He and G. Valencia, Phys. Lett. B779, 52 (2018),
arXiv:1711.09525 [hep-ph]
\bibitem{Kalinowski:1990ba}
J. Kalinowski, Phys. Lett. B245, 201 (1990).
\bibitem{Hou:1992sy}
W.-S. Hou, Phys. Rev. D48, 2342 (1993).
\bibitem{Kosnik:2012dj}
N. Kosnik, Phys. Rev. D86, 055004 (2012),
arXiv:1206.2970 [hep-ph]
\bibitem{Biswas:2018iak}
A. Biswas, A. Shaw, and A. K. Swain, (2018),
arXiv:1811.08887 [hep-ph].
\bibitem{Crivellin:2012ye}
A. Crivellin, C. Greub, and A. Kokulu, Phys. Rev.
D86, 054014 (2012), arXiv:1206.2634 [hep-ph].
\bibitem{Crivellin:2013wna}
A. Crivellin, A. Kokulu, and C. Greub, Phys. Rev.
D87, 094031 (2013), arXiv:1303.5877 [hep-ph].

\bibitem{Celis:2012dk}
A. Celis, M. Jung, X.-Q. Li, and A. Pich, JHEP 01,
054 (2013), arXiv:1210.8443 [hep-ph].
\bibitem{Ko:2012sv}
 P. Ko, Y. Omura, and C. Yu, JHEP 03, 151 (2013),
arXiv:1212.4607 [hep-ph].
\bibitem{Crivellin:2015hha}
A. Crivellin, J. Heeck, and P. Stoffer, Phys. Rev. Lett.
116, 081801 (2016), arXiv:1507.07567 [hep-ph].
\bibitem{Dhargyal:2016eri}
L. Dhargyal, Phys. Rev. D93, 115009 (2016),
arXiv:1605.02794 [hep-ph].
\bibitem{Chen:2017eby}
C.-H. Chen and T. Nomura, Eur. Phys. J. C77, 631
(2017), arXiv:1703.03646 [hep-ph].
\bibitem{Iguro:2017ysu} S. Iguro and K. Tobe, Nucl. Phys. B925, 560 (2017),
arXiv:1708.06176 [hep-ph]
\bibitem{Martinez:2018ynq} 
R. Martinez, C. F. Sierra, and G. Valencia, (2018),
arXiv:1805.04098 [hep-ph].
\bibitem{Biswas:2018jun}
A. Biswas, D. K. Ghosh, A. Shaw, and S. K. Patra,
(2018), arXiv:1801.03375 [hep-ph].
\bibitem{Greljo:2018tzh}
A. Greljo, J. Martin Camalich, and J. D. Ruiz-Alvarez, ´
Phys. Rev. Lett. 122, 131803 (2019), arXiv:1811.07920
[hep-ph].

\bibitem{Endo:2021lhi}
M.~Endo, S.~Iguro, T.~Kitahara, M.~Takeuchi, and R.~Watanabe, 
{JHEP {\bfseries 02} (2022)106}
arXiv:2111.04748 [hep-ph].

\bibitem{Iguro:2020keo}
S.~Iguro, M.~Takeuchi, and R.~Watanabe,
{Eur.\  Phys.\  J.\ C {\bfseries 81} (2021) 406}
arXiv:2011.02486[hep-ph].
\bibitem{Leljak:2019eyw}
D. Leljak, B. Melic, and M. Patra,
JHEP 05 (2019) 094 arXiv:1901.08368[hep-ph].



 








 







\end{thebibliography}
\end{document}